\documentclass[runningheads]{llncs}
\usepackage[T1]{fontenc}
\usepackage{graphicx}
\usepackage{lipsum}
\usepackage{pgfplots}
\usepackage{amsmath}
\usepackage{svg}
\usepackage{subcaption}
\usepackage{booktabs} 
\usepackage{tabularx}
\usepackage{hyperref}
\usepackage[percent]{overpic}
\pgfplotsset{compat=1.18}
\begin{document}
\title{Conditional Fetal Brain Atlas Learning for Automatic Tissue Segmentation}
\titlerunning{Conditional Atlas Learning in Fetal MRI}
%
\author{Johannes Tischer\inst{1,2,4}\orcidID{0009-0004-0870-0914} \and
Patric Kienast \inst{3}\orcidID{0000-0001-6407-9561} \and
Marlene Stümpflen \inst{3}\orcidID{0000-0001-9763-9422} \and
Gregor Kasprian \inst{3}\orcidID{0000-0003-3858-3347} \and
Georg Langs \inst{1,2,4}\orcidID{0000-0002-5536-6873} \and
Roxane Licandro \inst{1,2,4}\orcidID{0000-0001-9066-4473}}
\authorrunning{J. Tischer et al.}
%
\institute{Computational Imaging Research Lab, Department of Biomedical Imaging and Image-guided Therapy, Medical University of Vienna, Austria \href{mailto:johannes.tischer@meduniwien.ac.at}{johannes.tischer@meduniwien.ac.at} \and Comprehensive Center for Artificial Intelligence in Medicine, Medical University of Vienna, Austria \and Division of General Radiology, Department of Biomedical Imaging and Image-guided therapy, Medical University Vienna, Austria \and Christian Doppler Laboratory for Mathematical Modelling and Simulation of Next-Generation Medical Ultrasound Devices, Austria}
\maketitle              
\begin{abstract} Magnetic Resonance Imaging (MRI) of the fetal brain has become a key tool for studying brain development \textit{in vivo}. Yet, its assessment remains challenging due to variability in brain maturation, imaging protocols, and uncertain estimates of Gestational Age (GA). To overcome these, brain atlases provide a standardized reference framework that facilitates objective evaluation and comparison across subjects by aligning the atlas and subjects in a common coordinate system. In this work, we introduce a novel deep-learning framework for generating continuous, age-specific fetal brain atlases for real-time fetal brain tissue segmentation. The framework combines a direct registration model with a conditional discriminator. Trained on a curated dataset of 219 neurotypical fetal MRIs spanning from 21 to 37 weeks of gestation. The method achieves high registration accuracy, captures dynamic anatomical changes with sharp structural detail, and robust segmentation performance with an average Dice Similarity Coefficient (DSC) of 86.3\,$\%$ across six brain tissues. Furthermore, volumetric analysis of the generated atlases reveals detailed neurotypical growth trajectories, providing valuable insights into the maturation of the fetal brain. This approach enables individualized developmental assessment with minimal pre-processing and real-time performance, supporting both research and clinical applications. The model code is available at \url{https://github.com/cirmuw/fetal-brain-atlas}

\end{abstract}

\section{Introduction}
While ultrasound remains the primary imaging modality in prenatal care, recent advances have expanded the utility of fetal MRI, offering complementary diagnostic insights~\cite{ISUOG2023}.
However, the assessment of the fetal brain poses significant challenges, including fetal motion, inter-individual variability, such as morphological differences~\cite{Gholipour2017}\cite{Habas2010b}, the (dis-)appearance of brain structures over time~\cite{stuempflen2023}, and varying image intensities of the same structures across GA~\cite{Habas2010b}. Additional inconsistency in image intensity arises from manufacturer-specific differences in scanner hardware and image acquisition protocols~\cite{Gholipour2017}. Consequently, reliable inter- and intra-subject comparisons remain challenging, requiring profound medical expertise. \\
The introduction of time-varying fetal brain atlases~\cite{Ciceri2024} offers a population-weighted mean representation of the developing brain, providing an objective reference system. By registering individual scans to a normative atlas, prior anatomical or functional knowledge can be propagated from the atlas to previously unseen cases. This facilitates consistent multi-modal mapping, intensity normalization, and segmentation, which in turn, support advanced analysis techniques, such as volumetric assessment and surface-based measurement.\\

\noindent\textbf{Related work.} Pioneering work has constructed spatiotemporal fetal brain atlases by optimizing pairwise and group-wise registration methods~\cite{Ciceri2024}. Followed by further enhancements through larger cohorts, integration of multi-modal data~\cite{Uus2023_Atlas}, and improved anatomical labeling~\cite{Uus2023_BOUNTI}. Recent advances in deep learning have expanded the potential of atlas construction beyond traditional approaches, enabling more robust feature learning~\cite{dannecker2024}, generalization~\cite{abulnaga2025multimorph}, and real-time inference~\cite{Li2022}. Dalca et al.~\cite{Dalca2019} proposed AtlasMorph, 
an atlas construction framework for adult brains that incorporates a CNN-based registration approach. An extension of AtlasMorph by Dey et al.~\cite{Dey2021} included a projection discriminator network and improved conditional embedding, thereby generating anatomically sharper templates in adult brains. First adaptations to fetal brain atlas construction focused on the integration of segmentation maps~\cite{Li2022}, anatomical constraints to enhance atlas consistency and quality~\cite{Pei2021}, and network adaptations such as Tranformers~\cite{Zhang2024}. More recently, alternative strategies emerged using Implicit Neural Representations (INRs) and GroupBlock layers. INRs have been employed to improve the temporal continuity of brain atlases~\cite{chen2022} and to integrate pathological features~\cite{dannecker2024}, while GroupBlock layers, as proposed in MultiMorph~\cite{abulnaga2025multimorph}, facilitating the construction of group-specific templates within minutes. \\
Despite advancements in the field, current atlas construction methods have several limitations. Many approaches rely on discrete time points, affine alignment, or initialization with anatomical priors, which can restrict temporal resolution, neglect individual anatomical variability, or introduce bias. Additionally, these methods often struggle to capture the complex non-linear morphological changes that occur in the fetal brain, especially during the third trimester.
\\As demonstrated in adult brain studies~\cite{Dey2021}, conditional Generative Adversarial Networks (cGANs) provide improved, sharper representations of brain anatomy and are capable of modeling diverse morphological features. \\

\noindent\textbf{Contribution.} Extending prior work~\cite{Dalca2019}\cite{Dey2021}, we propose a fetal brain atlas learning approach that addresses existing limitations. Specifically, we introduce a conditional atlas with a continuous, age-appropriate representation of the fetal brain. The anatomically detailed templates are enhanced by a discriminator network, especially in the last trimester when rapid gyrification and sulcation occur. The method yields real-time, automatic segmentation of six anatomical labels in previously unseen subjects. It encompasses a minimal pre-processing approach that eliminates the need for affine alignment or prior anatomical knowledge during both training and inference. Overall, we provide a comprehensive quantitative evaluation of the segmentation accuracy and the quality of the generated templates.

\section{Methodology}
\begin{figure}[!t]
    \centering
    \includegraphics[width=12cm]{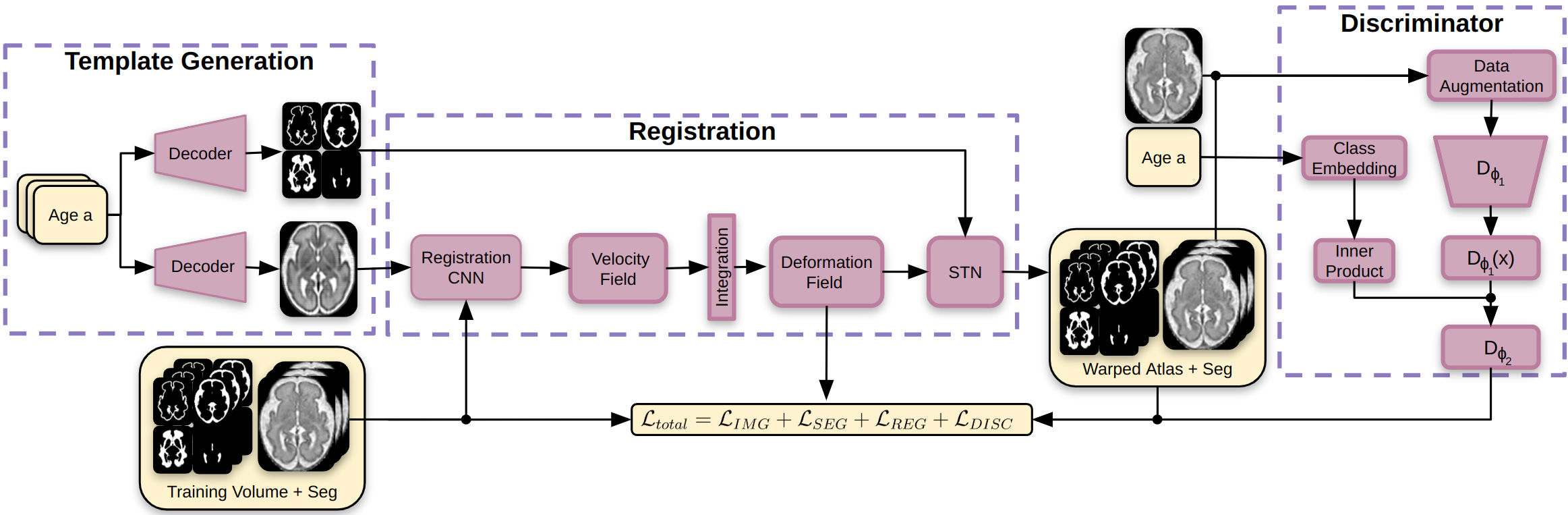}
    \caption{The model architecture of the proposed conditional atlas learning network incorporates two networks: 1.) The template generation framework, including age-appropriate structural representation and its associated anatomical labels. 2.) The U-net-based registration framework with the additional projection discriminator.}
    \label{fig:architecture}
\end{figure}

The proposed architecture, illustrated in Fig.~\ref{fig:architecture}, consists of two neural networks: 1.) the \textit{Template Generation Network} focusing on generating atlas templates with corresponding tissue maps parametrizable by the input condition (e.g. age) and 2.) a \textit{Registration Network}, which takes as input the template generation network's predictions in conjunction with a fetal brain MR image and provides as output the transformed template in the individual's image space with corresponding label maps tailored to the individual's anatomy. The entire framework is trained end-to-end using an adversarial setup, where a discriminator encourages the generation of realistic representations of the developing fetal brain. \\

\noindent\textbf{Template Generation Network.} This network consists of two identical decoder streams. While the first stream is trained to generate a structural representation of the brain conditioned on a single attribute as proposed in~\cite{Dalca2019}, the second stream creates the corresponding segmentation map as a one-hot representation (see Fig.~ \ref{fig:architecture}). The conditional embedding in the template generator is achieved through Feature-wise Linear Modulation (FiLM) \cite{perez2017}, defined as:
\[
FiLM(\textbf{F}_{i,c}|\gamma_{i,c}, \beta_{i,c}) = \gamma_{i,c} \textbf{F}_{i,c} + \beta_{i,c}
\]
where \( \gamma \) and \( \beta \) are scale and transformation parameters of the feature map \( \textbf{F} \) from the \( i \)-th layer and \( c \)-th feature, both learned from the condition \( a \). Unlike methods based on concatenation, FiLM requires only two parameters for each feature map, thus it does not scale with image size. Moreover, FiLM is capable of generalizing and learning complex connections from limited data, enabling rapid adaptations during training~\cite{Dey2021}\cite{perez2017}. \\

\noindent\textbf{Registration Network.} Consists of a U-Net based registration network~\cite{Dalca2019} with the task to align an image pair, in our setup the generated template and an image sample, both sharing the same condition. The output of the registration network is a stationary velocity field $v_{i}$. After integration of the velocity field the deformation field $\Phi$ is retained $\Phi=\int_{t=0}^{t=1}v(\Phi^{t})dt$. Finally, the generated templates are warped into the subject space by a Spatial Transformer Network (STN)~\cite{Jaderberg2018}. The template segmentation map is warped by the same deformation field into the subject space (see Fig.~\ref{fig:architecture}). \\

\noindent \textbf{Discriminator Network.} The condition is integrated into the discriminator using the projection method~\cite{Miyato2018}. Unlike traditional cGANs that concatenate conditional vectors with input features, this approach projects the condition onto the intermediate feature space using an inner product~\cite{Miyato2018}. The discriminator employs downsampling residual blocks, followed by ReLU, global sum pooling, and a linear projection that is combined via inner product with a learned class embedding~\cite{Miyato2018}. This method preserves spatial information while it converges faster, with stable training and high-quality output even for small datasets~\cite{Miyato2018}\cite{sauer2021}. \\

\noindent \textbf{Loss function.} The model is trained end-to-end, enabling simultaneous optimization of the atlas construction and template-to-subject registration. The optimization is guided by a loss function (see Eq.~\ref{eq:loss}) with four components: image similarity loss $\mathcal{L}_{IMG}$, segmentation overlap loss $\mathcal{L}_{SEG}$, regularization loss $\mathcal{L}_{REG}$ applied on the deformation field, and a discriminator loss $\mathcal{L}_{DISC}$.
\begin{equation}\label{eq:loss}
    \mathcal{L}_{TOTAL} = \mathcal{L}_{IMG} + \mathcal{L}_{SEG} + \mathcal{L}_{REG} + \mathcal{L}_{DISC}
\end{equation}
The image similarity loss promotes spatial alignment using a localized Normalized Cross-Correlation (NCC) loss, computed bidirectionally: comparing the template warped into the subject space with the subject image, and vice versa, using the inverse deformation field $\Phi^{-1}$ to warp the subject into the template space.
The segmentation loss is defined as the Mean Squared Error (MSE) between the one-hot encoded labels of the warped subject and the fixed segmentation map, thereby providing additional anatomical guidance. The regularization loss follows the formulation in \cite{Dalca2019}, penalizing both the magnitude and spatial gradients of the deformation field. This promotes the generation of age-appropriate templates while preserving accurate individual registrations. The discriminator loss encourages the generated template to resemble real anatomical images and intensity distributions. It is computed via an adversarial training objective~\cite{sauer2021}.

\section{Experimental Setup}
\begin{figure}[!t]
\noindent
\begin{minipage}[t]{0.65\textwidth}
    \begin{tikzpicture}
    \node[anchor=north west] at (-1.3,3.7) {\textbf{a)}};
    \definecolor{myteal}{RGB}{46, 95, 127}
    \definecolor{myrosa}{RGB}{234, 209, 220}
    \definecolor{mypurple}{RGB}{213, 166, 189}
    \begin{axis}[
        ybar stacked,
        bar width=6pt,
        ylabel={Number of subjects},
        xlabel={Gestational week},
        xmin=20,
        xmax=38,
        xtick={21, 22, 23, 24, 25, 26, 27, 28, 29, 30, 31, 32, 33, 34, 35, 36, 37},
        xticklabel style={rotate=90, anchor=east}, 
        ymin=0,
        ymax=16,
        ymajorgrids,
        width=\textwidth,
        height=5cm,
        axis lines=left, 
        tick align=outside, 
        clip=false,
        legend style={
            at={(0.98,0.93)}, anchor=north east,
            draw=none,
            fill=none,
            font=\small,
        },
    ]
    \addplot[fill=mypurple] coordinates {
        (21,1) (22,1) (23,0) (24,1) (25,4) (26,3) (27,2)
        (28,0) (29,0) (30,0) (31,0) (32,1) (33,0)
        (34,0) (35,1) (36,0) (37,1) 
    };
    \addlegendentry{3\,T}
    \addplot[fill=myrosa] coordinates {
        (21,8) (22,14) (23,12) (24,13) (25,11) (26,12) (27,12)
        (28,13) (29,9) (30,13) (31,12) (32,8) (33,9)
        (34,10) (35,4) (36,8) (37,2) 
    };
    \addlegendentry{1.5\,T}
    \end{axis}
    \end{tikzpicture}
\end{minipage}%
\begin{minipage}[t]{0.34\textwidth}
    \vspace{-35pt}
    \hspace{5pt}
    \raisebox{55pt}[0pt][0pt]{%
        \begin{overpic}[width=0.32\textwidth]{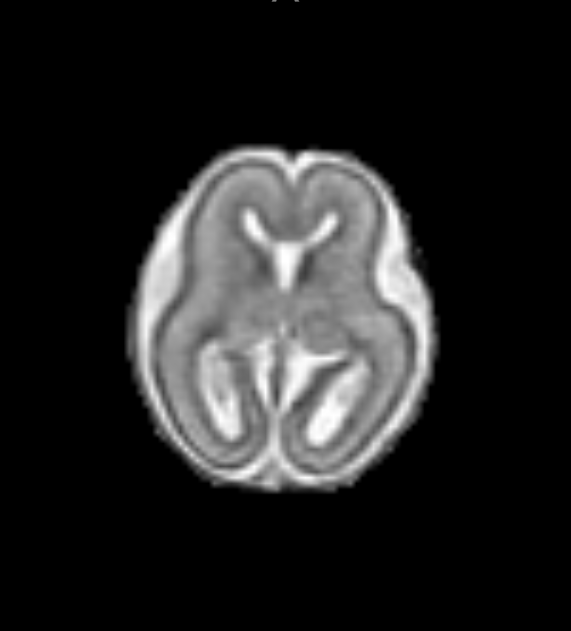}
            \put(-20,115){\textbf{b)}}
            \put(5,90){\color{white}\tiny 21}
        \end{overpic}%
    }%
    \raisebox{55pt}[0pt][0pt]{%
        \begin{overpic}[width=0.32\textwidth]{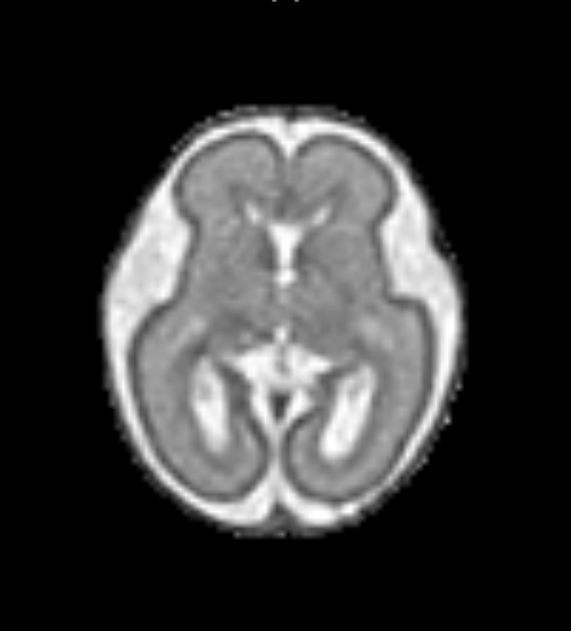}
            \put(0,90){\color{white}\tiny 24}
        \end{overpic}%
    }%
    \raisebox{55pt}[0pt][0pt]{%
        \begin{overpic}[width=0.32\textwidth]{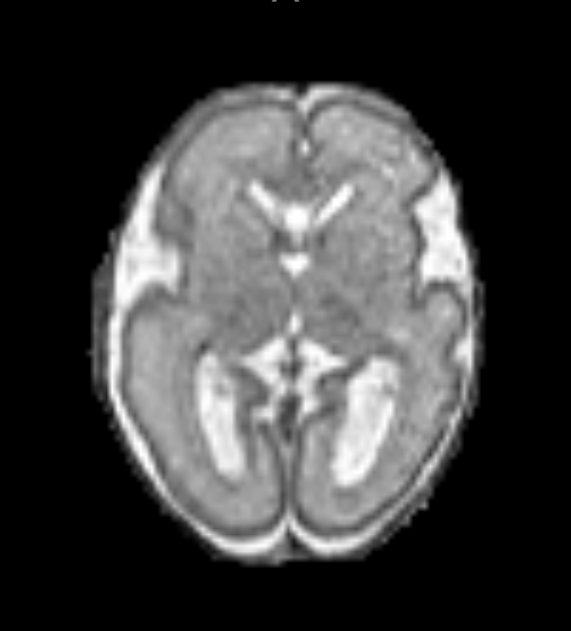}
            \put(0,90){\color{white}\tiny 27}
        \end{overpic}%
    }%

    \hspace{-6.2pt}
    \raisebox{25pt}[0pt][0pt]{
        \begin{overpic}[width=0.32\textwidth]{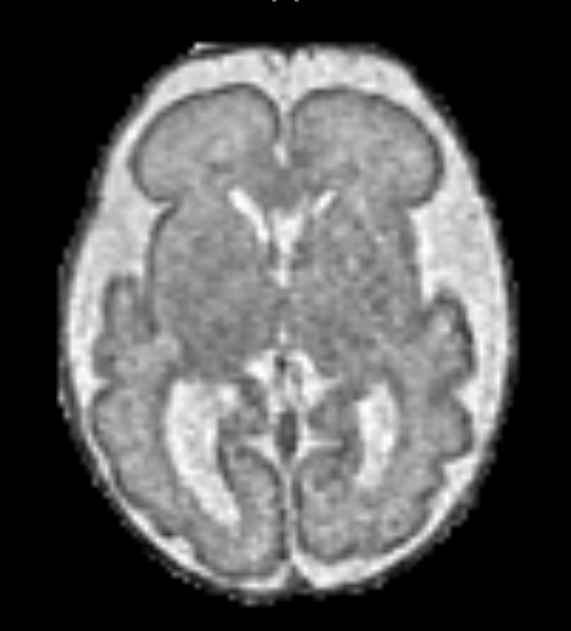}
            \put(5,90){\color{white}\tiny 30}
        \end{overpic}%
    }%
    \raisebox{25pt}[0pt][0pt]{%
        \begin{overpic}[width=0.32\textwidth]{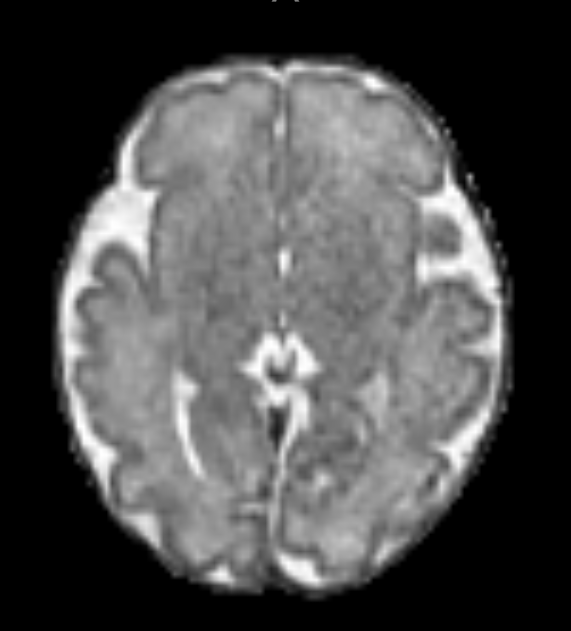}
            \put(0,90){\color{white}\tiny 33}
        \end{overpic}%
    }%
    \raisebox{25pt}[0pt][0pt]{%
        \begin{overpic}[width=0.32\textwidth]{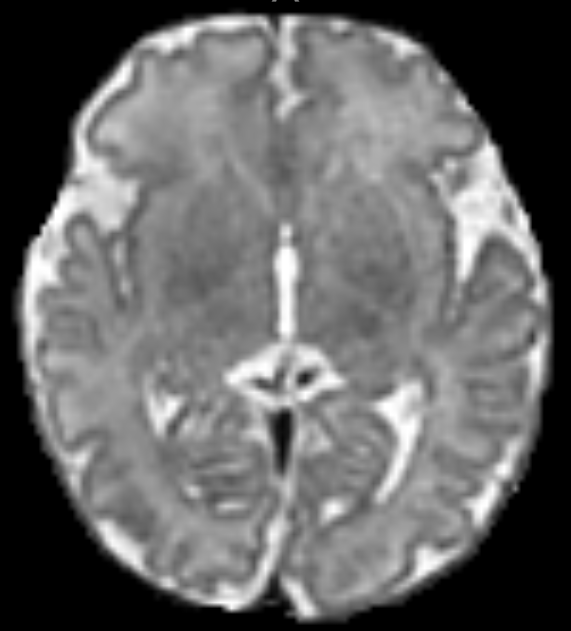}
            \put(0,90){\color{white}\tiny 37}
        \end{overpic}%
    }%
\end{minipage}
\caption{a) The histogram illustrates the heterogeneous age distribution of the neurotypical dataset divided in 1.5 and 3\,Tesla acquisitions. b) Exemplary SVR samples across different GA.}
\label{fig:histogram}
\end{figure}
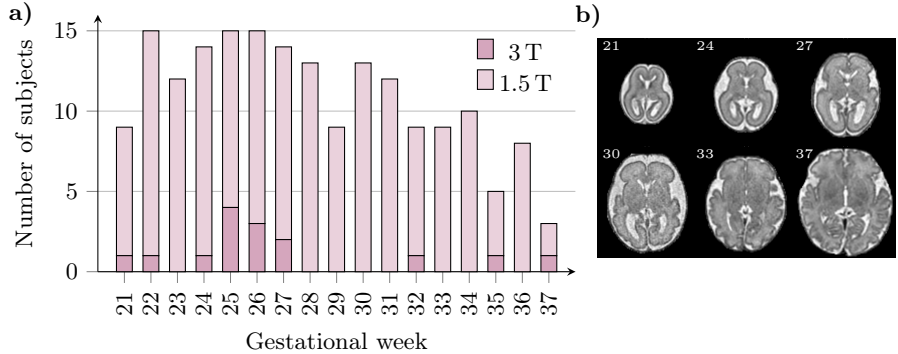

\textbf{Dataset.}
The dataset consists of 219 retrospectively curated, in-house fetal MRI cases from neurotypical pregnancies. The study was conducted in accordance with the Declaration of Helsinki (2013) and approved by the ethics board of the Medical University of Vienna (No. 2032/2022). \textit{In vivo} fetal MRI was performed following the ISUOG Practice Guidelines \cite{ISUOG2023} and includes T2-weighted scans from clinical exams conducted from 2012 to 2022 using either a 1.5\,T (Philips Ingenia/Intera, Best, The Netherlands) or a 3\,T (Philips Achieva) scanner. Image acquisition parameters are given in Table \ref{tab:img_acquisition}. Subjects were selected based on the absence of pathological findings, as verified independently by two experienced radiologists (M.S. and P.K.), each with over five years of specialization in fetal MRI. The age distribution ranges from 21 to 37 Gestational Weeks (GWs), as shown in Fig.~\ref{fig:histogram}a.\\

\begin{table}[!t]
    \centering
    \caption{Fetal MRI acquisition parameters, including echo time $T_E$, repetition time $T_R$, slice thickness, field of view (FOV), and image resolution for field strengths of 1.5\,T and 3\,T.}
    \begin{tabularx}{\textwidth}{@{\extracolsep{\fill}}cccccc}
    \toprule
     & $T_E$ & $T_R$ & slice thickness & FOV & image resolution (x,y,z) \\
     & [ms] & [ms] & [mm] & [mm] & [mm] \\
    \midrule
    1.5\,$T$ & 0.126 & 14.699 & 3.559 & 249.844 & 0.771 / 0.771 / 3.826 \\
    std $\pm$ & 0.023 & 6.332 & 0.664 & 44.808 & 0.124 / 0.124 / 0.937 \\
    3\,$T$ & 0.181 & 7.811 & 2.877 & 262.408 & 0.805 / 0.805 / 2.946 \\
    std $\pm$ & 0.037 & 4.445 & 1.253 & 44.787 & 0.152 / 0.152 / 1.407 \\
    \bottomrule
    \end{tabularx}
    \label{tab:img_acquisition}
\end{table}

\noindent\textbf{Preprocessing.} The raw data captured during clinical routines underwent multiple preprocessing steps. First, the DICOM files were converted to NIfTI format using the BIDS framework, allowing for the automatic identification of T2-weighted scans of the fetal brain. Next, the scans were preprocessed using NeSVoR \cite{NeSVor}, a INR-based Slice-to-Volume Reconstruction (SVR) method specifically designed for fetal brain imaging. This process included bias field correction and automatic image quality assessment to identify the five best scans, resulting in an 1\,mm, motion-corrected, isotropic volume. After reconstruction, the image intensities were normalized to the range [0, 1] and resized to 128x128x96 (see exemplary samples in Fig.~\ref{fig:histogram}b).
In the second step, segmentation maps corresponding to the SVR volumes, which serve as ground truth, were automatically generated using the BOUNTI toolbox \cite{Uus2023_BOUNTI}. The resulting 19 anatomical labels, covering both hemispheres, were merged into six brain regions stored as one-hot representations: external cerebrospinal fluid (eCSF), cortical gray matter (cGM), total white matter (tWM), ventricles (Ven), deep gray matter (dGM), and brainstem (BS).\\

\noindent\textbf{Baselines.} The proposed work is compared against three baseline approaches: \\
\textit{1. ANTs}~\cite{Avants2011} atlas construction is performed using \textsf{antsMultivariateTemplateConstruction} (v2.4.4), a coarse-to-fine iterative approach that improves alignment through multi-resolution downsampling and Gaussian smoothing. Non-affine registration uses the SyN model with default parameters for comparability with standard practices. As the data is already preprocessed, no bias correction or rigid alignment is applied. Segmentation maps are generated by fusing anatomical labels of all subjects sharing the same condition~\cite{Wang2013}. The registration of the test dataset was performed using both \textsf{SyN} and \textsf{SyNQuick}, applying the same parameters as those used during the template construction. \\ 
\textit{2. MultiMorph}~\cite{abulnaga2025multimorph}, which introduces a feed-forward neural network with a GroupBlock layer and centrality enforcement, enabling the rapid generation of unbiased, subgroup-specific atlases. Age-specific fetal templates are constructed using the pre-trained model weight of MultiMorph and our training samples for each age subgroup. Following the approach in~\cite{abulnaga2025multimorph}, training samples are used to construct the atlases, while for the test samples, only the deformation fields are predicted. This allows atlas segmentations to be transferred to the test samples. \\
\textit{3. nnUNet}~\cite{isensee_nnu-net_2020} (v2) is a self-supervised segmentation method capable of dynamically adapting model configurations and training parameters. The training is initiated with a 3D UNet using default parameters~\cite{isensee_nnu-net_2020}. For reasons of time efficiency and fair comparison, only a single fold is trained and used for inference. \\

\noindent \textbf{Training Setup.} The dataset is split into 185 training volumes and 34 test volumes, with two volumes per age. To address the heterogeneous distribution in the training set, the sampling rate of underrepresented classes is increased. Two model implementations are trained, in \textit{Ours-Disc} the discriminator is included, while in \textit{Ours-Reg} the optimization is performed solely by the registration framework. Both models are trained for 300 epochs using a batch size of 8. For optimization, we use ADAM with $\beta_1$=0.0, $\beta_2$=0.9, $\epsilon=10^{-7}$, a generator learning rate of $10^{-4}$, and a discriminator learning rate of $3$$\times$$10^{-4}$. The training takes approximately 3 days on an NVIDIA A100 GPU. In \textit{Ours-Disc} we use an imbalanced training schedule between the generator and discriminator (1:5 ratio). Similar to \cite{Dey2021}, the input class labels are normalized to a range of 0 to 1, resulting in an artifact-free generation of brain images. The discriminator training loop incorporates data augmentation by applying random flips and translations to both real and fake samples before feeding the transformed data into the discriminator. Training the discriminator on augmented samples avoids overfitting, hence enhancing its generalization capability \cite{sauer2021}. The final hyperparameters used are defined by $\lambda_{IMG}$=1.0, $\lambda_{SEG}$=0.5, $\lambda_{DEF}$=1.0, $\lambda_{GRAD}$=0.5, and $\lambda_{DISC}=0.5$.

\begin{table}[!t]
    \centering
    \small
    \caption{Quantitative evaluation of atlas construction and segmentation performance: \textit{Our-Reg} uses registration without a discriminator; \textit{Our-Disc} adds adversarial training to improve atlas quality. $t_N$ denotes the training time, and $t_i$ denotes the total inference time for the test dataset.}
    \begin{tabularx}{\textwidth}{@{\extracolsep{\fill}}lcccccccc}
        \toprule
        Method & $t_N$ & $t_i$ & $|J_\phi|$ & ||Def||  & EFC & DSC  & HD \\
        & [min] & [min] & ($\uparrow$) & ($\downarrow$) & ($\downarrow$) & [$\%$] ($\uparrow$) & [mm] ($\downarrow$) \\
        \midrule
        SyNQuick\,\cite{Avants2011} & 1290 & 20  & \textbf{1.000$\pm$0.0} & \textbf{692$\pm$234} & 0.345 & 80.5$\pm$10.3 & 1.66$\pm$0.29 \\
        SyN\,\cite{Avants2011} & 1290 & 980 & \textbf{1.000$\pm$0.0} & 909$\pm$290 & 0.347 & 87.2$\pm$6.1 & 1.25$\pm$0.17 \\
        MultiMorph\,\cite{abulnaga2025multimorph}  & \textbf{0} & 20 & \textbf{1.000$\pm$0.0} & 861$\pm$358 &\textbf{0.292} & 82.9$\pm$5.6 & 2.14$\pm$0.92 \\
        nnUnet\,\cite{isensee_nnu-net_2020} & 3580 & 1 & - & - & - & \textbf{96.9$\pm$0.0} & \textbf{0.99 $\pm$ 0.13} \\
        Ours-Reg & 4020 & \textbf{0.5} & 0.998$\pm$0.0 & 807$\pm$148 & 0.344 & 84.1$\pm$1.9 & 1.71$\pm$0.23\\
        Ours-Disc &  4020 & \textbf{0.5} & 0.995$\pm$0.0 & 775$\pm$240 & 0.326 & 86.3$\pm$5.2 & 1.60$\pm$0.20 \\
        \bottomrule
    \end{tabularx}
    \label{tab:metrics}
\end{table}

\section{Results}

\subsection{Fetal Brain Atlas Construction}
The registration performance is assessed through following metrics: training $t_N$ and inference time $t_i$, the Jacobian determinant $|J_{\phi}|$ to identify local or non-invertible deformations, and the average deformation norm $||\text{Def}||$, which measures the geometric distance between the template and individual subjects. Additionally, the image quality of the generated templates, once warped into the subject space, is quantified using the Entropy Focus Criteria (EFC) \cite{Atkinson1997}. Segmentation outcomes are analyzed by comparing test labels to the predicted labels transformed into the subject space using the DSC for overall correspondence and the 95th percentile of the Hausdorff Distance (HD), with particular attention given to the segmentation contours. The results for all baseline methods and our proposed approach are summarized in Table \ref{tab:metrics}.\\
\noindent In Fig.~\ref{fig:atlas_templates}a brain templates are illustrated from 21 to 37 GWs generated by ANTs, MultiMorph and \textit{OursDisc}. Across all methodologies brain size increases with GA. Moreover, brain asymmetry and the growing morphological complexity, marked by the development of numerous gyri and sulci, becomes increasingly pronounced as GA progresses. In contrast to other methods, our approach demonstrates a strong gradient in cGM representation, particularly during the last trimester. \\
\renewcommand{\arraystretch}{0} 
\setlength{\tabcolsep}{0pt}     
\setlength{\fboxrule}{0.5pt} 
\definecolor{eCSF}{RGB}{255,127,14}
\definecolor{WM}{RGB}{31,119,180}
\definecolor{dGM}{RGB}{140,86,75}
\definecolor{Brainstem}{RGB}{23,190,207}
\definecolor{cGM}{RGB}{44,160,44}
\definecolor{Ventricles}{RGB}{214,39,40}

\begin{figure}[!t]
    \centering
    \begin{tabular}{c@{}c@{}c@{}c@{}c@{}c@{}c@{}c@{}c@{}c@{}c@{}}
        \raisebox{0cm}{\hspace{-0.5cm}\textbf{a)}} \hspace{0.3cm}GW & 21 & 23 & 25 & 27& 29 & 31 & 33 & 35 & 37 & \\
        \vspace{0.025cm} \\
        \raisebox{0.4cm}{\textbf{ANTs}}
        & \reflectbox{\rotatebox[origin=c]{90}{\includegraphics[height=1.1cm, trim=0.4cm 1.4cm 0.4cm 1.4cm, clip]{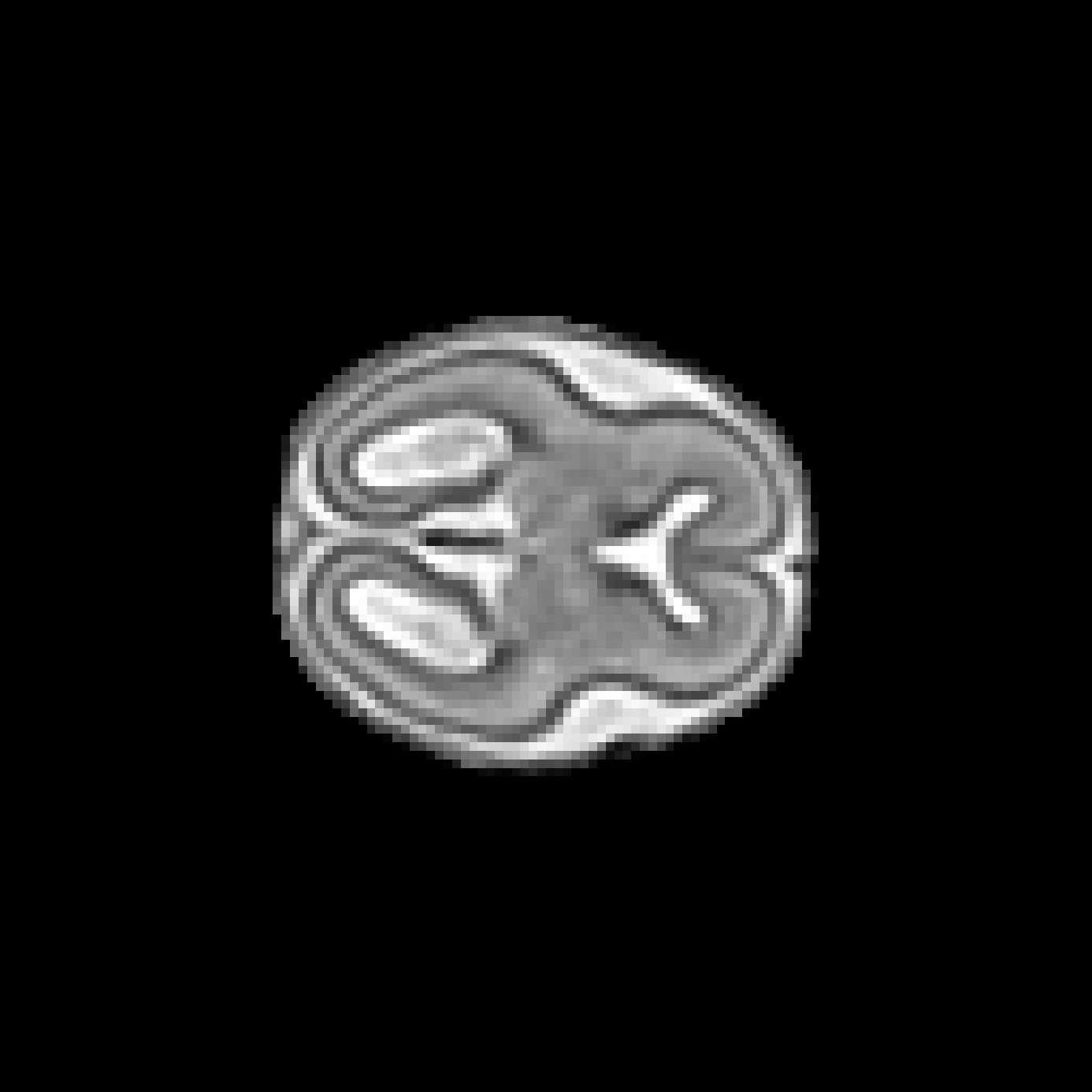}}}
        & \raisebox{-0.122cm}{\includegraphics[width=1.1cm, trim=1.4cm 0.4cm 1.4cm 0.4cm, clip]{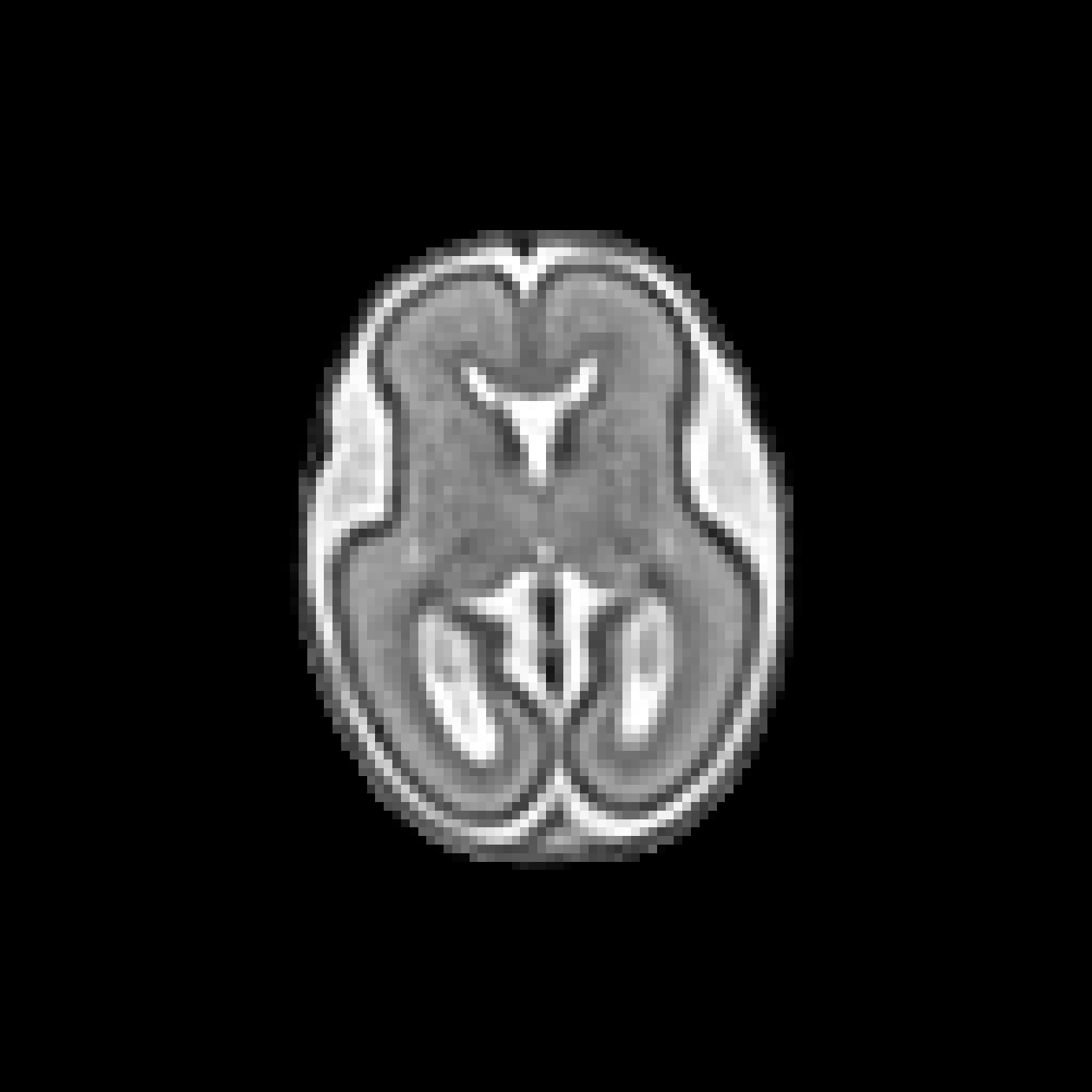}}
        & \reflectbox{\rotatebox[origin=c]{90}{\includegraphics[height=1.1cm, trim=0.4cm 1.4cm 0.4cm 1.4cm, clip]{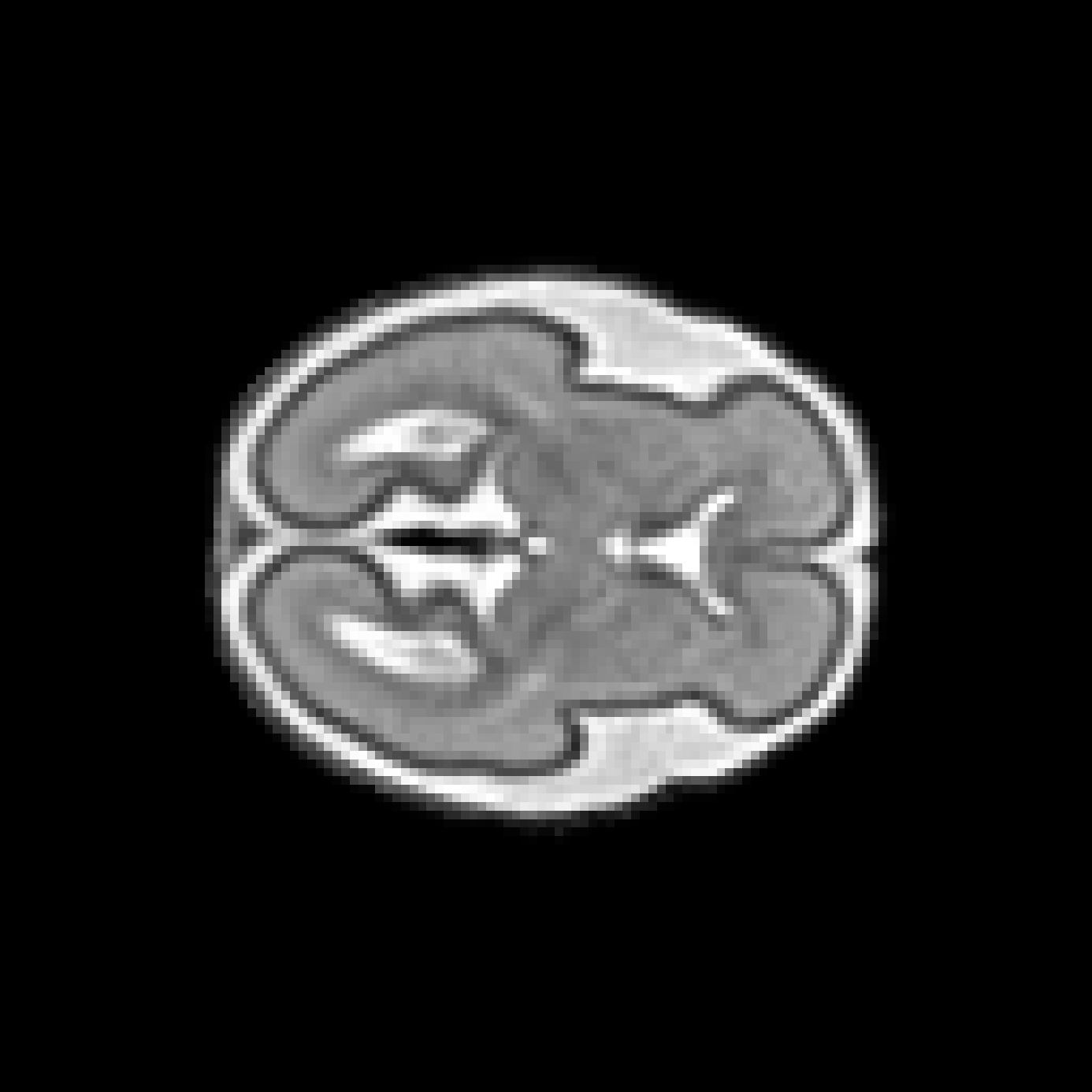}}}
        & \raisebox{-0.122cm}{\includegraphics[width=1.1cm, trim=1.4cm 0.4cm 1.4cm 0.4cm, clip]{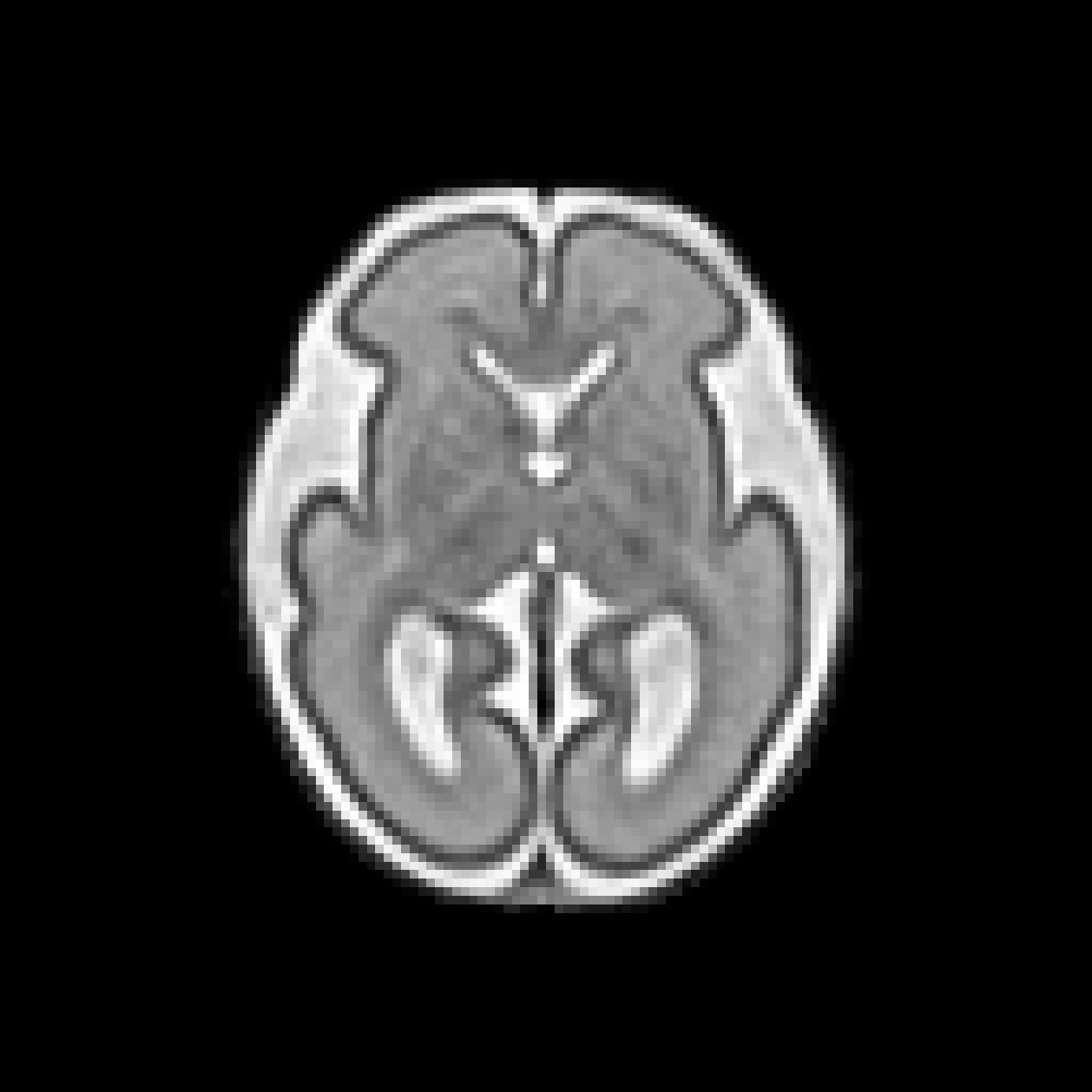}}
        & \reflectbox{\rotatebox[origin=c]{90}{\includegraphics[height=1.1cm, trim=0.4cm 1.4cm 0.4cm 1.4cm, clip]{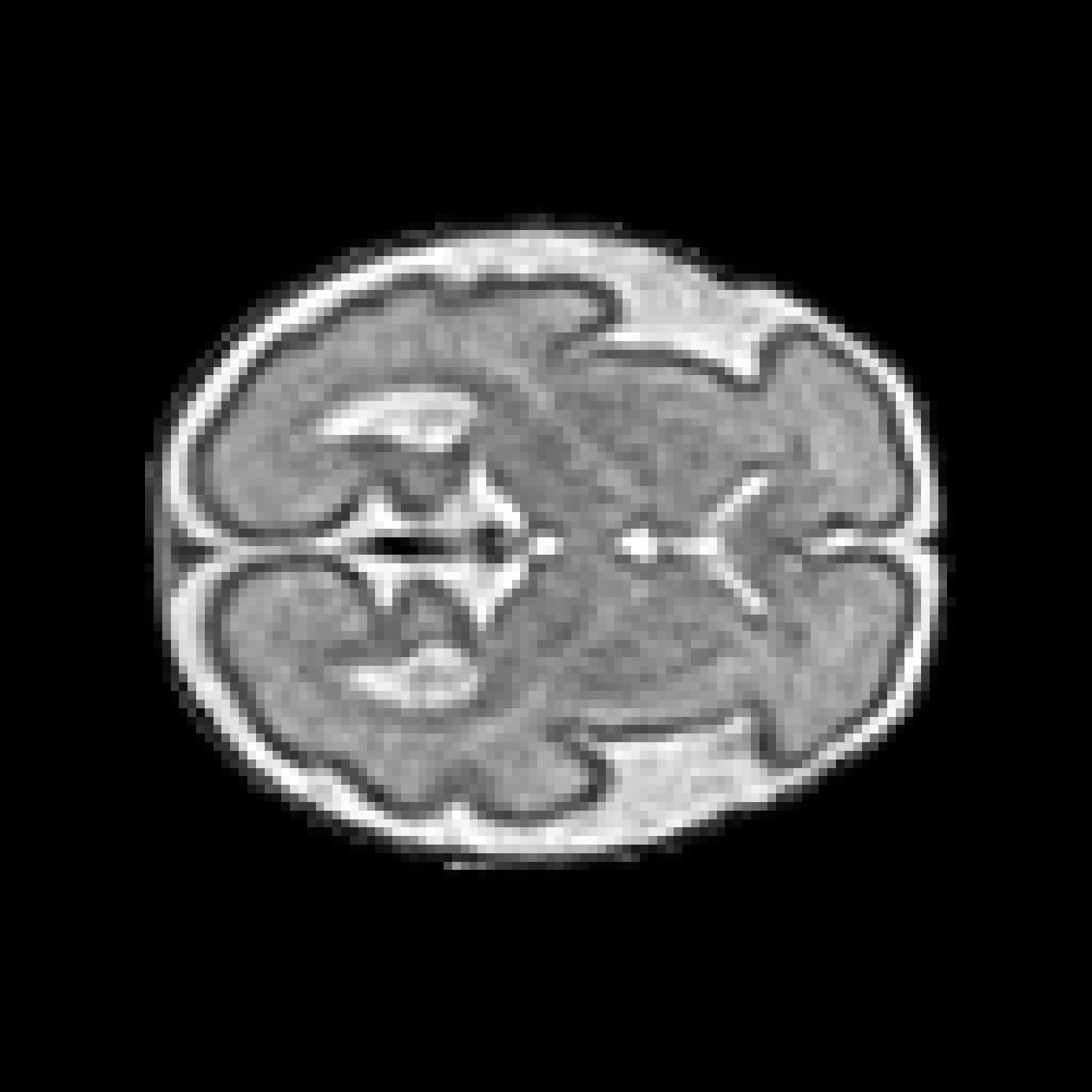}}}
        & \raisebox{-0.122cm}{\includegraphics[width=1.1cm, trim=1.4cm 0.4cm 1.4cm 0.4cm, clip]{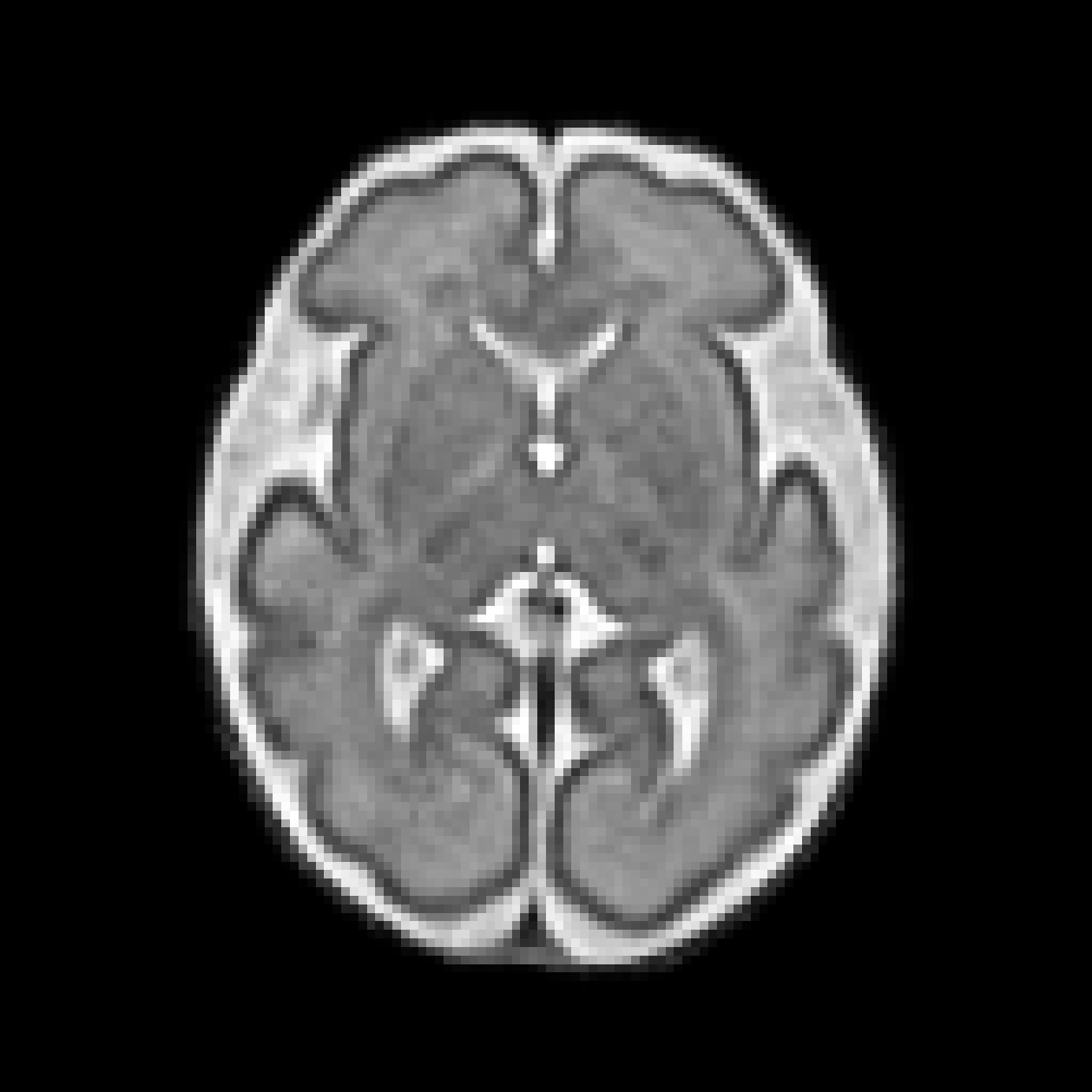}}
        & \reflectbox{\rotatebox[origin=c]{90}{\includegraphics[height=1.1cm, trim=0.4cm 1.4cm 0.4cm 1.4cm, clip]{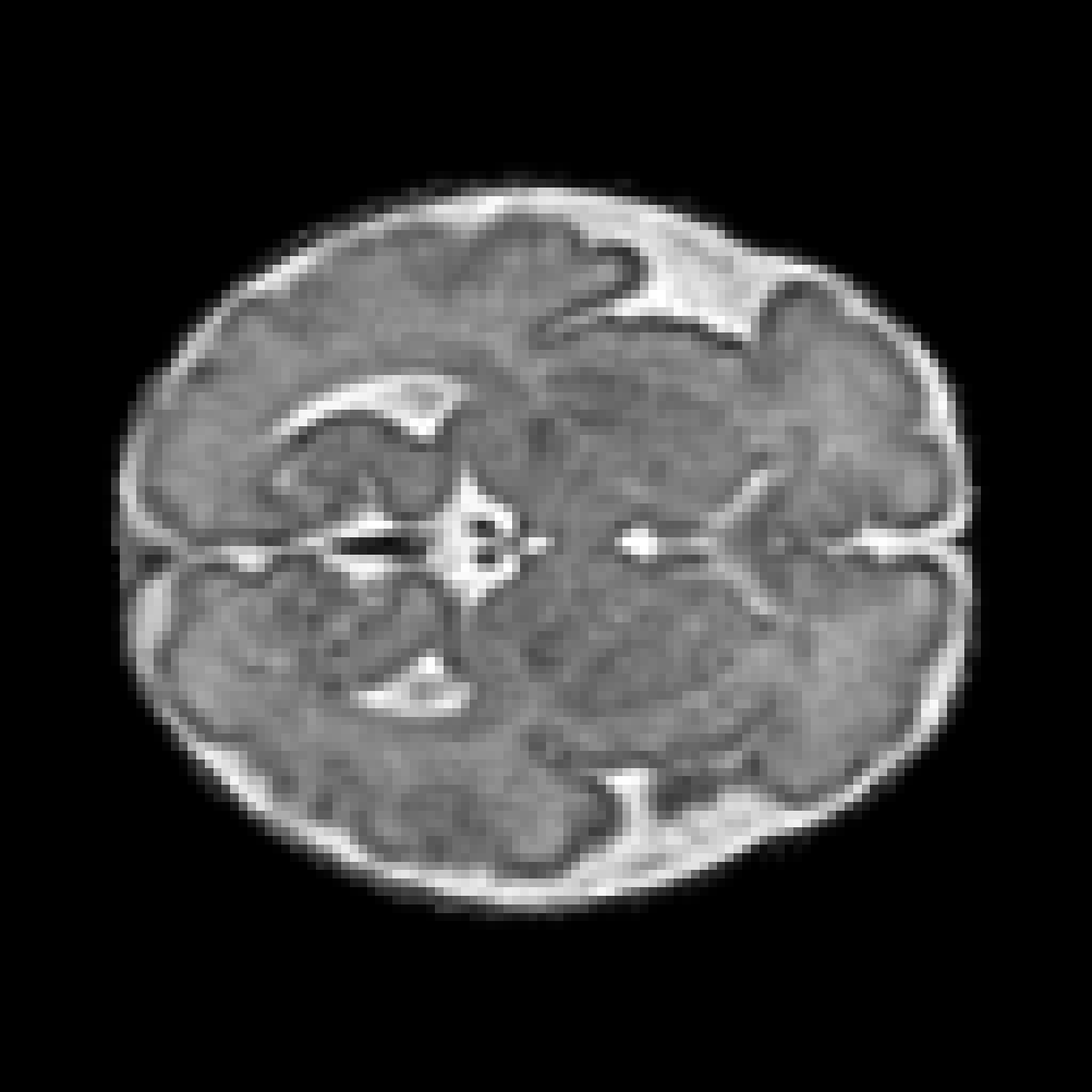}}}
        & \raisebox{-0.122cm}{\includegraphics[width=1.1cm, trim=1.4cm 0.4cm 1.4cm 0.4cm, clip]{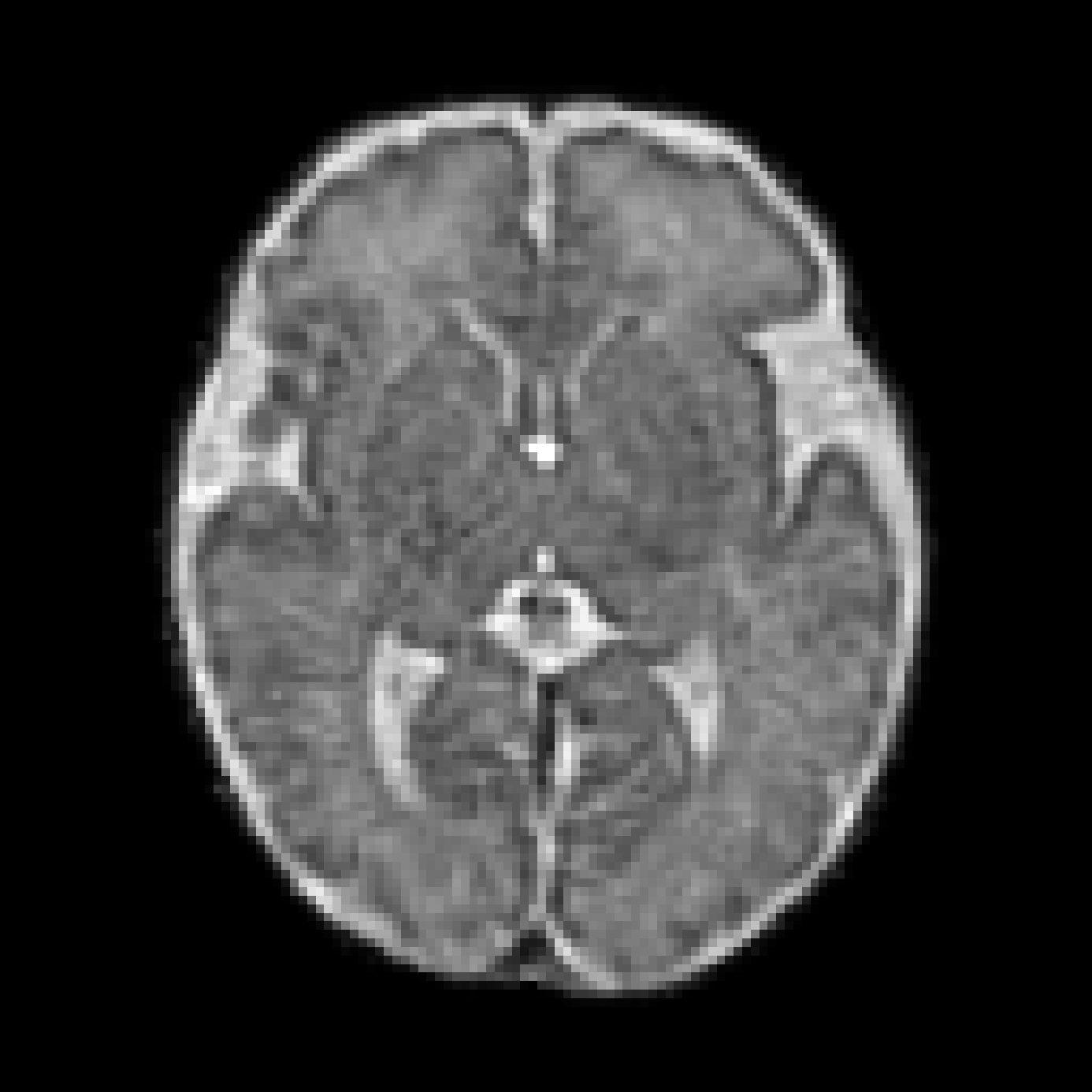}}
        & \reflectbox{\rotatebox[origin=c]{90}{\includegraphics[height=1.1cm, trim=0.4cm 1.4cm 0.4cm 1.4cm, clip]{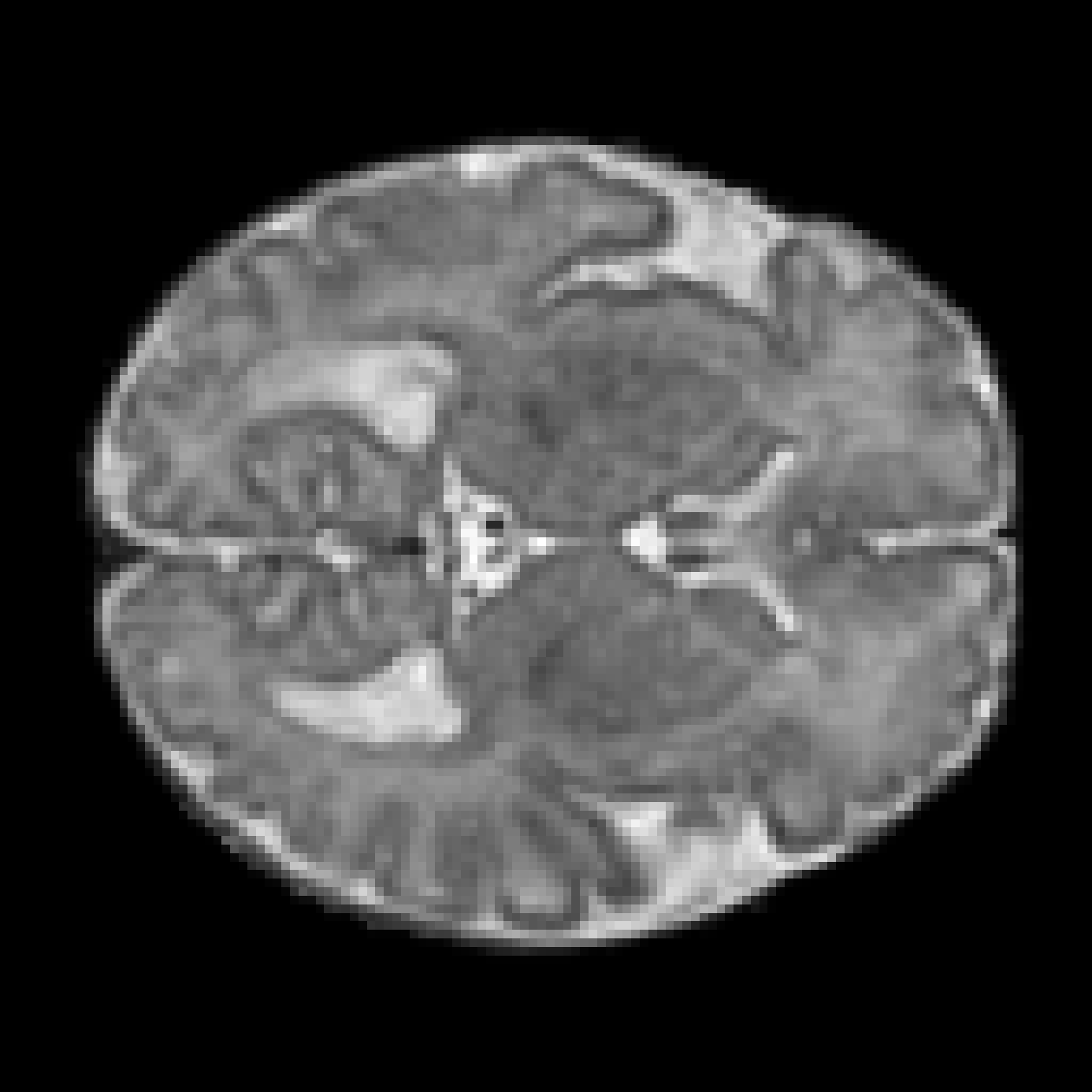}}} \\
        \raisebox{0.25cm}{\textbf{\shortstack{Multi\\Morph}}}
        & \reflectbox{\rotatebox[origin=c]{90}{\includegraphics[height=1.1cm, trim=0.4cm 1.4cm 0.4cm 1.4cm, clip]{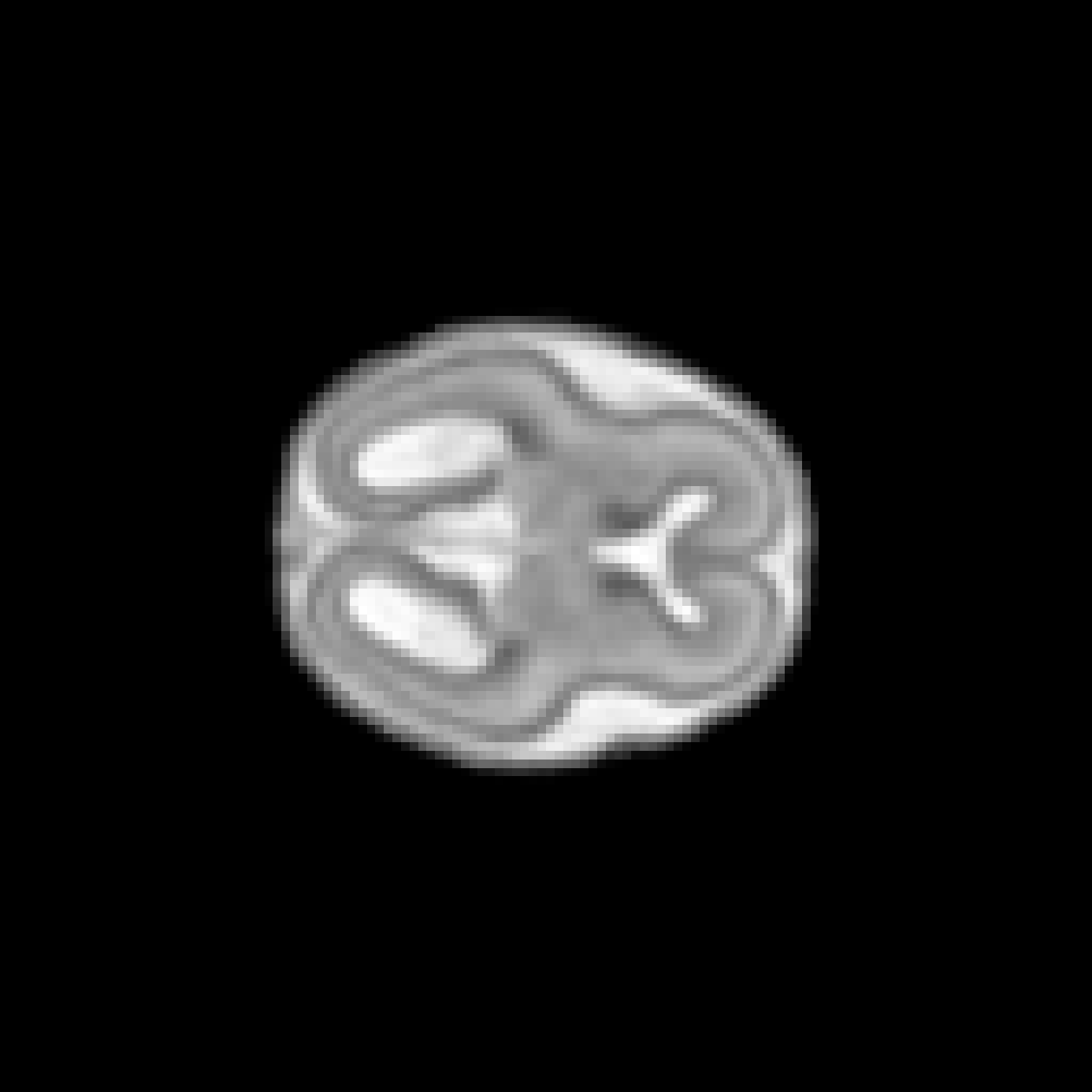}}}
        & \reflectbox{\rotatebox[origin=c]{90}{\includegraphics[height=1.1cm, trim=0.4cm 1.4cm 0.4cm 1.4cm, clip]{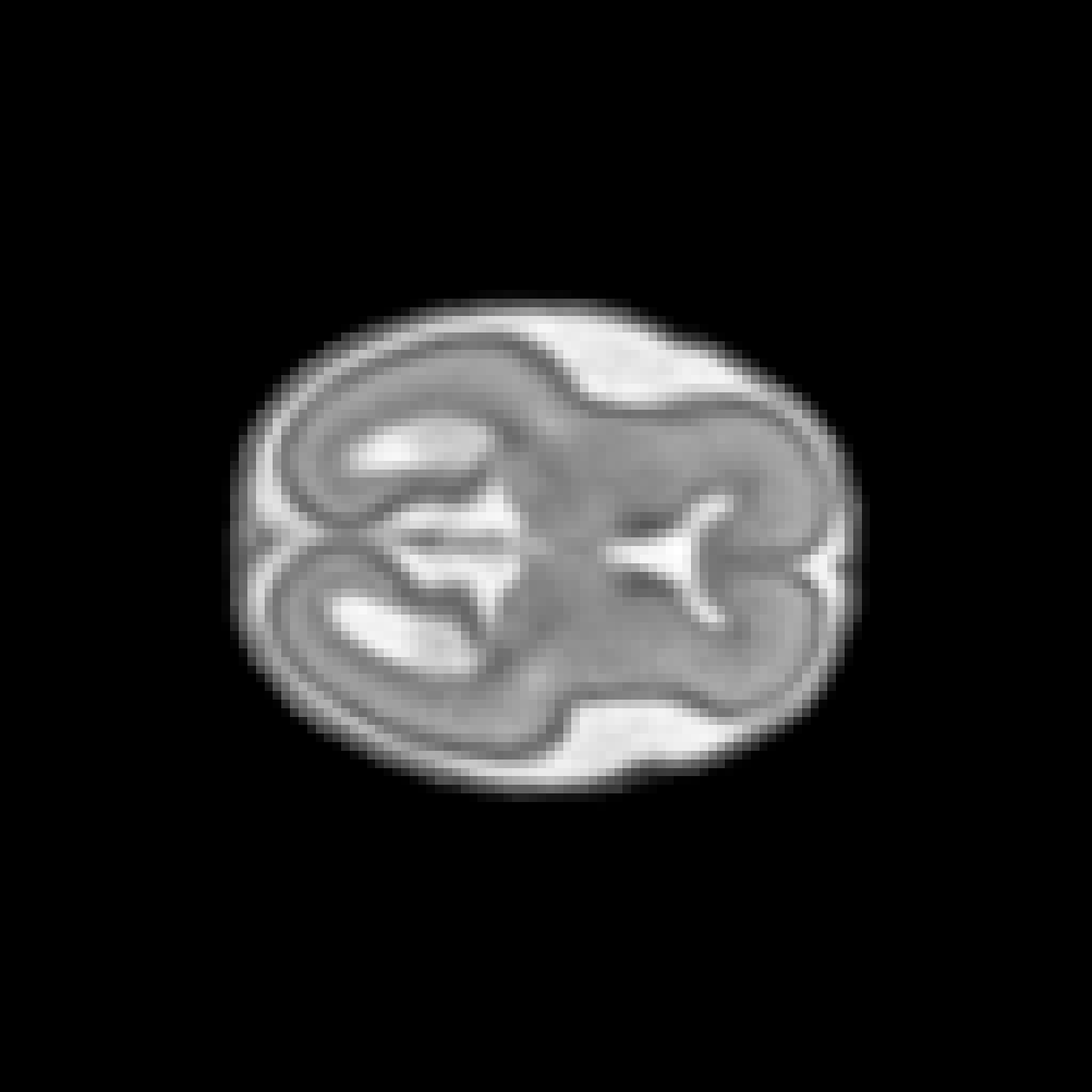}}}
        & \reflectbox{\rotatebox[origin=c]{90}{\includegraphics[height=1.1cm, trim=0.4cm 1.4cm 0.4cm 1.4cm, clip]{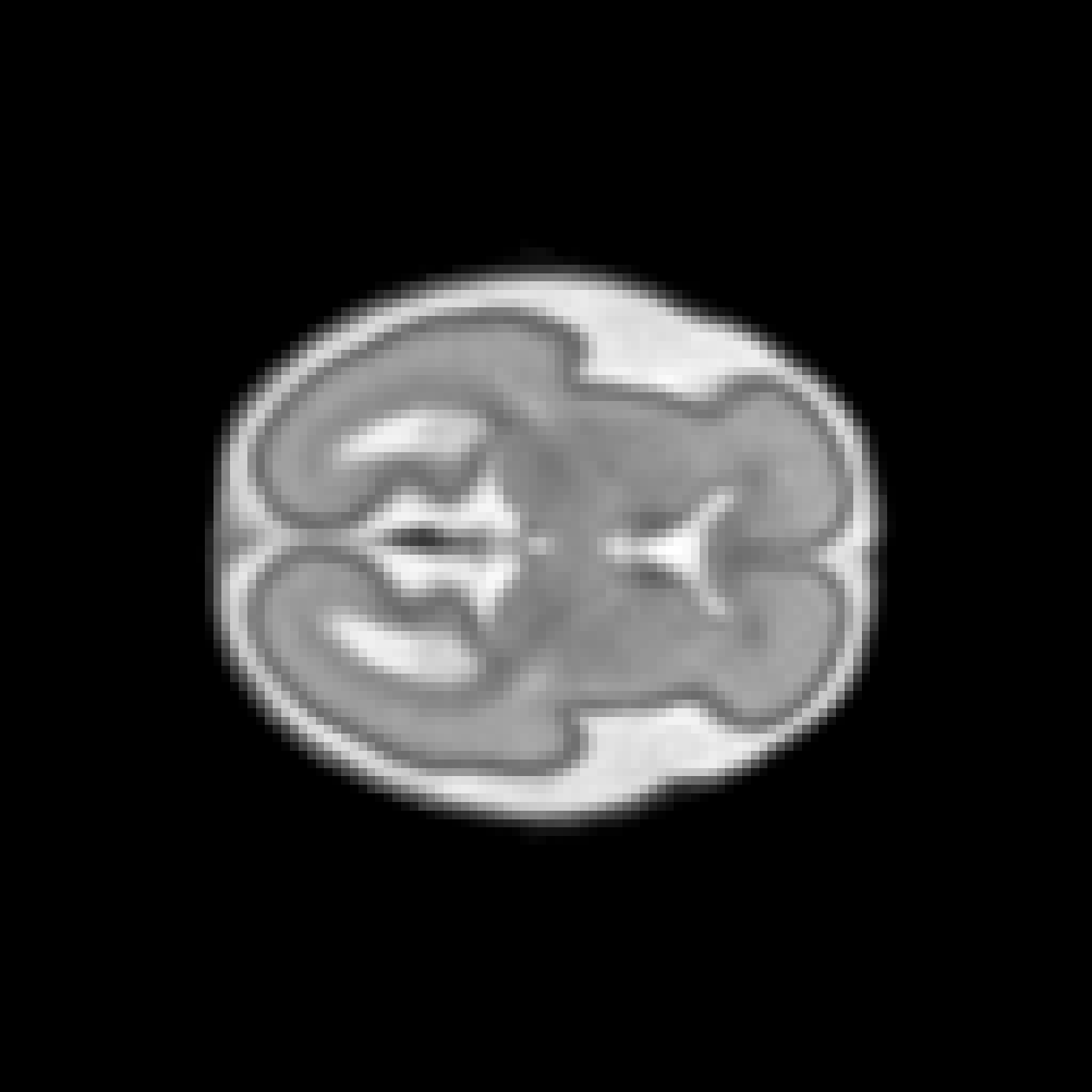}}}
        & \reflectbox{\rotatebox[origin=c]{90}{\includegraphics[height=1.1cm, trim=0.4cm 1.4cm 0.4cm 1.4cm, clip]{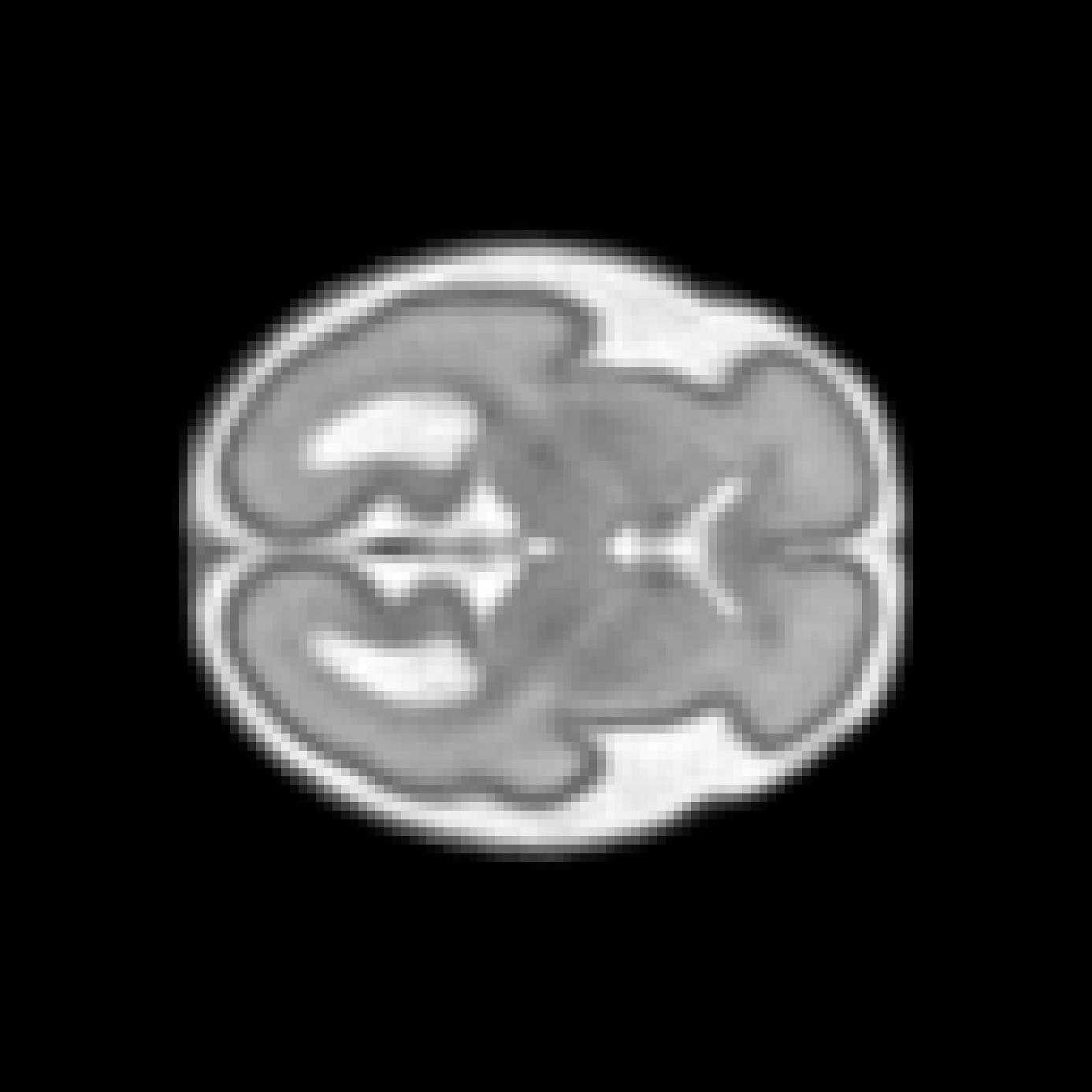}}}
        & \reflectbox{\rotatebox[origin=c]{90}{\includegraphics[height=1.1cm, trim=0.4cm 1.4cm 0.4cm 1.4cm, clip]{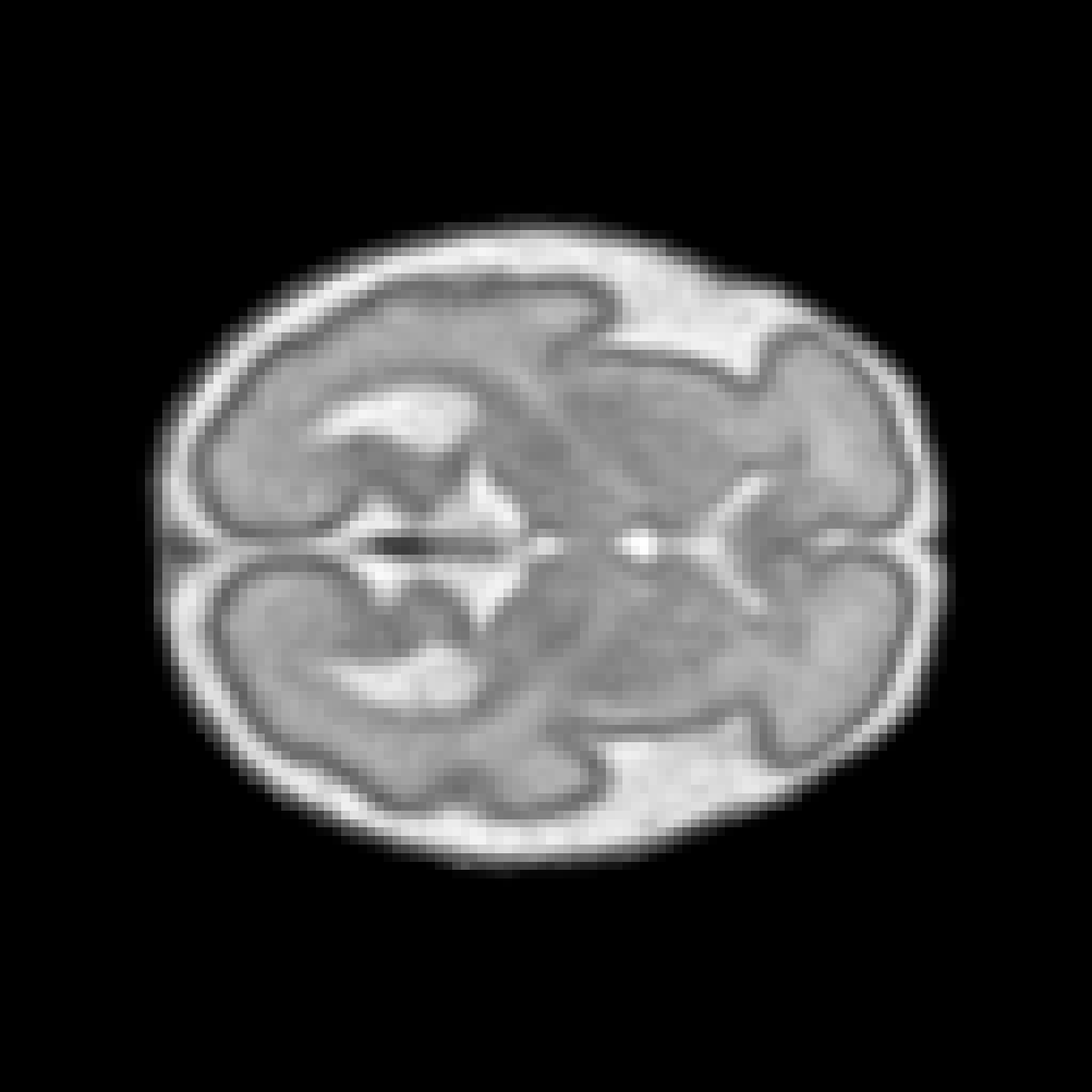}}}
        & \reflectbox{\rotatebox[origin=c]{90}{\includegraphics[height=1.1cm, trim=0.4cm 1.4cm 0.4cm 1.4cm, clip]{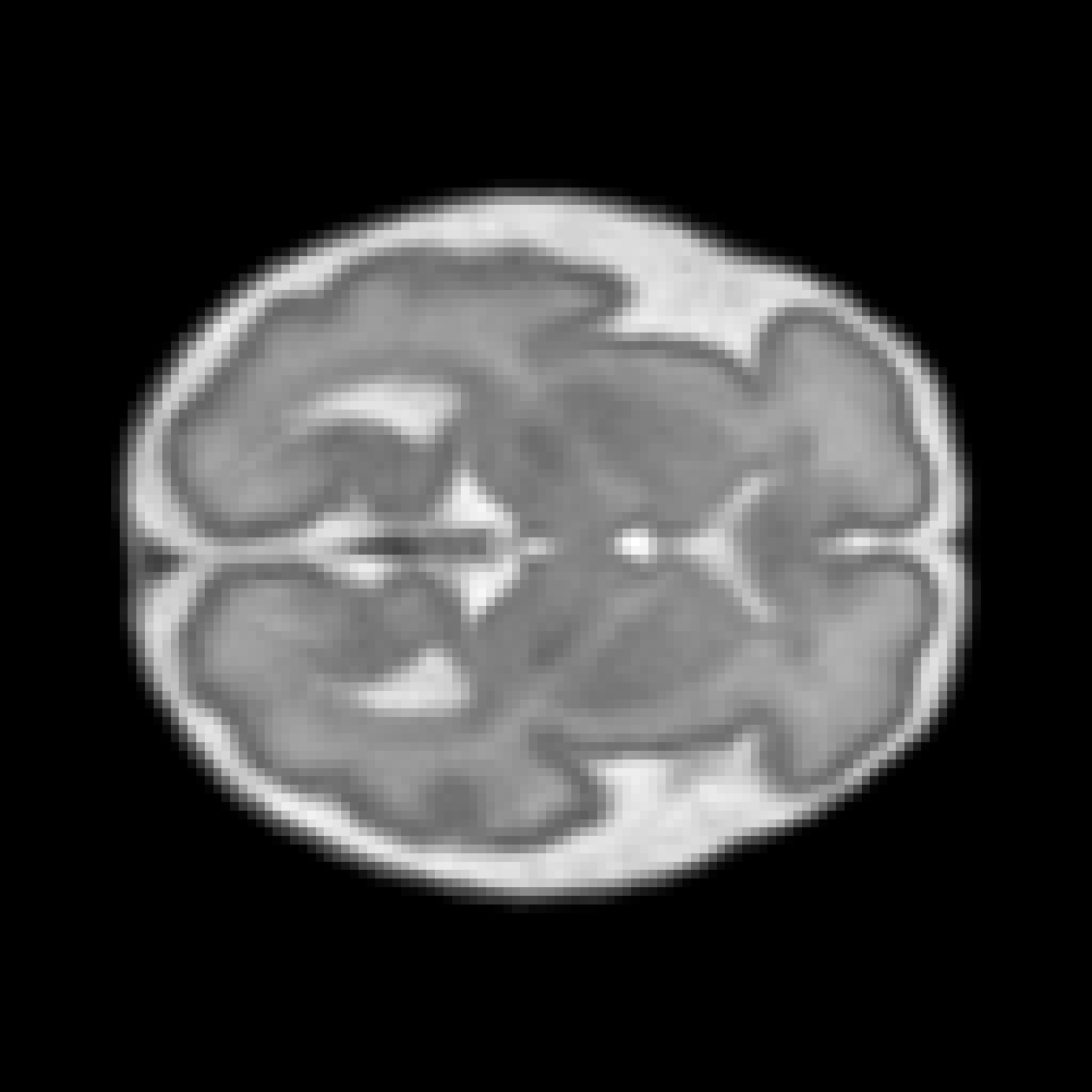}}}
        & \reflectbox{\rotatebox[origin=c]{90}{\includegraphics[height=1.1cm, trim=0.4cm 1.4cm 0.4cm 1.4cm, clip]{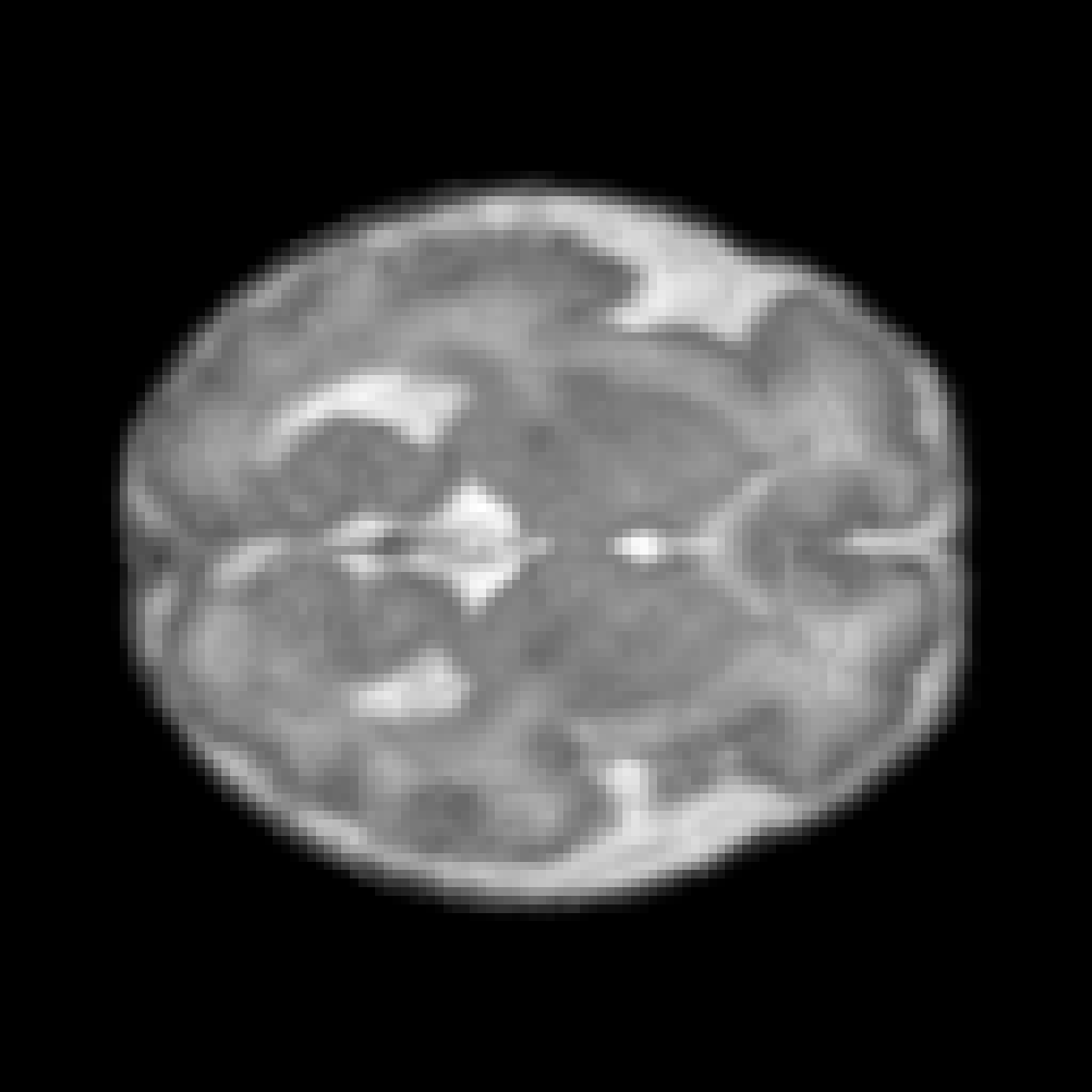}}}
        & \reflectbox{\rotatebox[origin=c]{90}{\includegraphics[height=1.1cm, trim=0.4cm 1.4cm 0.4cm 1.4cm, clip]{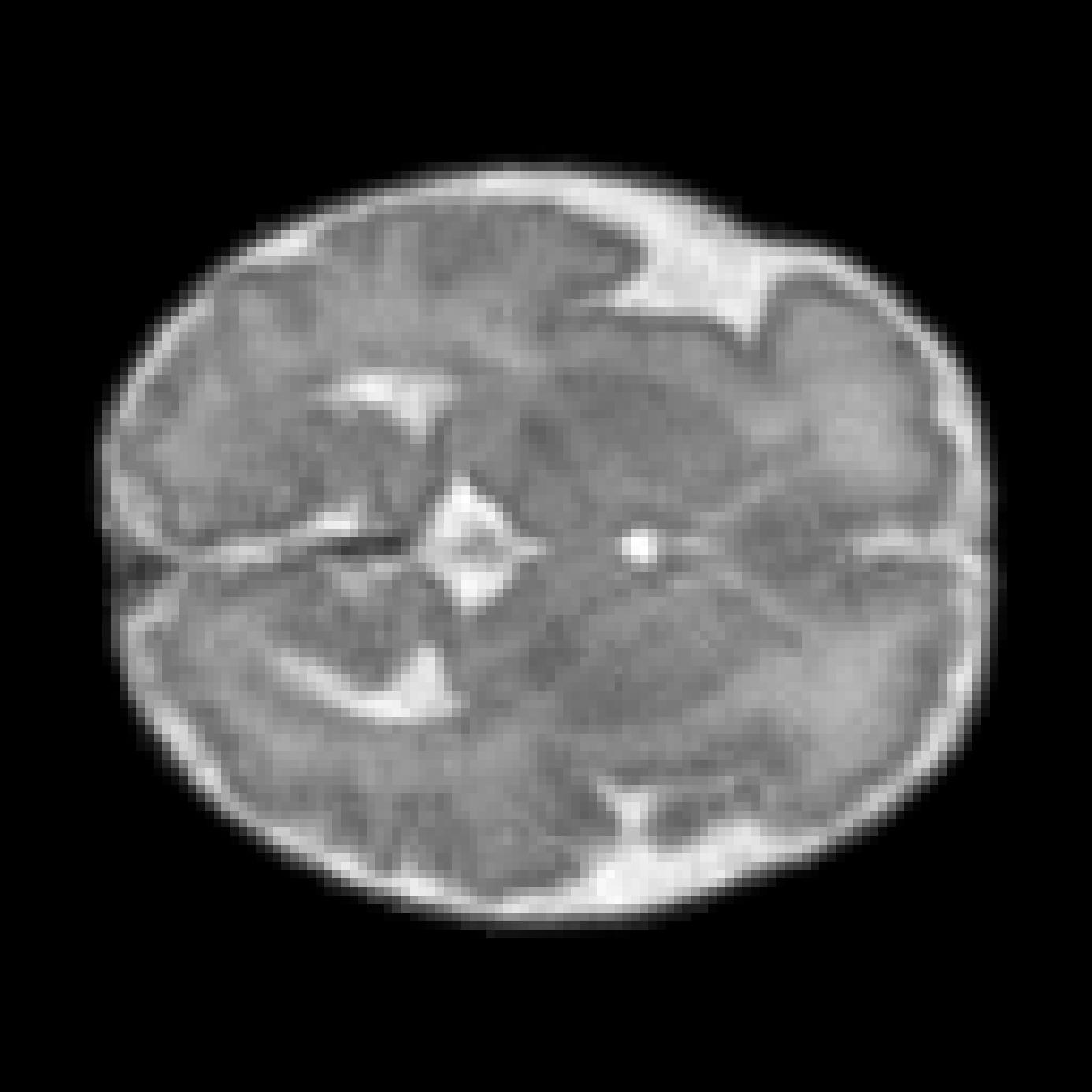}}}
        & \reflectbox{\rotatebox[origin=c]{90}{\includegraphics[height=1.1cm, trim=0.4cm 1.4cm 0.4cm 1.4cm, clip]{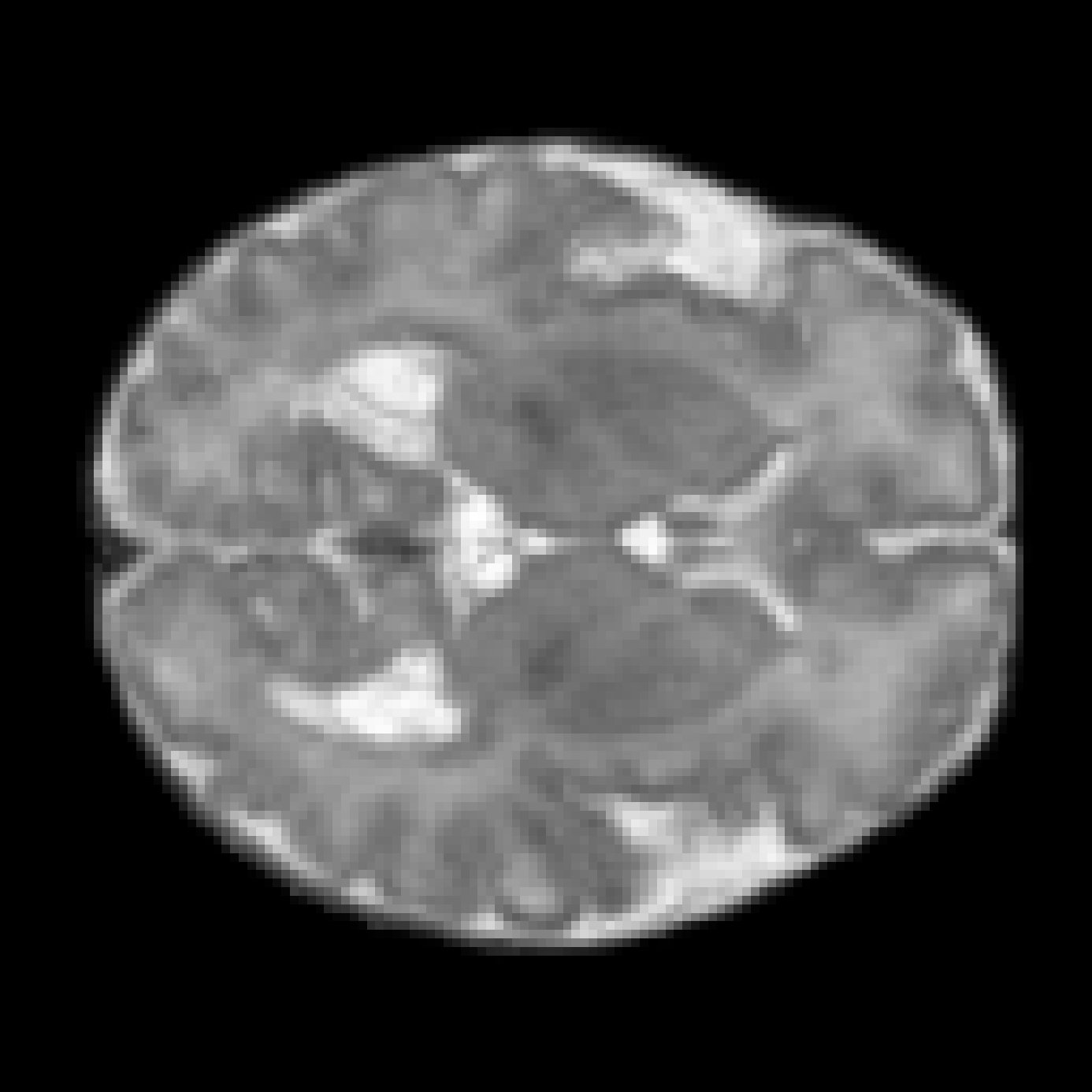}}} \\
        \raisebox{0.25cm}{\textbf{\shortstack{Ours\\Disc}}}
        & \reflectbox{\rotatebox[origin=c]{90}{\includegraphics[height=1.1cm, trim=0.4cm 1.4cm 0.4cm 1.4cm, clip]{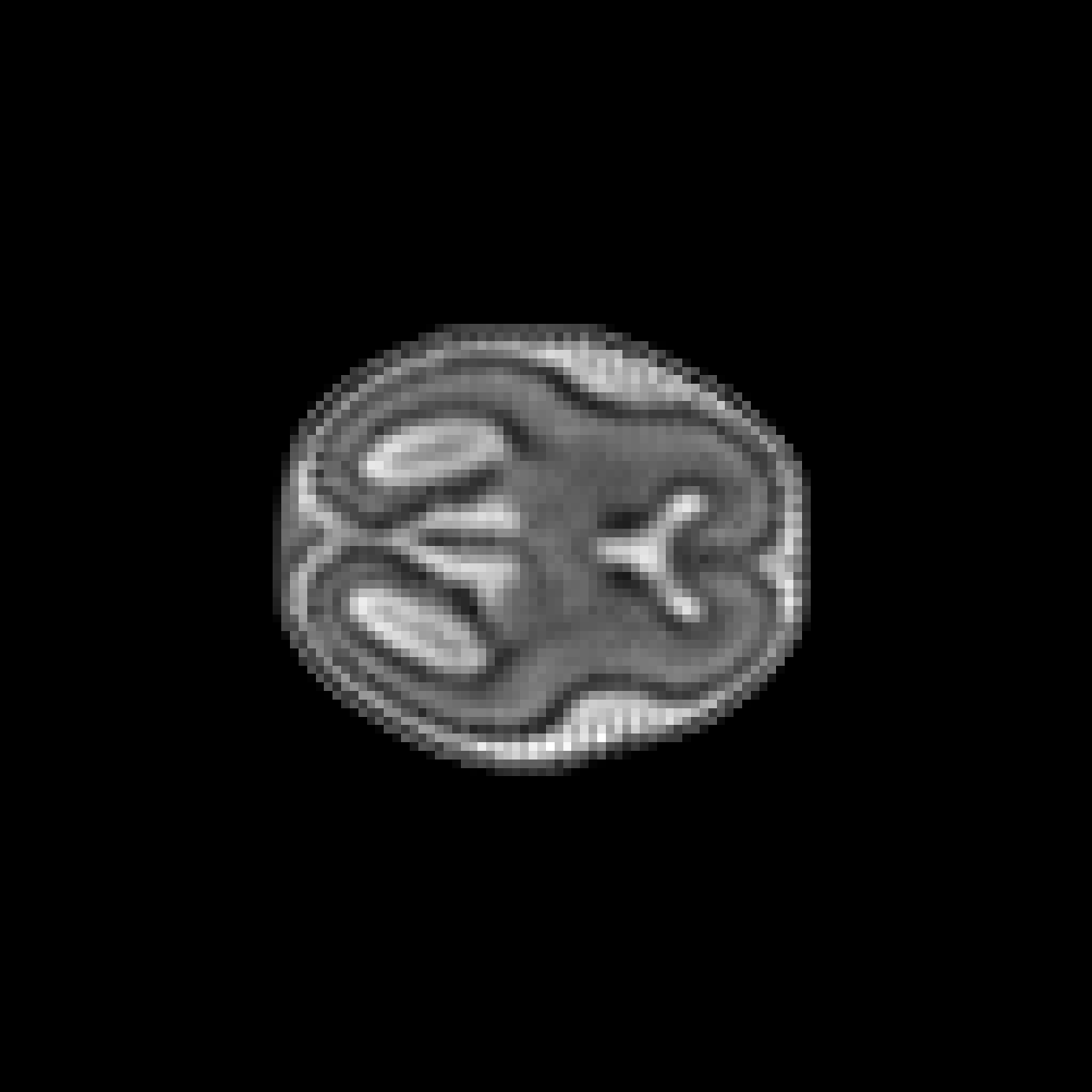}}}
        & \reflectbox{\rotatebox[origin=c]{90}{\includegraphics[height=1.1cm, trim=0.4cm 1.4cm 0.4cm 1.4cm, clip]{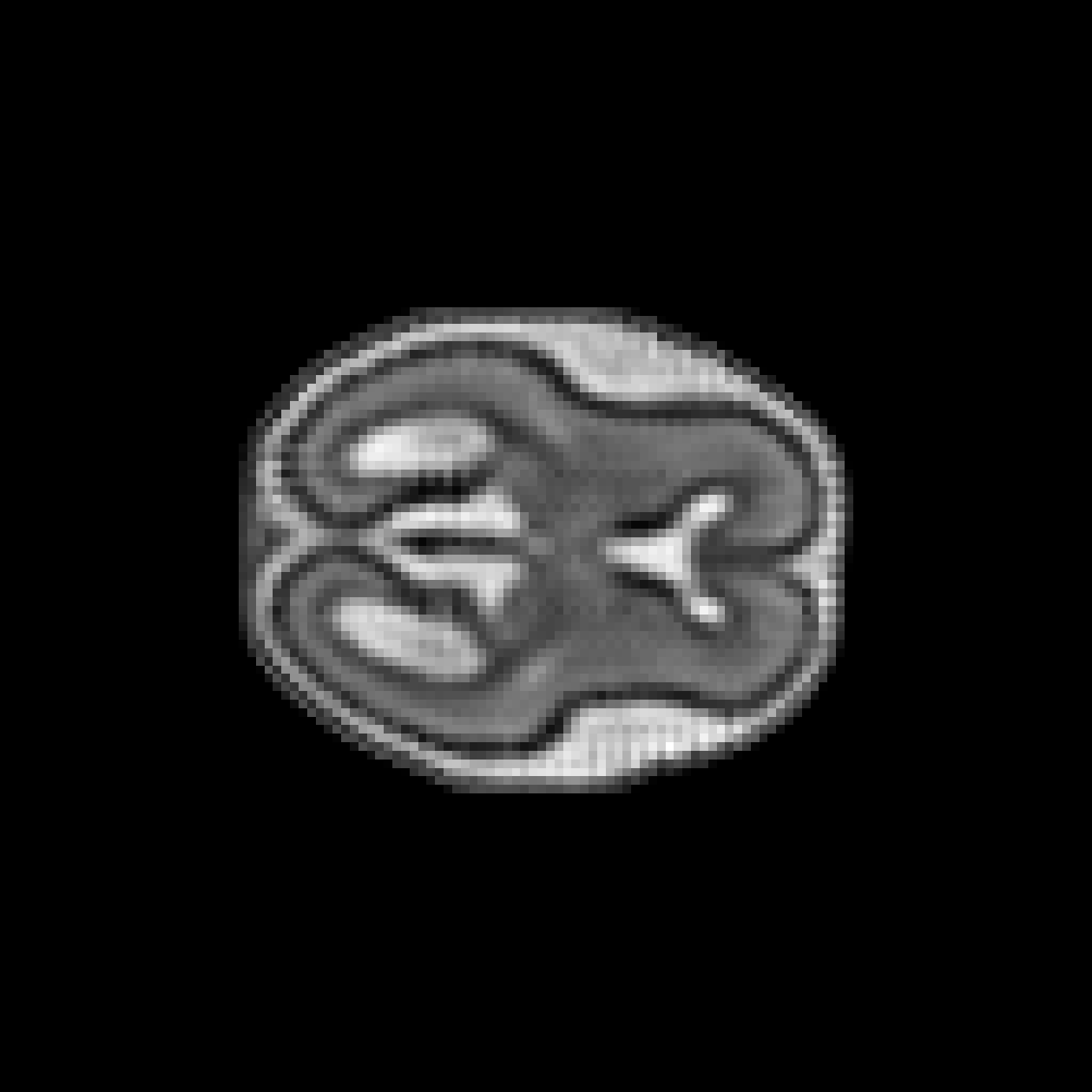}}}
        & \reflectbox{\rotatebox[origin=c]{90}{\includegraphics[height=1.1cm, trim=0.4cm 1.4cm 0.4cm 1.4cm, clip]{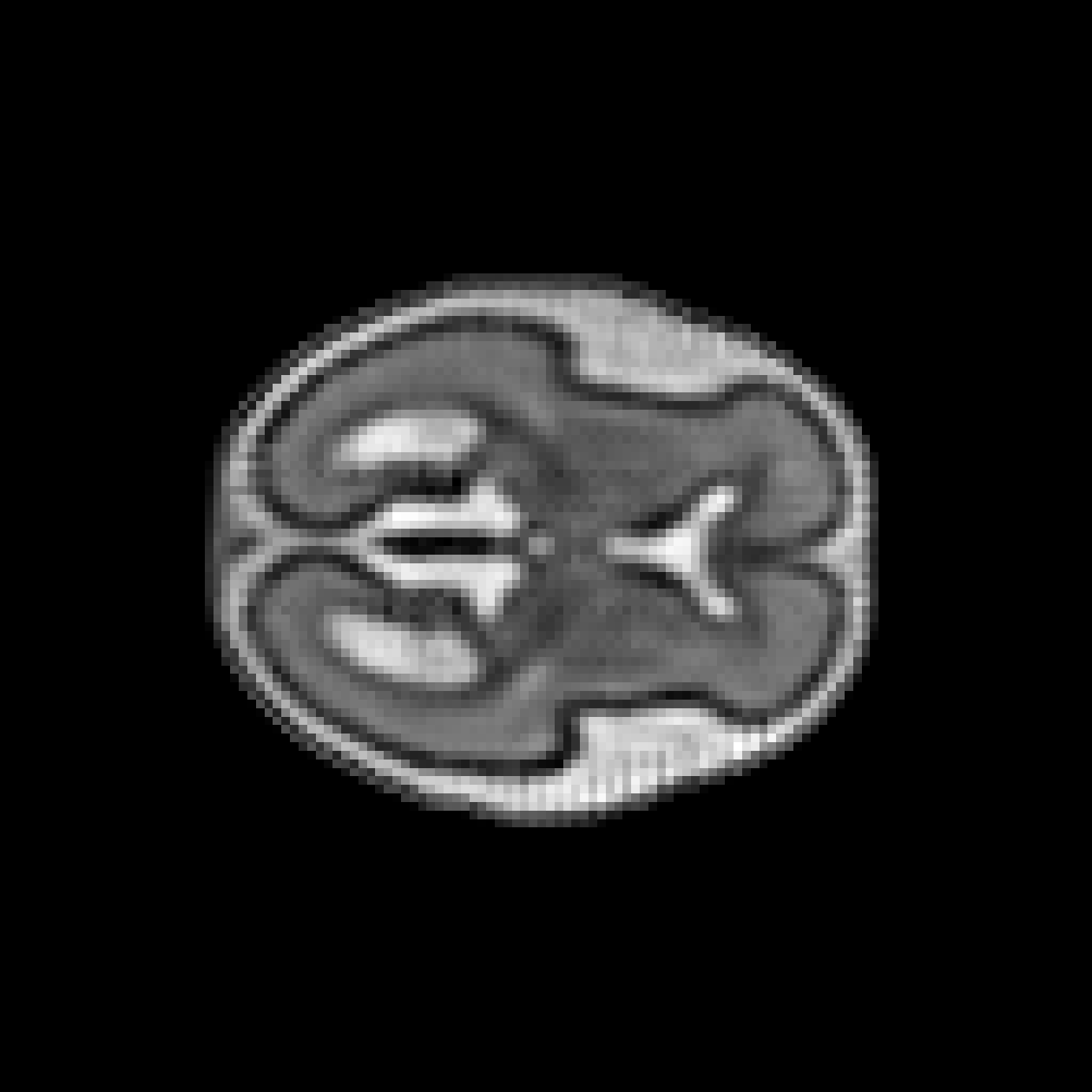}}}
        & \reflectbox{\rotatebox[origin=c]{90}{\includegraphics[height=1.1cm, trim=0.4cm 1.4cm 0.4cm 1.4cm, clip]{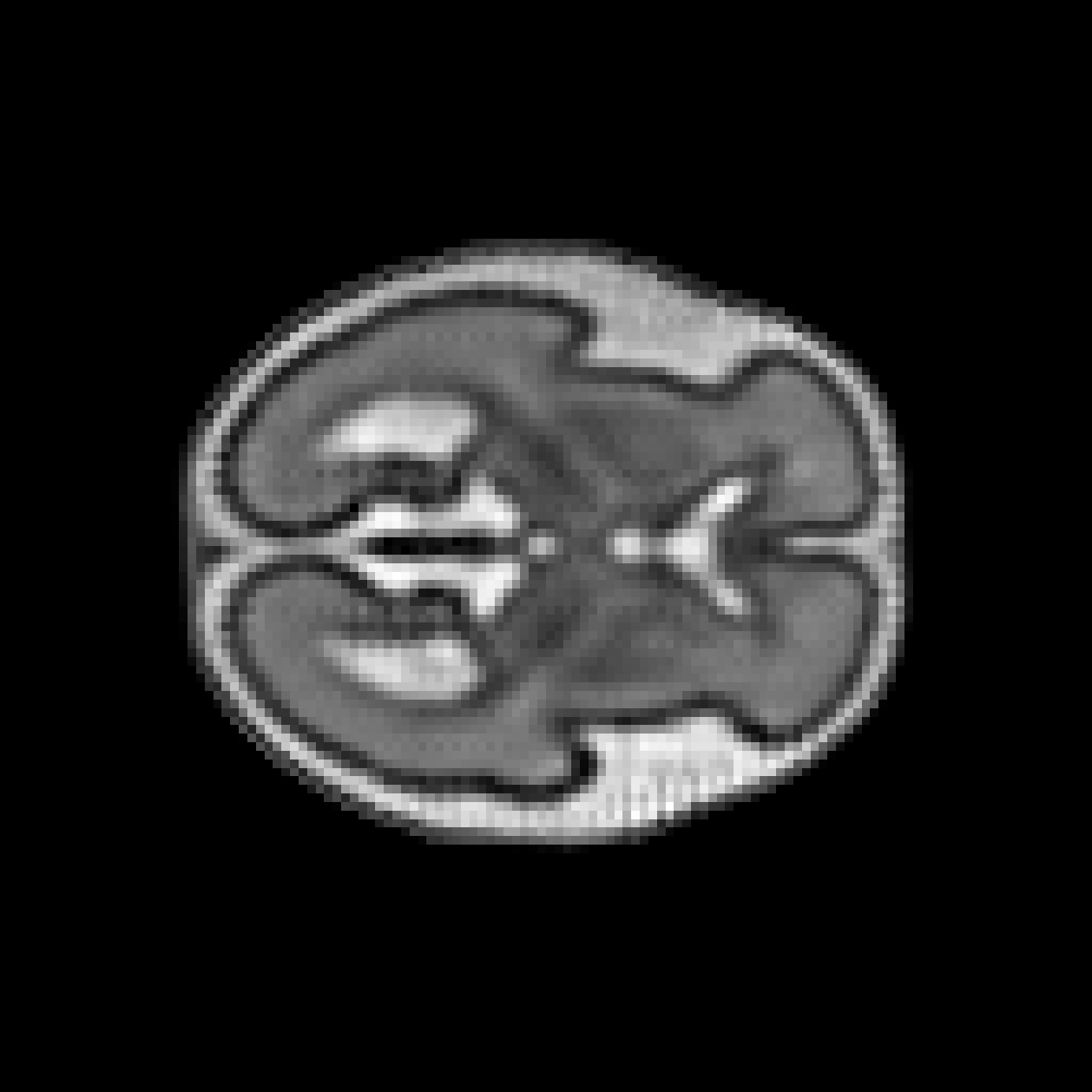}}}
        & \reflectbox{\rotatebox[origin=c]{90}{\includegraphics[height=1.1cm, trim=0.4cm 1.4cm 0.4cm 1.4cm, clip]{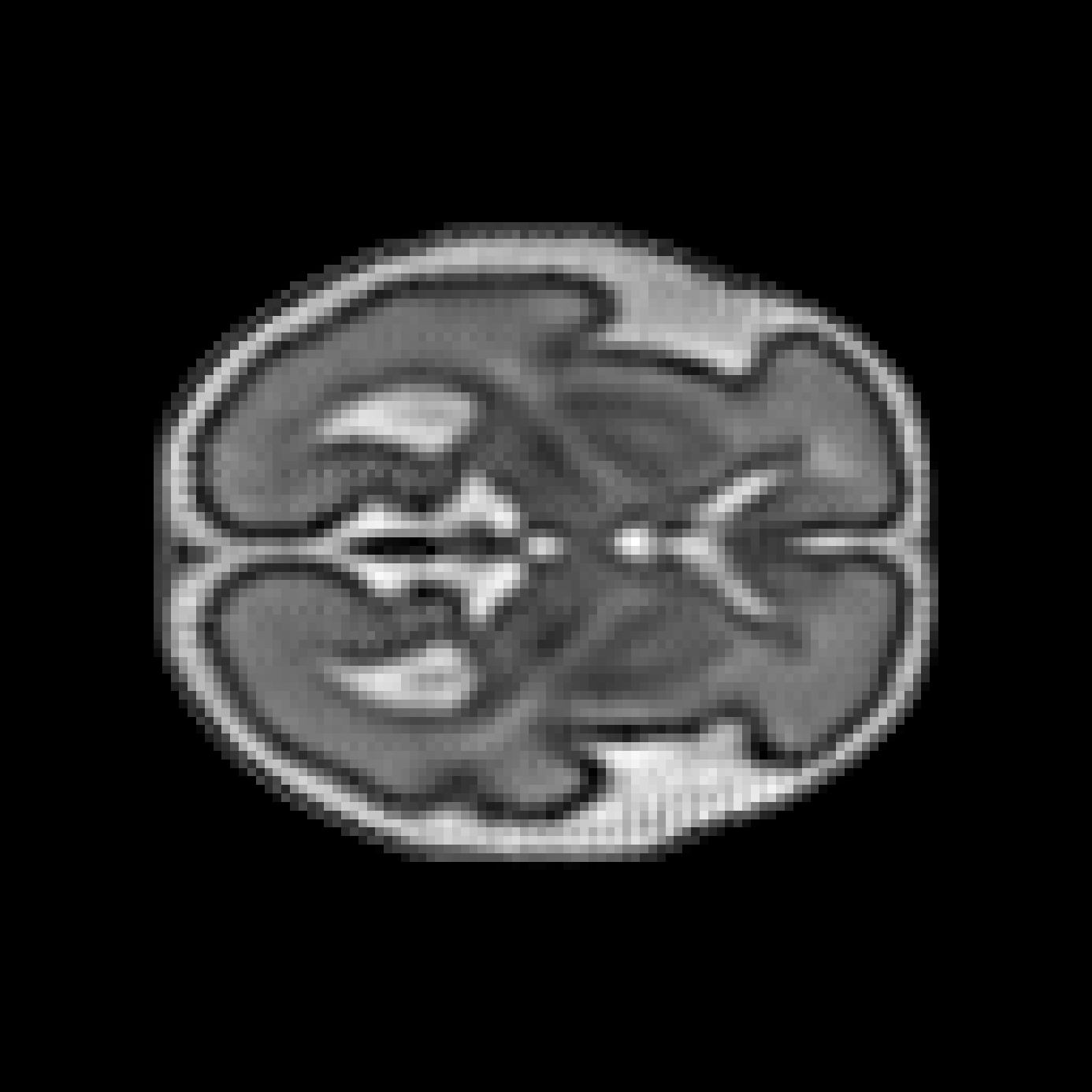}}}
        & \reflectbox{\rotatebox[origin=c]{90}{\includegraphics[height=1.1cm, trim=0.4cm 1.4cm 0.4cm 1.4cm, clip]{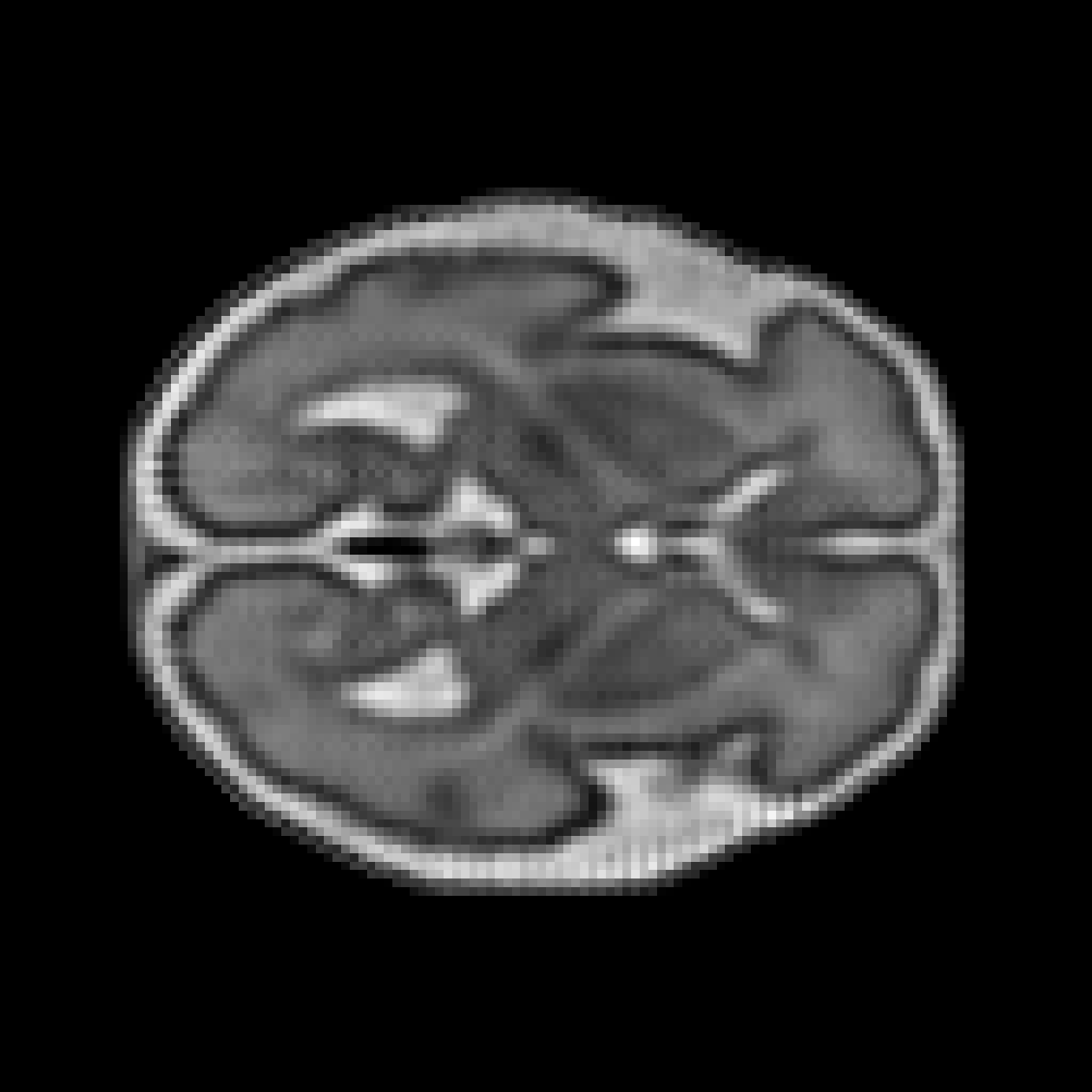}}}
        & \reflectbox{\rotatebox[origin=c]{90}{\includegraphics[height=1.1cm, trim=0.4cm 1.4cm 0.4cm 1.4cm, clip]{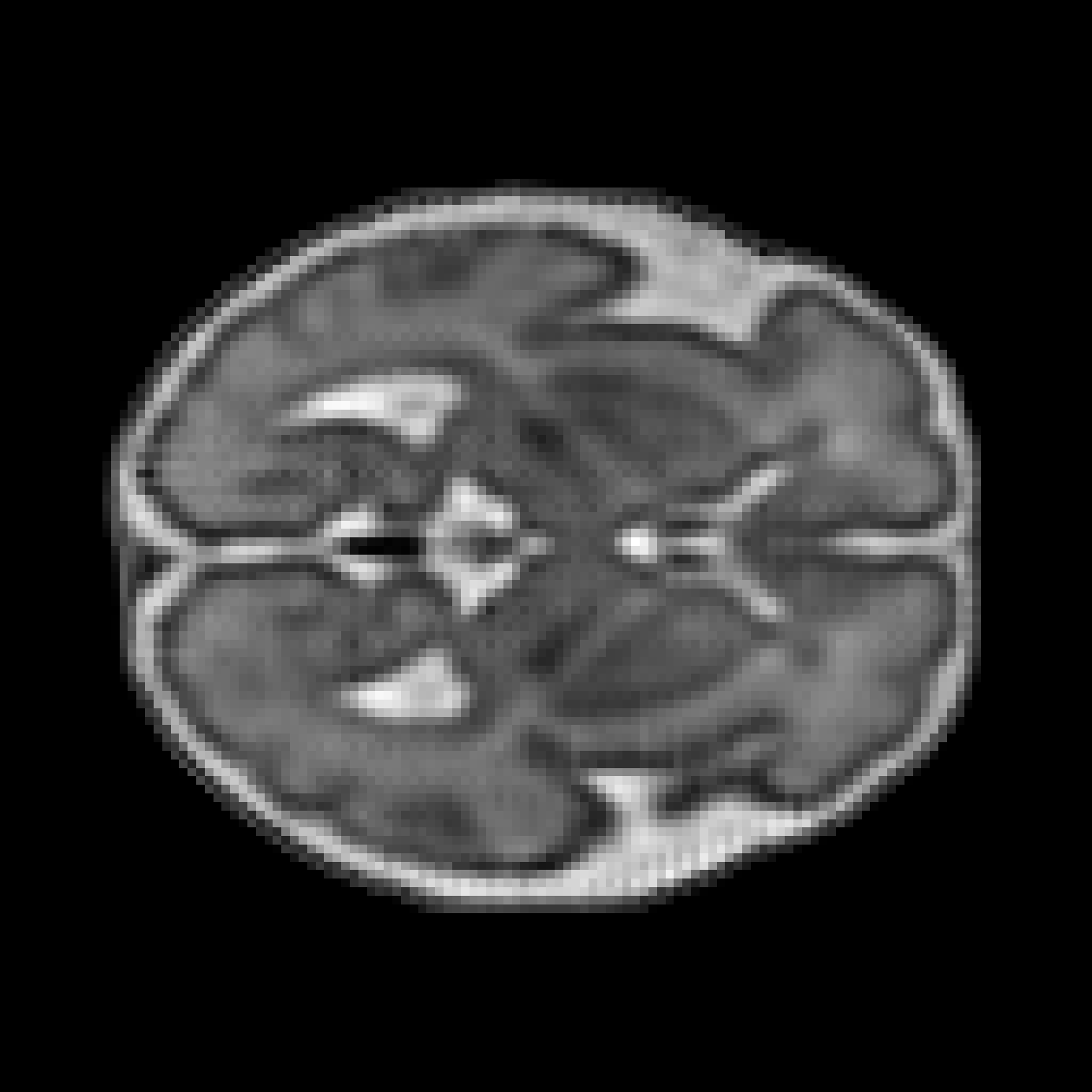}}}
        & \reflectbox{\rotatebox[origin=c]{90}{\includegraphics[height=1.1cm, trim=0.4cm 1.4cm 0.4cm 1.4cm, clip]{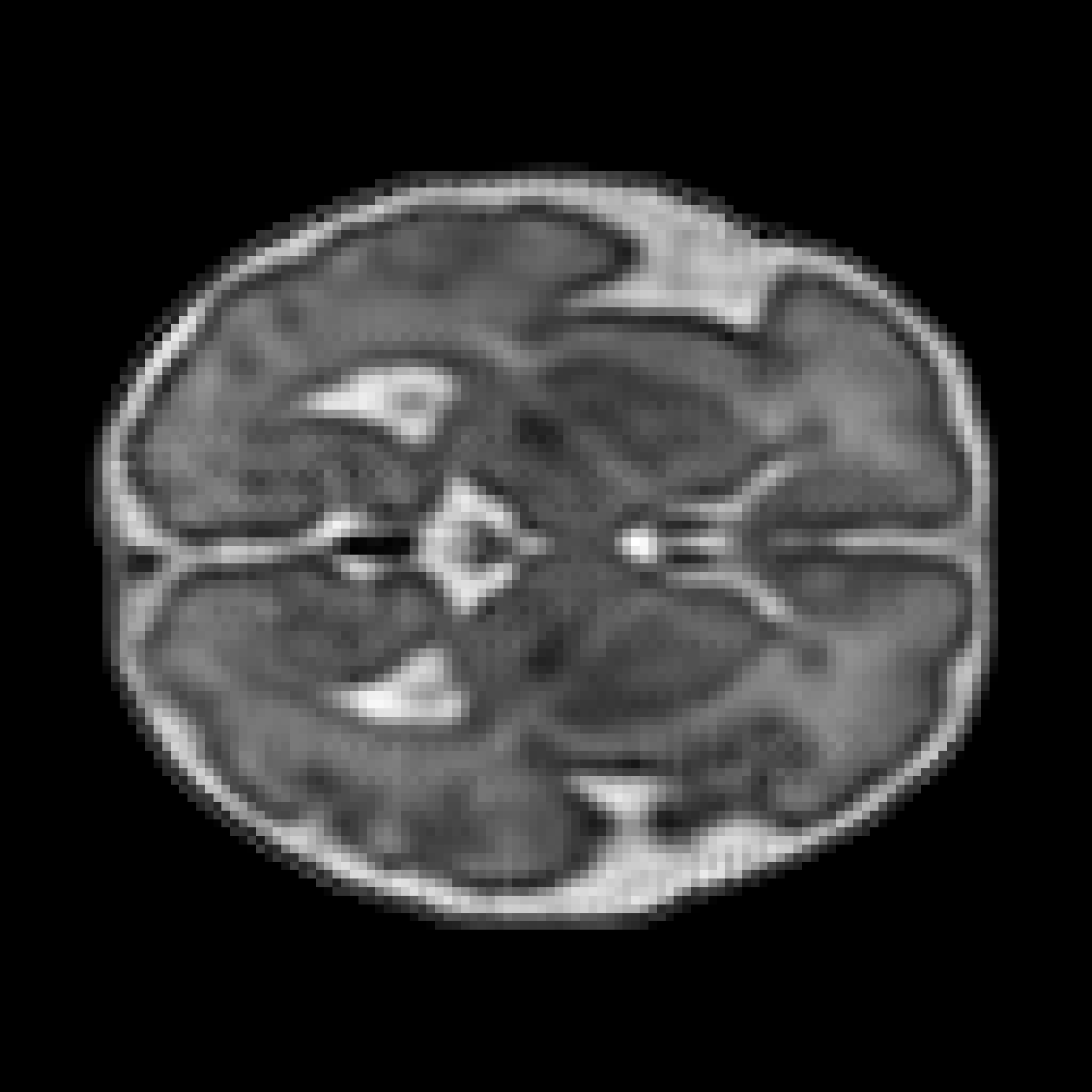}}}
        & \reflectbox{\rotatebox[origin=c]{90}{\includegraphics[height=1.1cm, trim=0.4cm 1.4cm 0.4cm 1.4cm, clip]{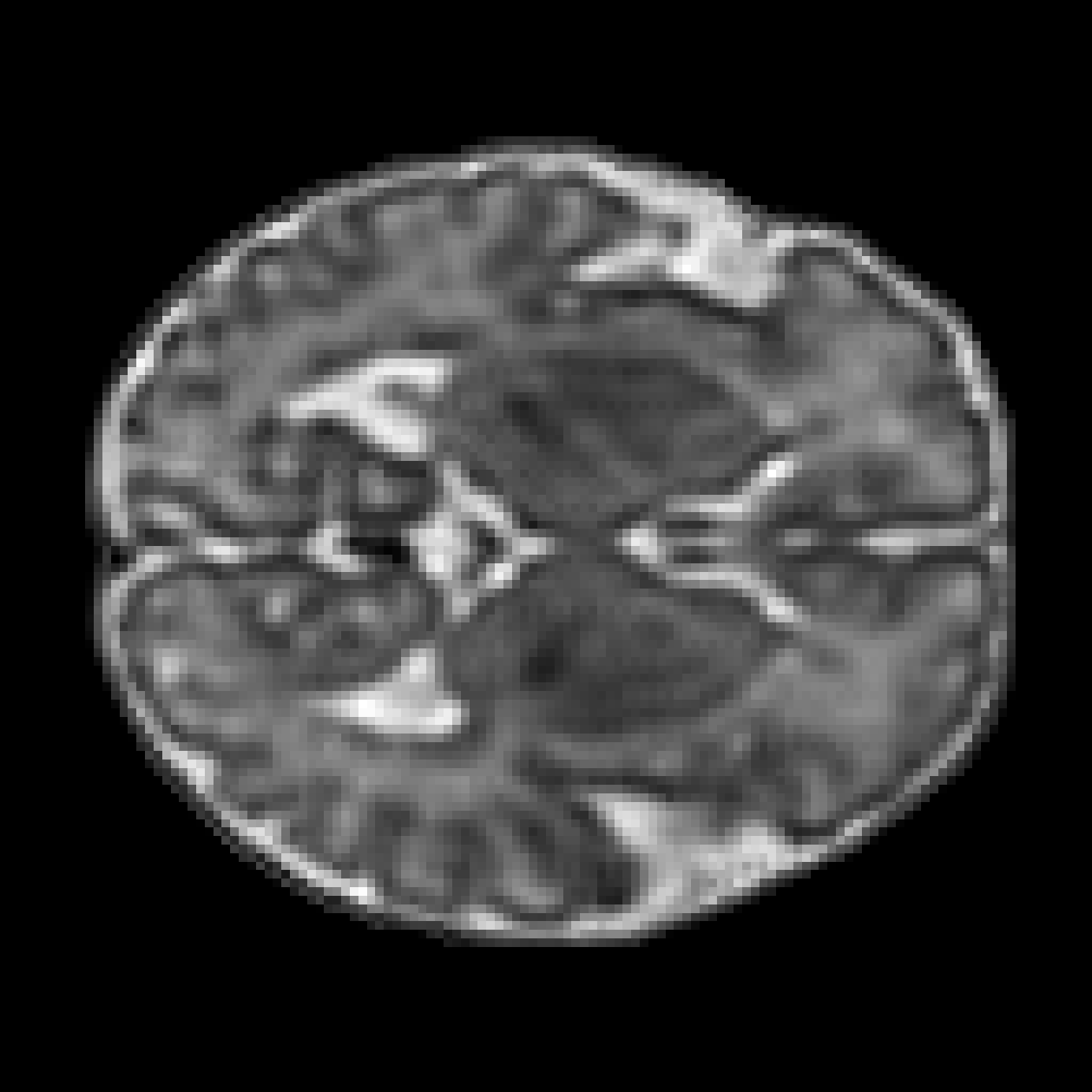}}}
    \end{tabular}
\begin{minipage}[t]{0.78\textwidth}
    \centering
    \begin{tabular}{c@{}c@{}c@{}c@{}c@{}c@{}c@{}}
        & & & & & & \\
        \raisebox{0cm}{\hspace{-1.5cm}\textbf{b)}} & Atlas & Fix & Warp & Atlas & Fix & Warp \\
        \raisebox{0.4cm}{\textbf{ANTs}} & \reflectbox{\rotatebox[origin=c]{90}{\includegraphics[height=1.1cm, trim=0.4cm 1.4cm 0.4cm 1.4cm, clip]{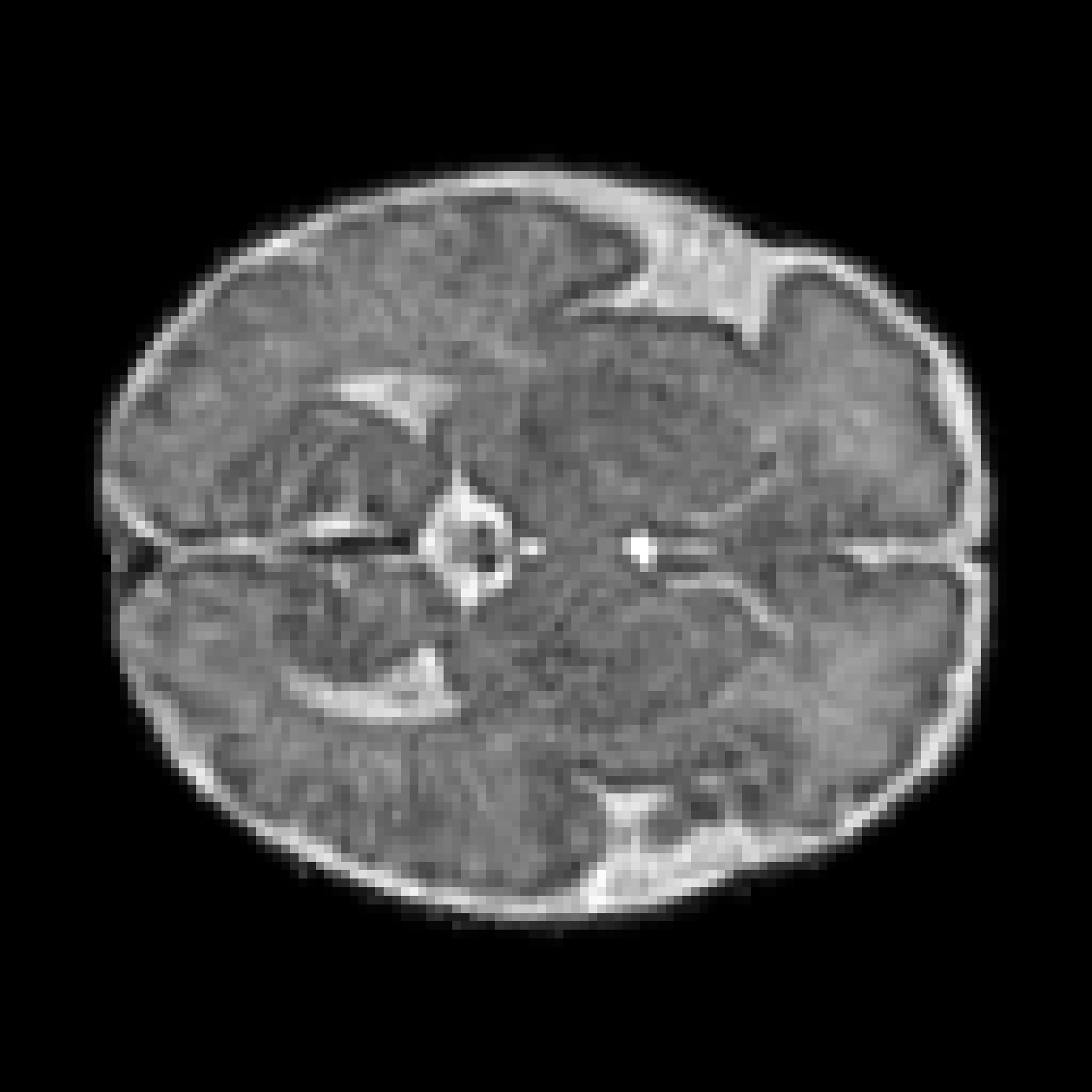}}}
        & \reflectbox{\rotatebox[origin=c]{90}{\includegraphics[height=1.1cm, trim=0.4cm 1.4cm 0.4cm 1.4cm, clip]{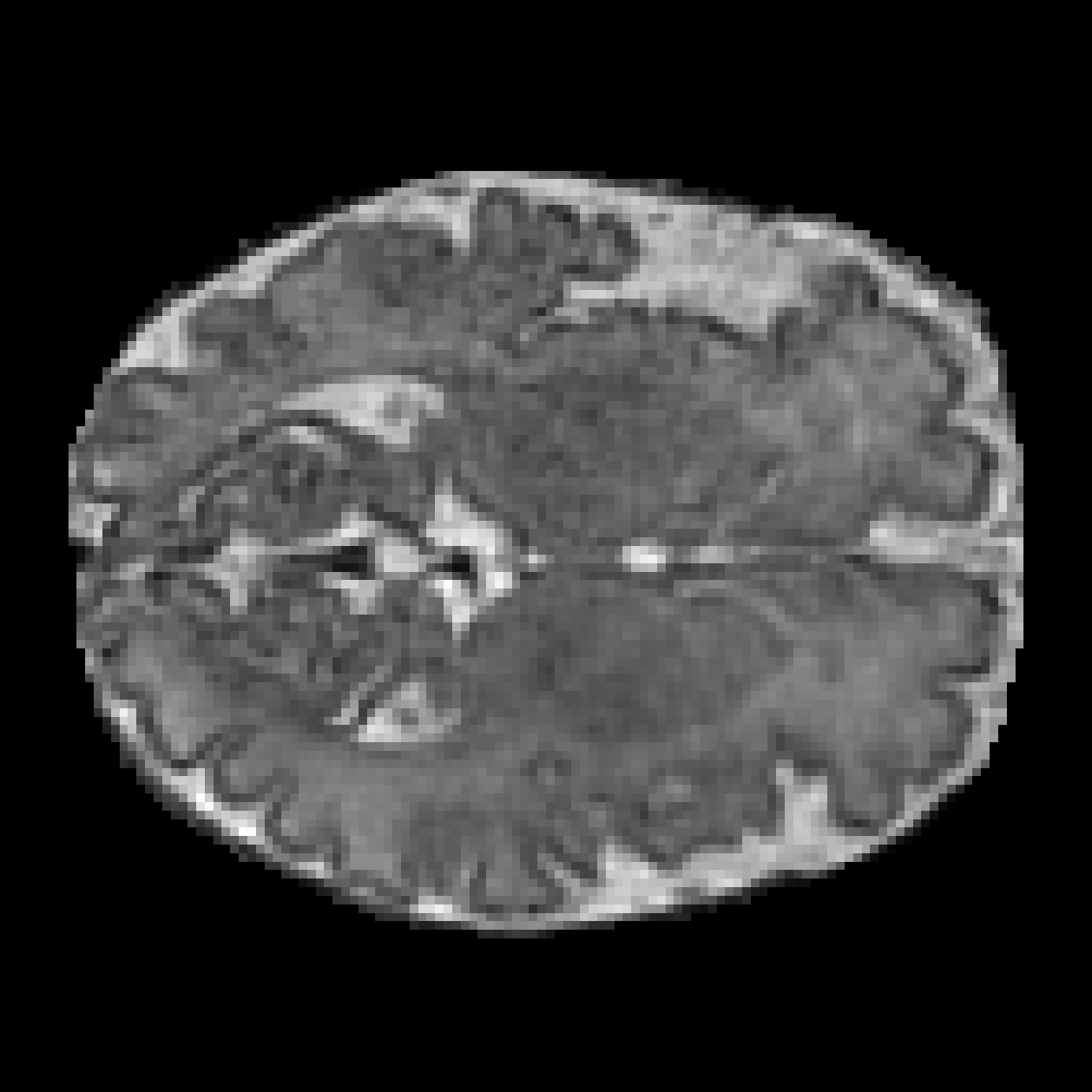}}}
        & \reflectbox{\rotatebox[origin=c]{90}{\includegraphics[height=1.1cm, trim=0.4cm 1.4cm 0.4cm 1.4cm, clip]{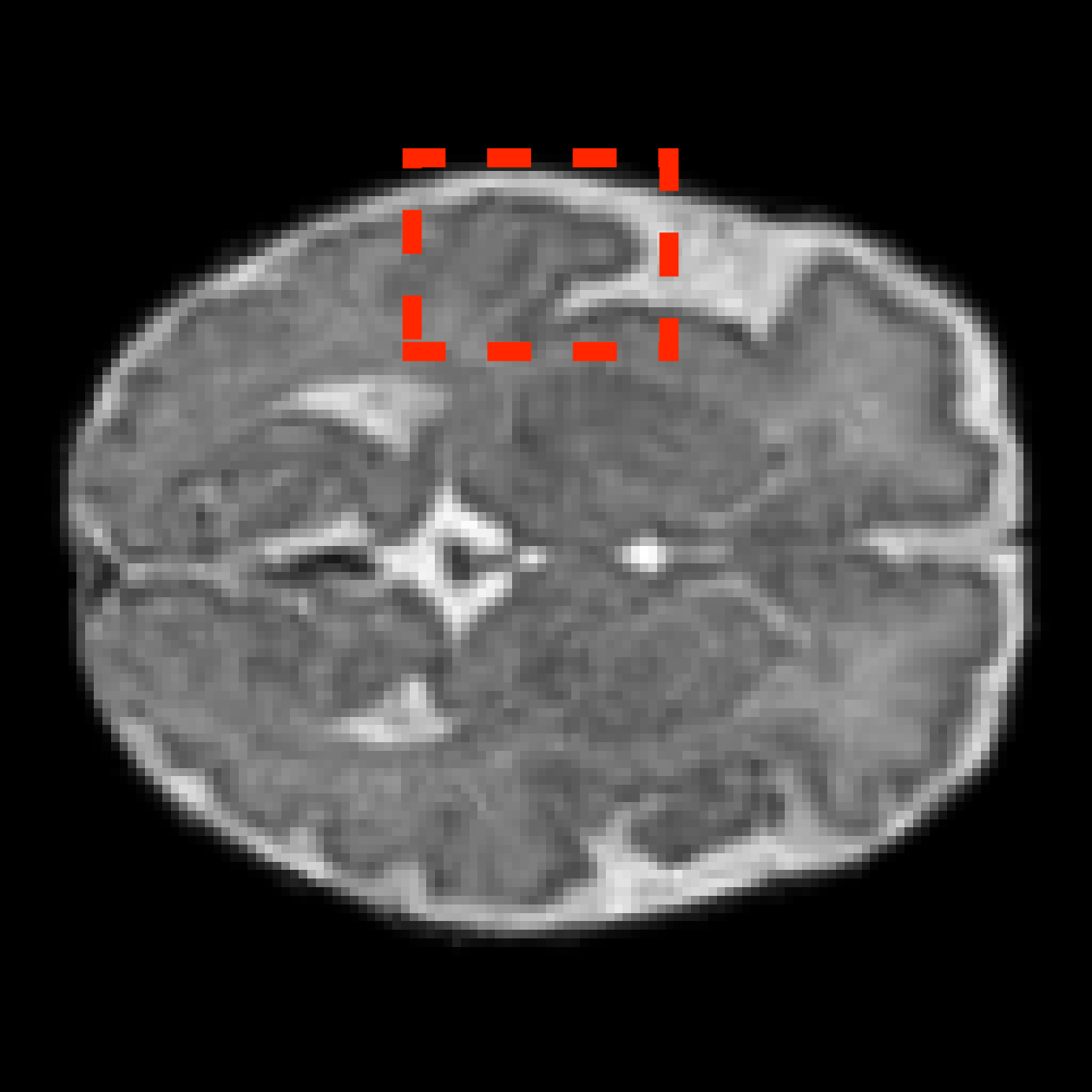}}}
        & \raisebox{-0.095cm}{\reflectbox{\rotatebox[origin=c]{90}{\includegraphics[height=1.29cm, trim=0.4cm 1.4cm 0.4cm 1.4cm, clip]{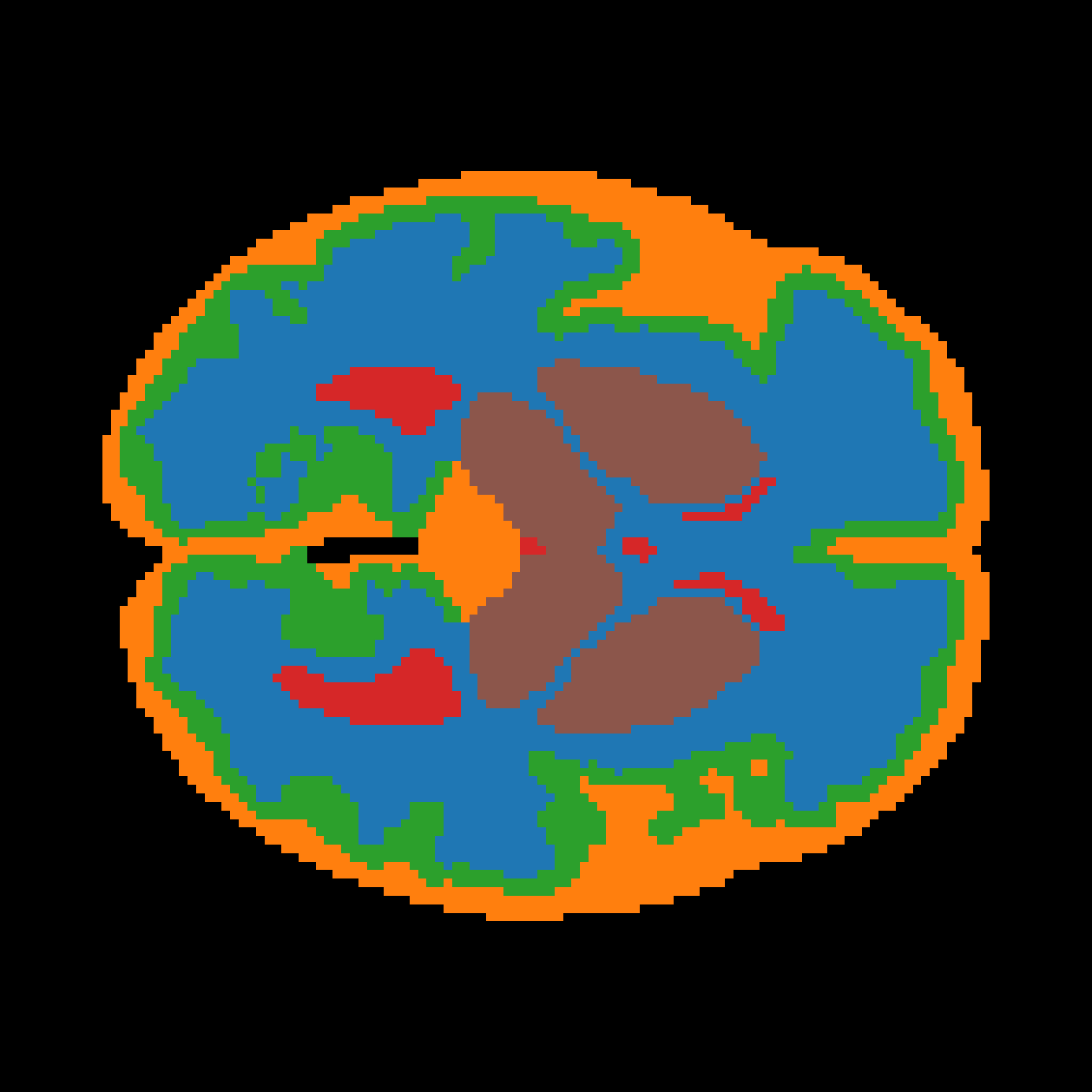}}}}
        & \raisebox{-0.095cm}{\reflectbox{\rotatebox[origin=c]{90}{\includegraphics[height=1.29cm, trim=0.4cm 1.4cm 0.4cm 1.4cm, clip]{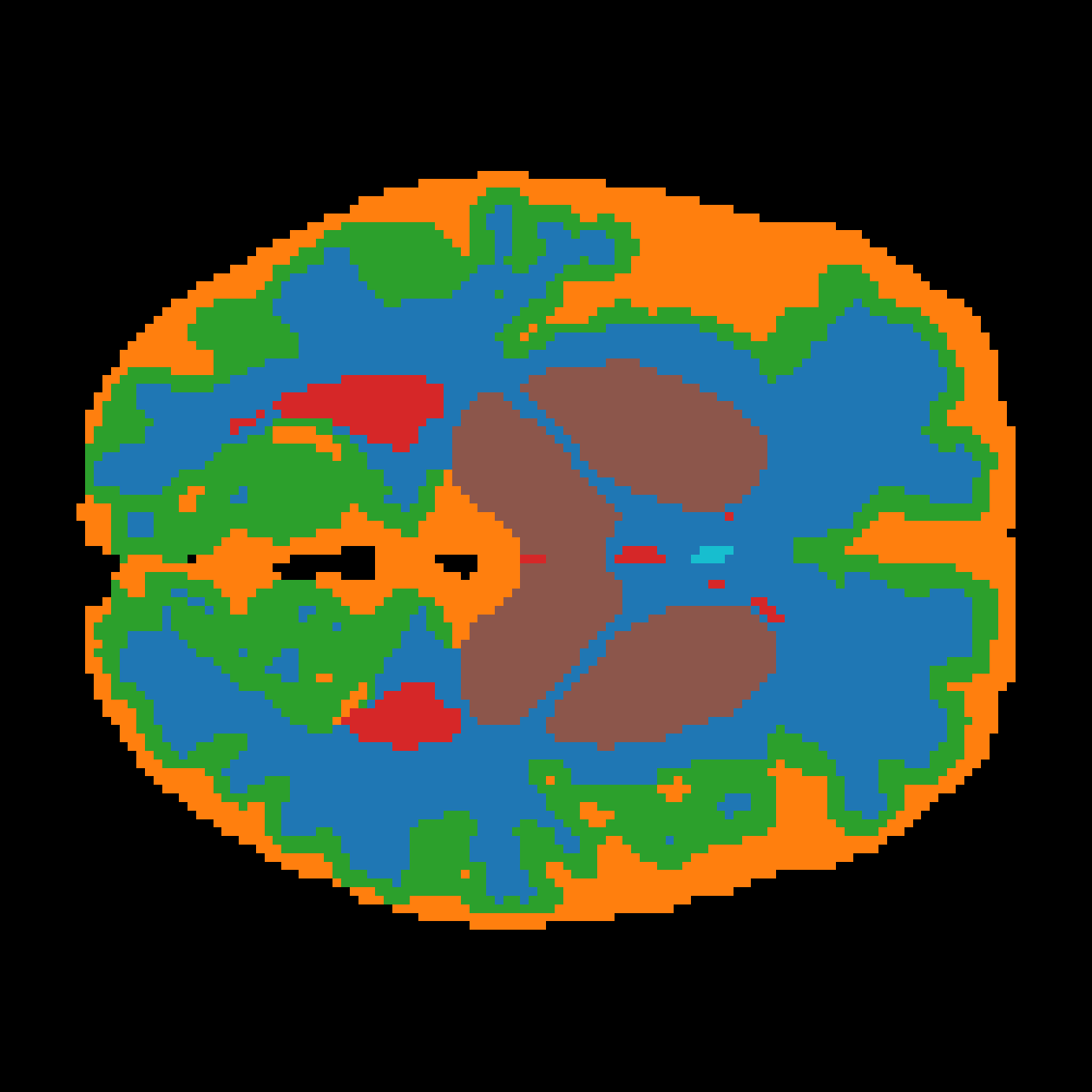}}}}
        & \raisebox{-0.095cm}{\reflectbox{\rotatebox[origin=c]{90}{\includegraphics[height=1.29cm, trim=0.4cm 1.4cm 0.4cm 1.4cm, clip]{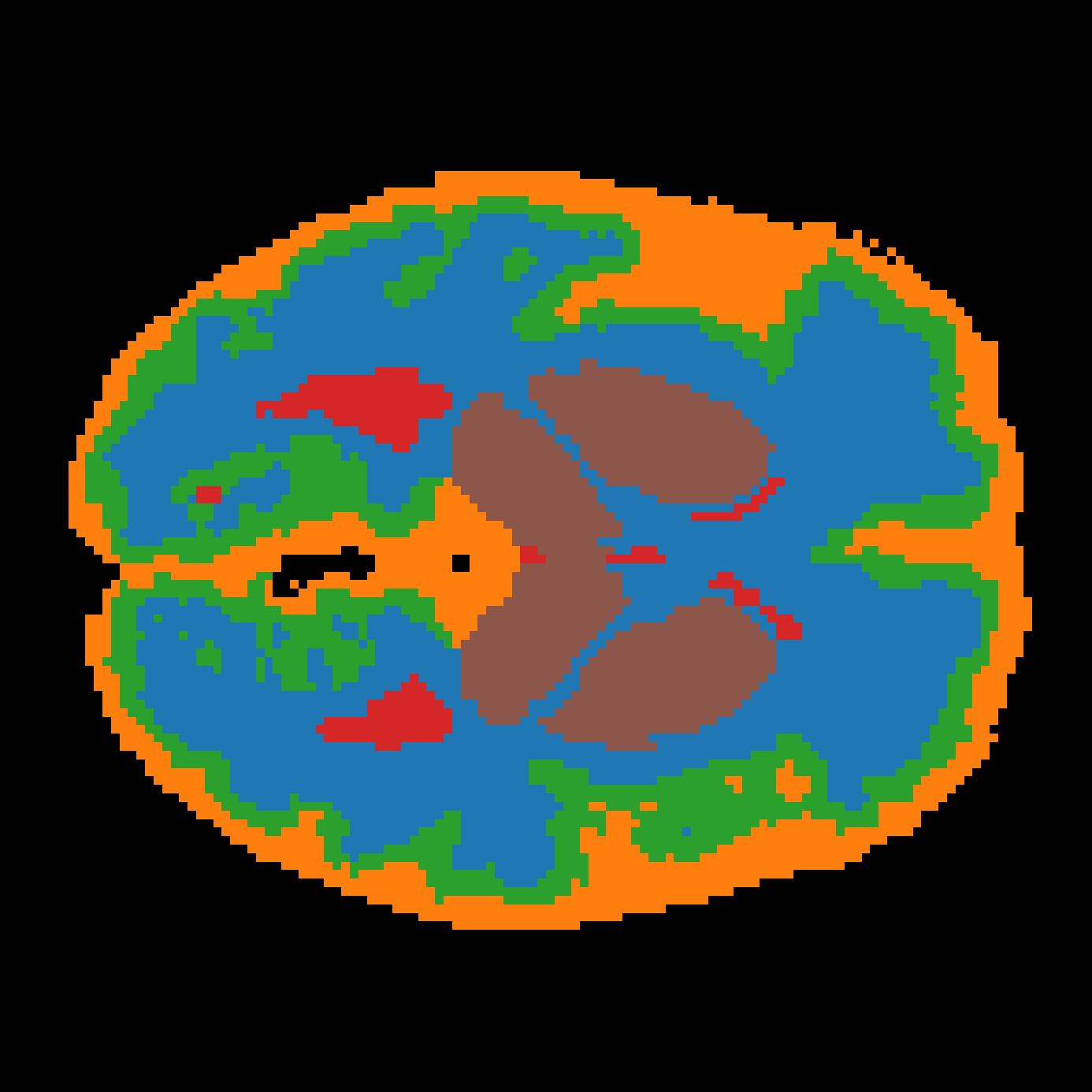}}}} \\
    
        \raisebox{0.25cm}{\textbf{\shortstack{Multi\\Morph}}} & \reflectbox{\rotatebox[origin=c]{90}{\includegraphics[height=1.1cm, trim=0.4cm 1.4cm 0.4cm 1.4cm, clip]{MultiMorph/atlas_35.png}}} 
        & \reflectbox{\rotatebox[origin=c]{90}{\includegraphics[height=1.1cm, trim=0.4cm 1.4cm 0.4cm 1.4cm, clip]{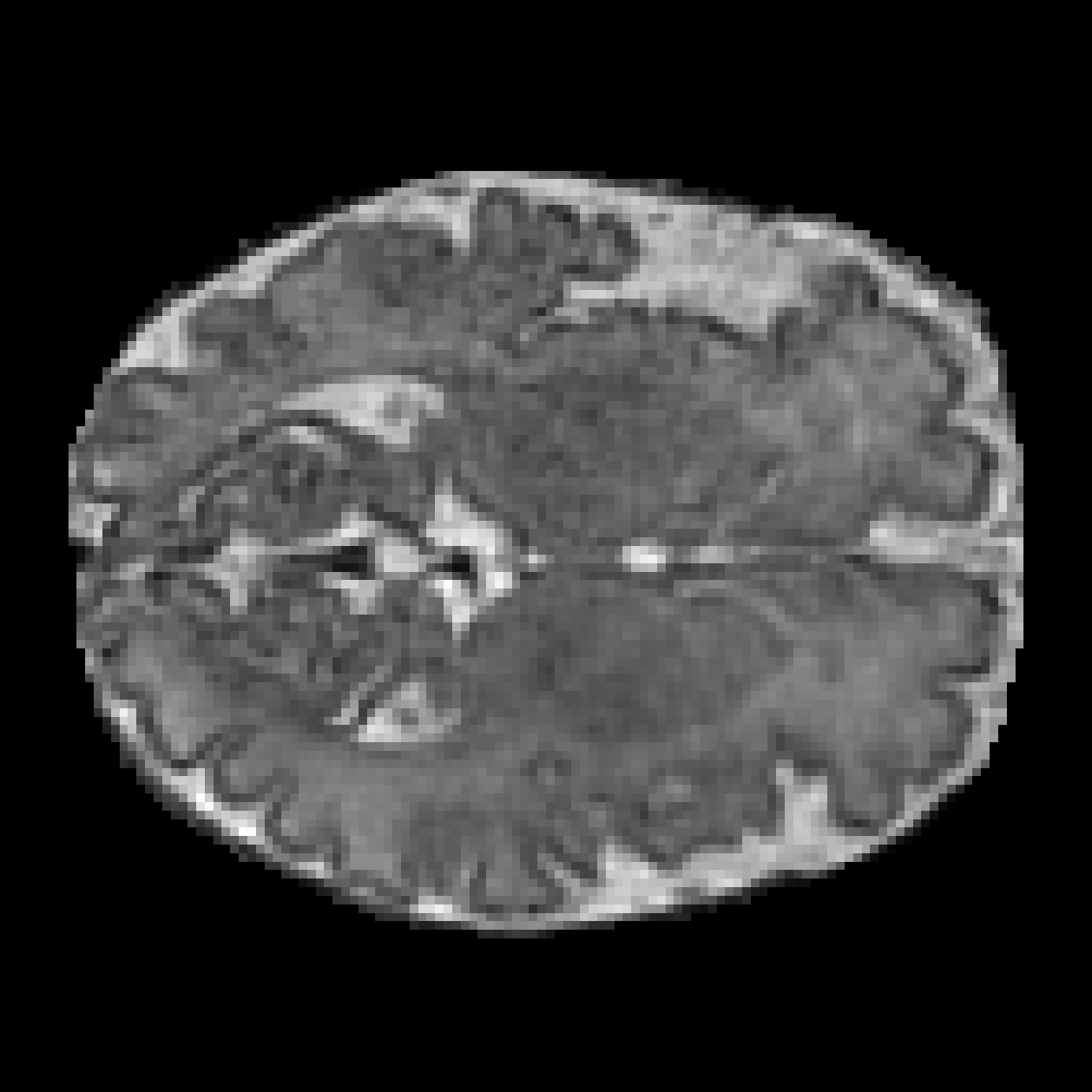}}} 
        & \reflectbox{\rotatebox[origin=c]{90}{\includegraphics[height=1.1cm, trim=0.4cm 1.4cm 0.4cm 1.4cm, clip]{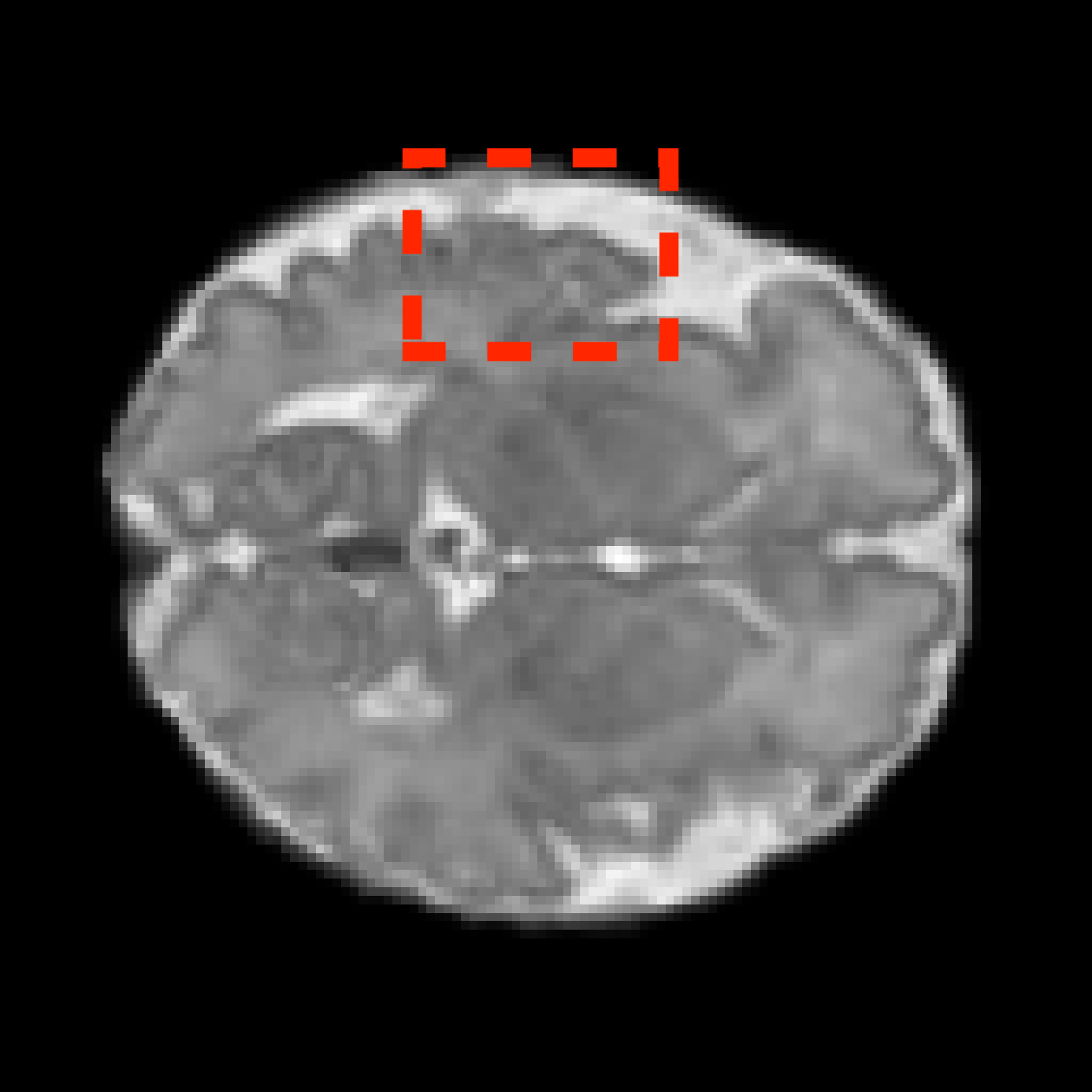}}} 
        & \raisebox{-0.095cm}{\reflectbox{\rotatebox[origin=c]{90}{\includegraphics[height=1.29cm, trim=0.4cm 1.4cm 0.4cm 1.4cm, clip]{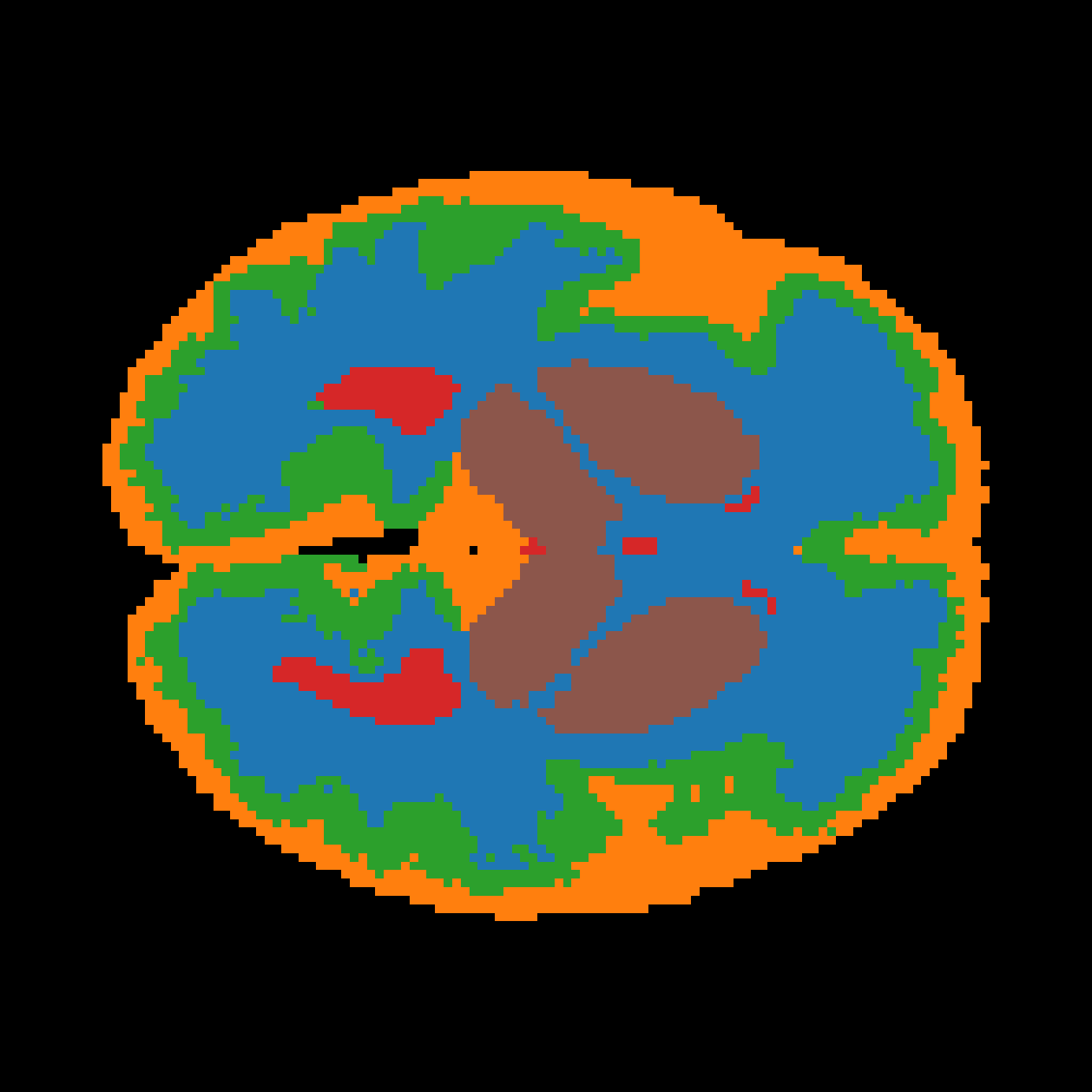}}}}
        & \raisebox{-0.095cm}{\reflectbox{\rotatebox[origin=c]{90}{\includegraphics[height=1.29cm, trim=0.4cm 1.4cm 0.4cm 1.4cm, clip]{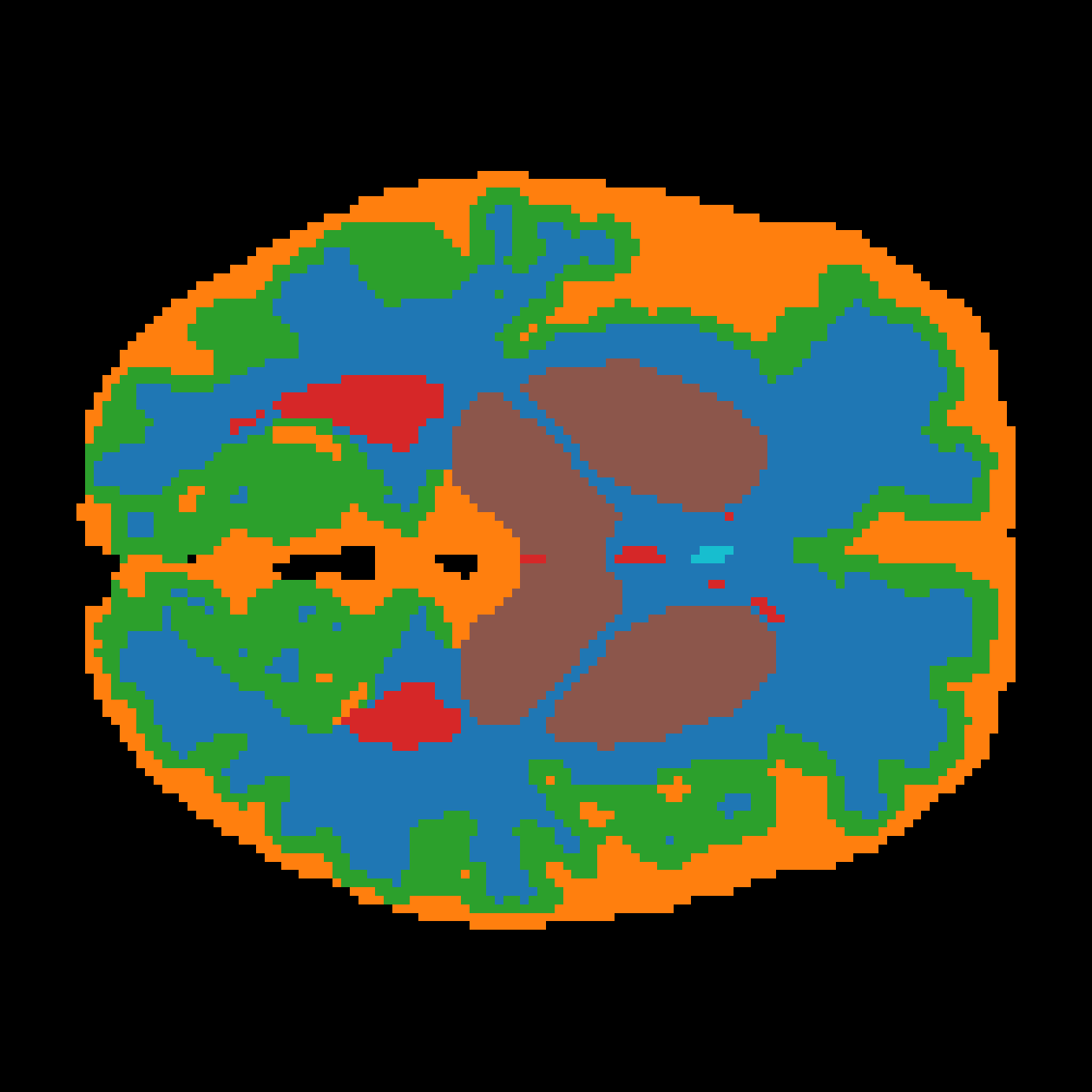}}}}
        & \raisebox{-0.095cm}{\reflectbox{\rotatebox[origin=c]{90}{\includegraphics[height=1.29cm, trim=0.4cm 1.4cm 0.4cm 1.4cm, clip]{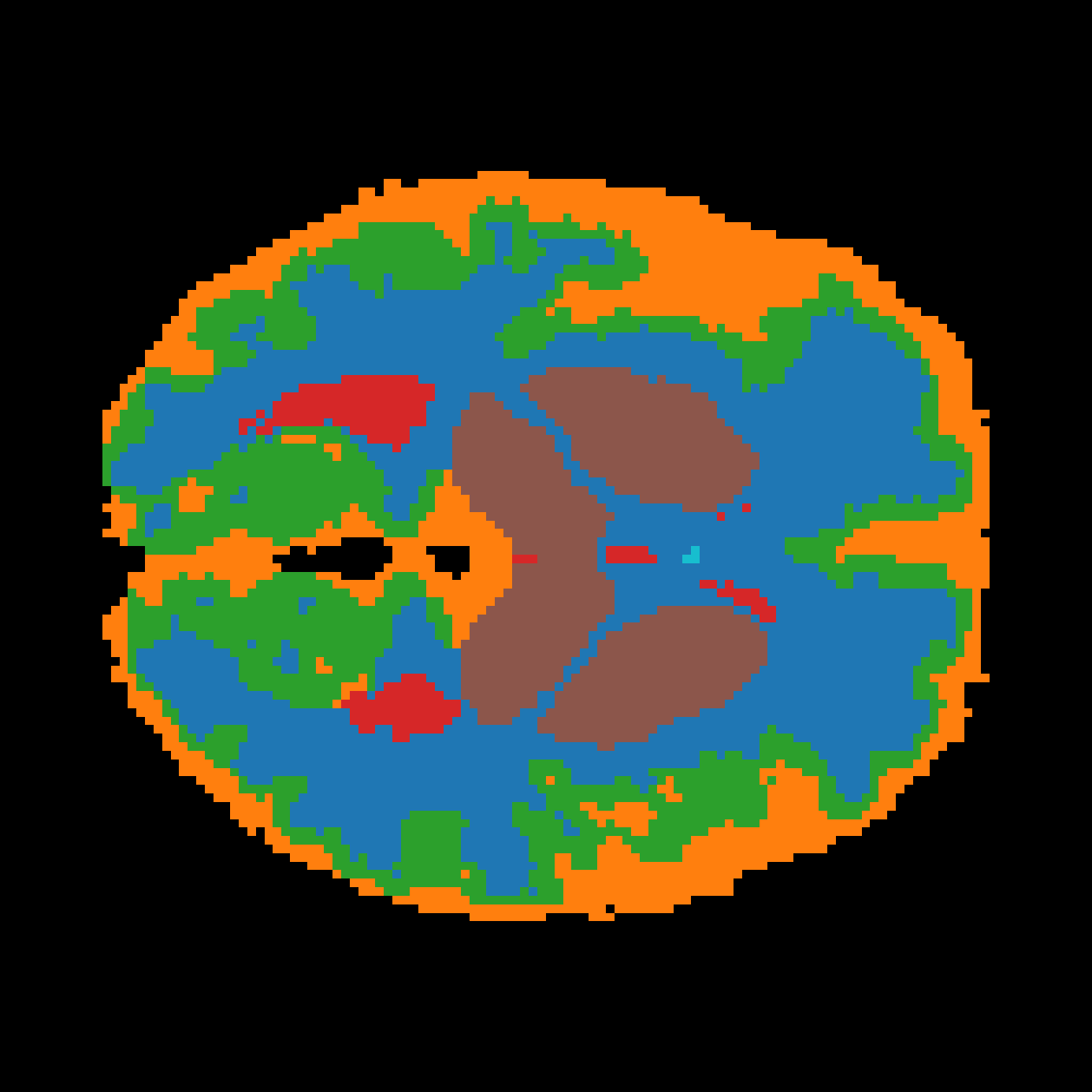}}}} \\

        \raisebox{0.25cm}{\textbf{\shortstack{Ours\\Disc}}} & \reflectbox{\rotatebox[origin=c]{90}{\includegraphics[height=1.1cm, trim=0.4cm 1.4cm 0.4cm 1.4cm, clip]{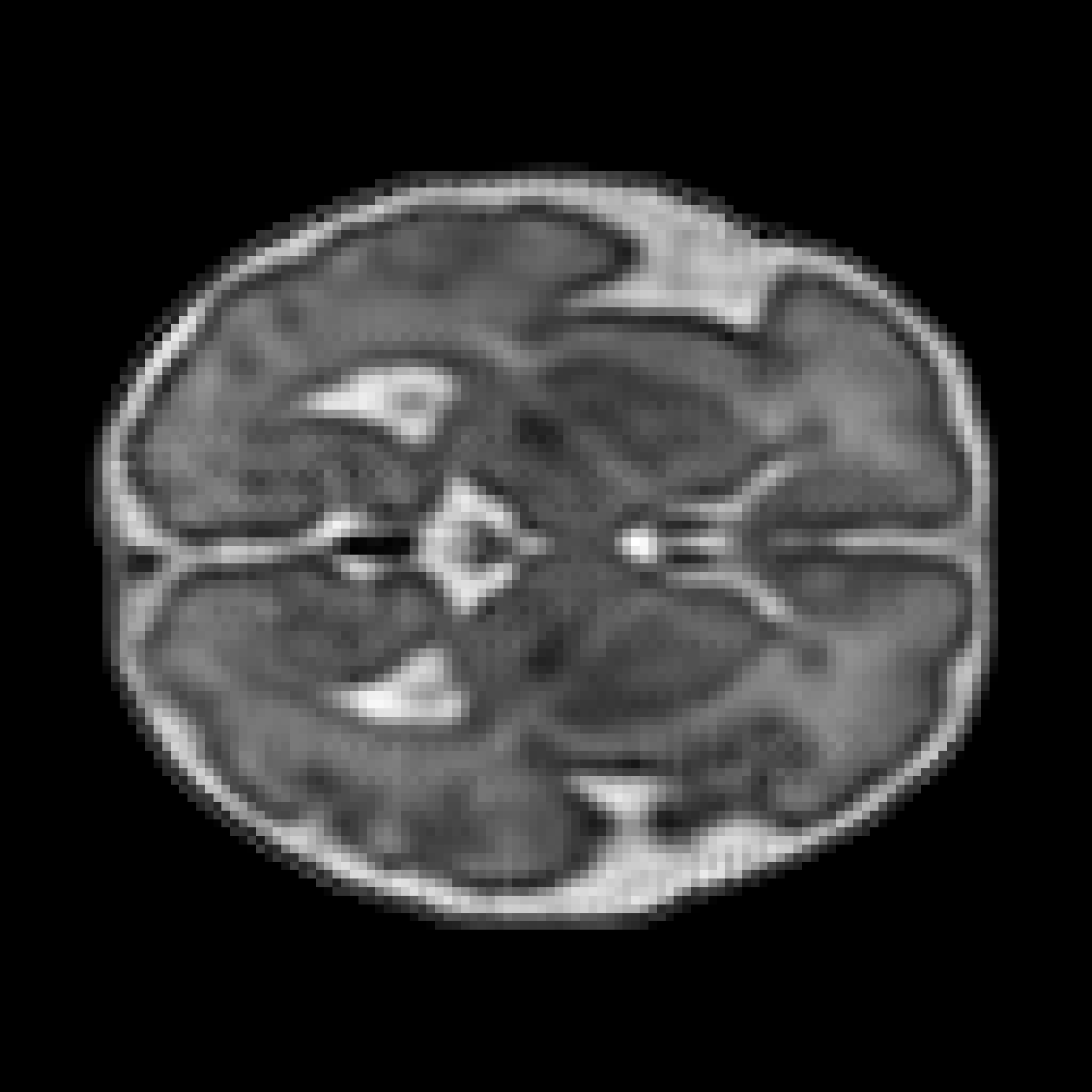}}} 
        & \reflectbox{\rotatebox[origin=c]{90}{\includegraphics[height=1.1cm, trim=0.4cm 1.4cm 0.4cm 1.4cm, clip]{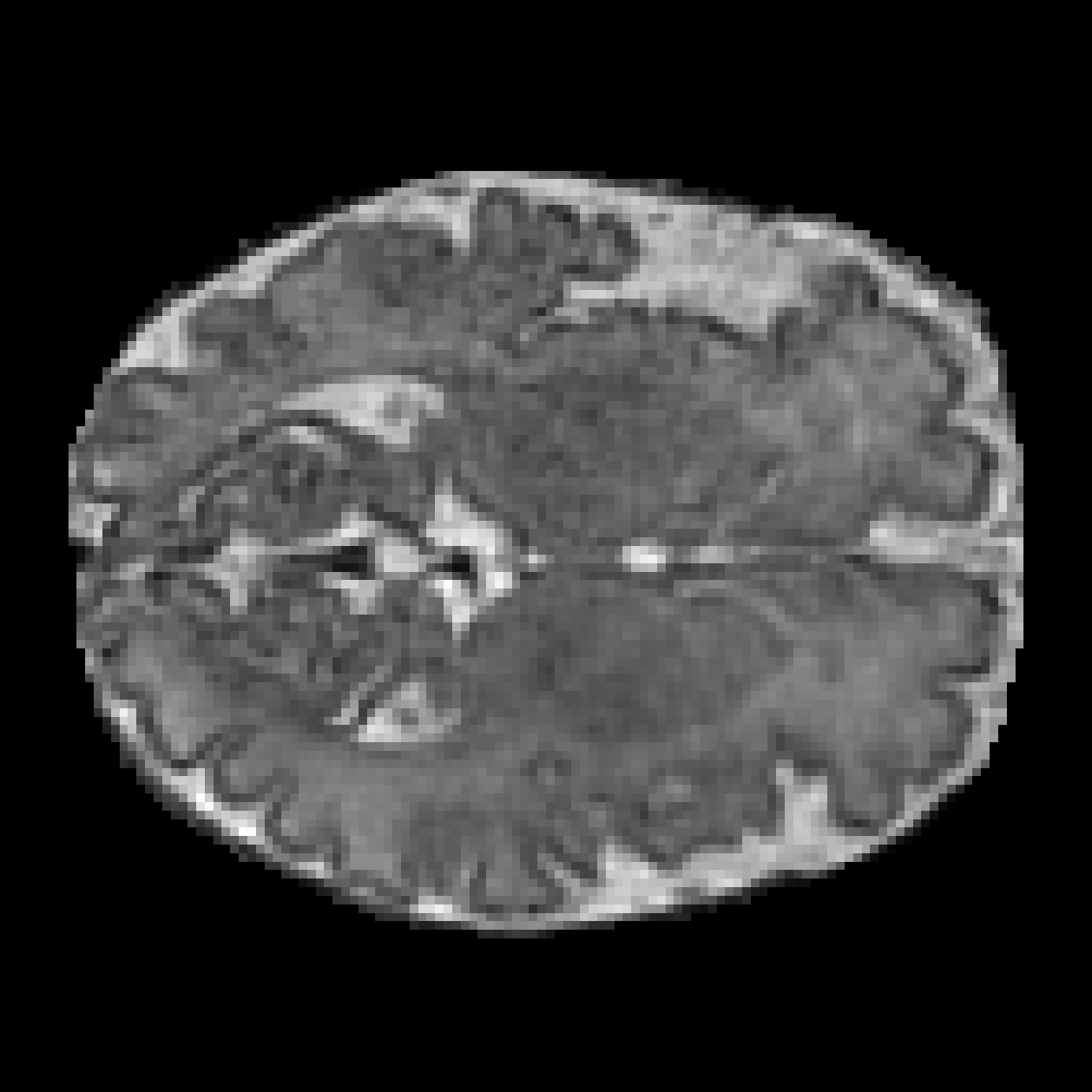}}}
        & \reflectbox{\rotatebox[origin=c]{90}{\includegraphics[height=1.1cm, trim=0.4cm 1.4cm 0.4cm 1.4cm, clip]{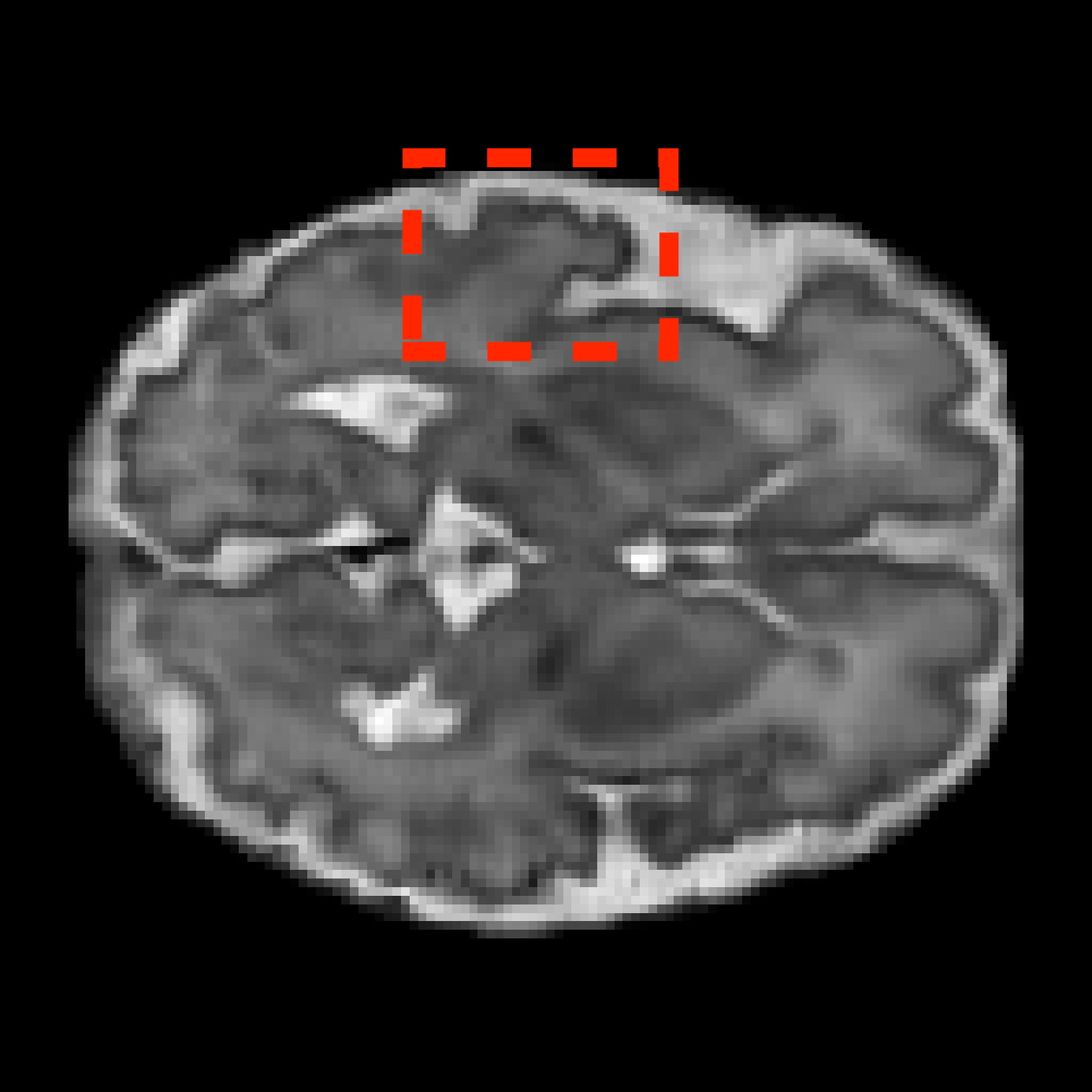}}}    
        & \raisebox{-0.095cm}{\reflectbox{\rotatebox[origin=c]{90}{\includegraphics[height=1.29cm, trim=0.4cm 1.4cm 0.4cm 1.4cm, clip]{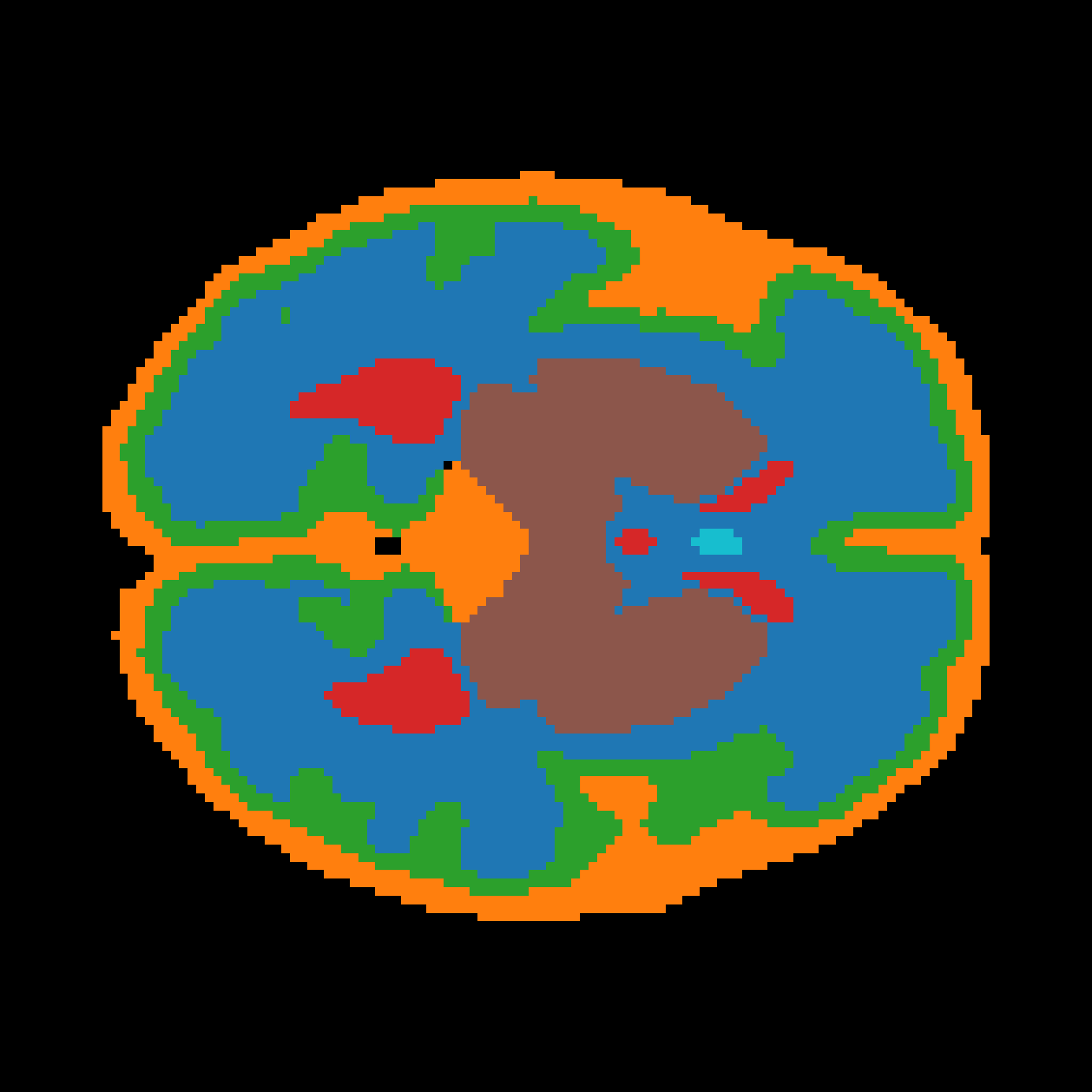}}}}
        & \raisebox{-0.095cm}{\reflectbox{\rotatebox[origin=c]{90}{\includegraphics[height=1.29cm, trim=0.4cm 1.4cm 0.4cm 1.4cm, clip]{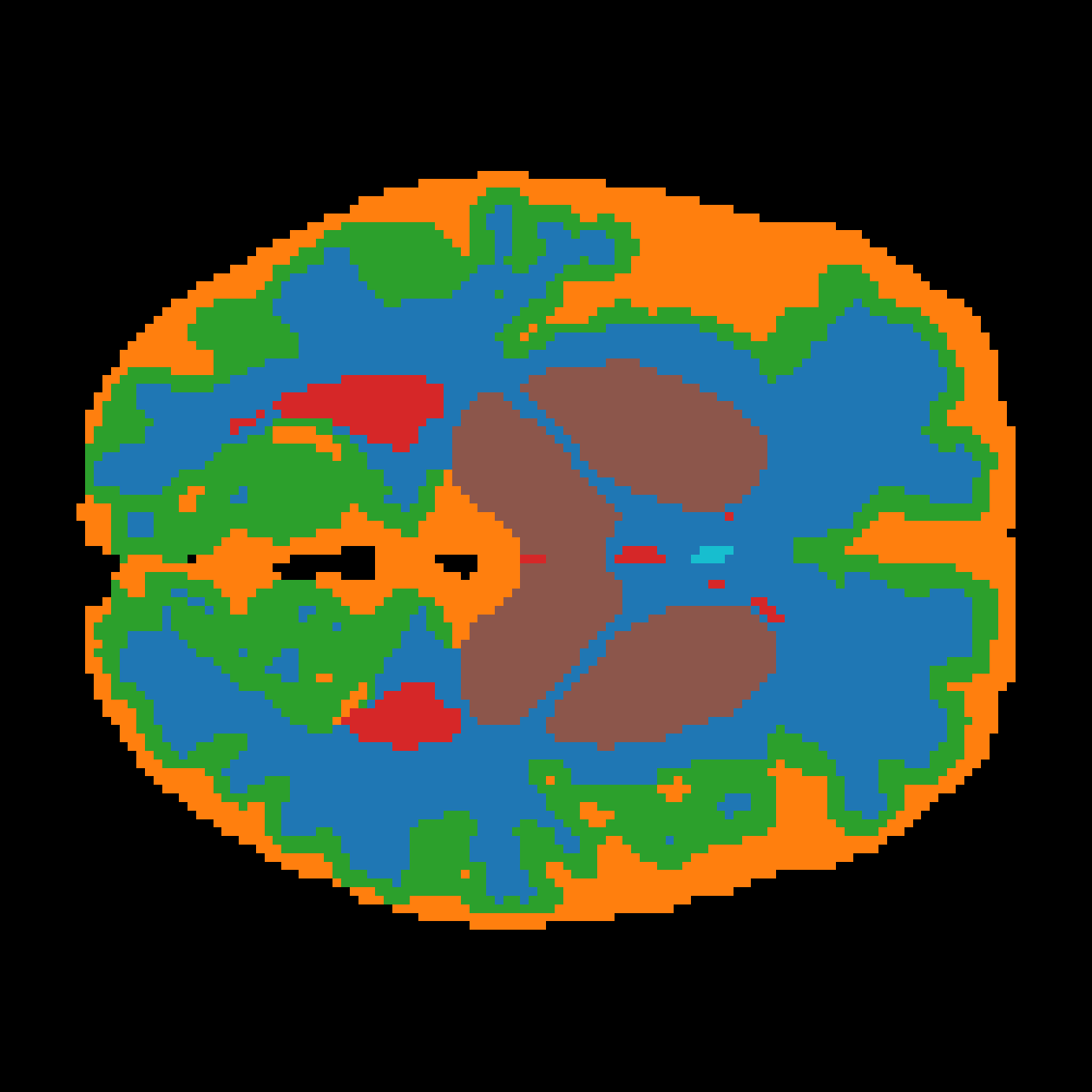}}}}
        & \raisebox{-0.095cm}{\reflectbox{\rotatebox[origin=c]{90}{\includegraphics[height=1.29cm, trim=0.4cm 1.4cm 0.4cm 1.4cm, clip]{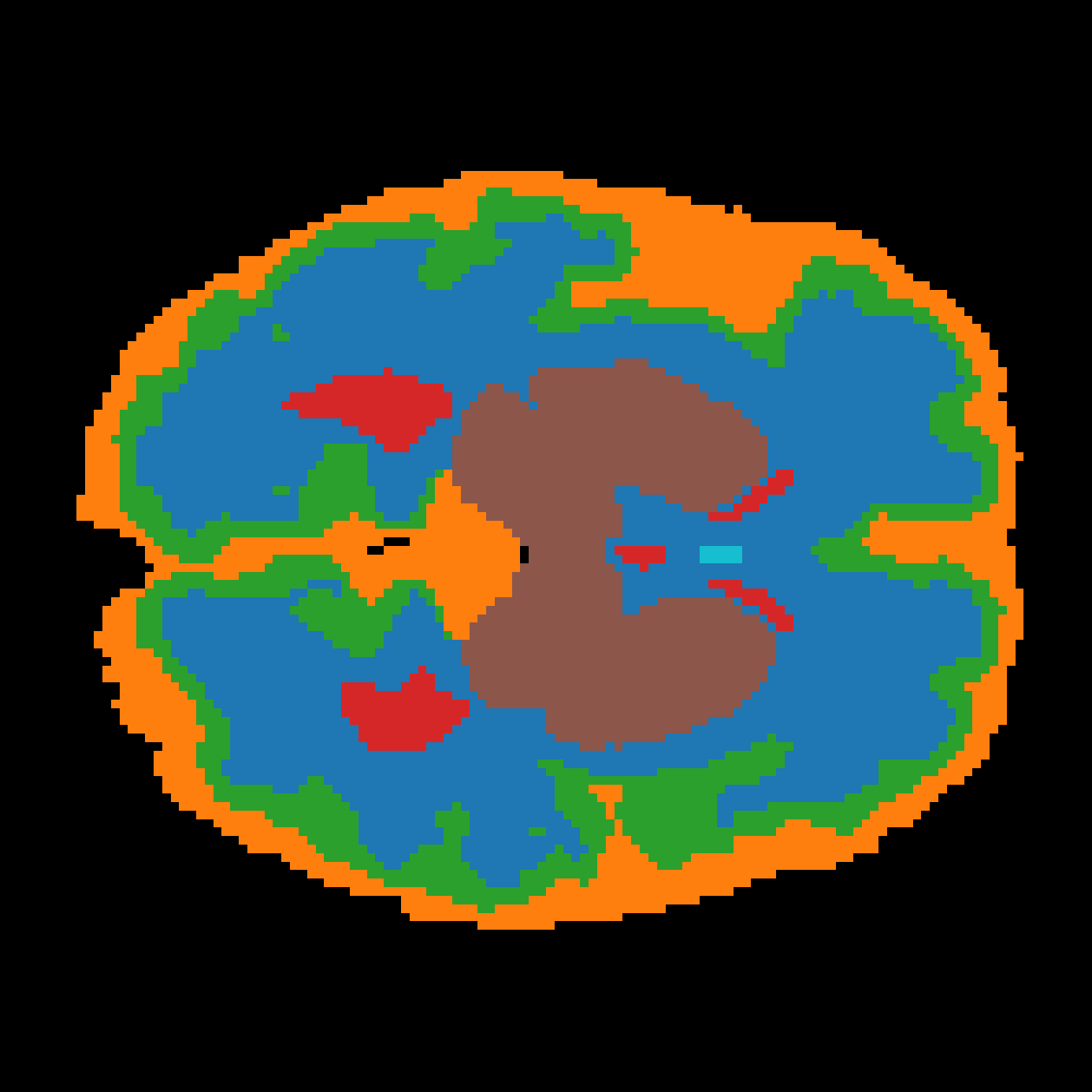}}}}
    \end{tabular}
    \end{minipage}%
    \hfill
    \begin{minipage}[t]{0.19\textwidth}
        \renewcommand{\arraystretch}{1.3} 
        \begin{tabular}{l@{}}
            \fcolorbox{black}{eCSF}{\color{eCSF}\rule{0.4cm}{0.2cm}}~eCSF \\
            \fcolorbox{black}{WM}{\color{WM}\rule{0.4cm}{0.2cm}}~Total WM \\
            \fcolorbox{black}{dGM}{\color{dGM}\rule{0.4cm}{0.2cm}}~Deep GM \\
            \fcolorbox{black}{Brainstem}{\color{Brainstem}\rule{0.4cm}{0.2cm}}~Brainstem \\
            \fcolorbox{black}{cGM}{\color{cGM}\rule{0.4cm}{0.2cm}}~cGM \\
            \fcolorbox{black}{Ventricles}{\color{Ventricles}\rule{0.4cm}{0.2cm}}~Ventricles
        \end{tabular}
    \end{minipage}
    \caption{a) Axial slices of the generated templates from the baseline approaches and our best-performing model. b) Visualization of the generated template (Atlas), test subject (Fix), warped atlas (Warp), and associated segmentation maps for a brain at 35\,GW. The red box highlights a region with pronounced individual cortical folding.}
    \label{fig:atlas_templates}
\end{figure}

\noindent \textbf{Segmentation Performance.} The nnUNet algorithm outperforms all evaluated methods, achieving a $96.9\pm0.0$ and the lowest HD of 0.99\,mm. Among atlas-based segmentation approaches, the ANTs SyN algorithm demonstrates the highest performance, with a DSC of $87.2\pm6.1$ and the lowest HD of 1.25\,mm. Our proposed method, \textit{Ours-Disc}, demonstrates competitive performance, achieving an average DSC of $86.3\pm5.2$, while maintaining a low average deformation norm and offering real-time segmentation. \\
Although nnUNet reaches high segmentation accuracy on neurotypical datasets, where image intensity profiles closely align with anatomical structure, its performance might degrade in pathological cases due to increased anatomical variability and limited training data. In contrast, \textit{Ours-Disc} offers greater adaptability to such challenging scenarios, as it can be retrained to generate condition-specific templates and perform segmentation jointly. Additionally, the incorporation of deformation fields facilitates spatial correspondence between template and subject, enabling interpretable anatomical comparisons. \\

\begin{figure}[!t]
    \hspace{0.4em}
    \begin{overpic}[width=0.5\linewidth]{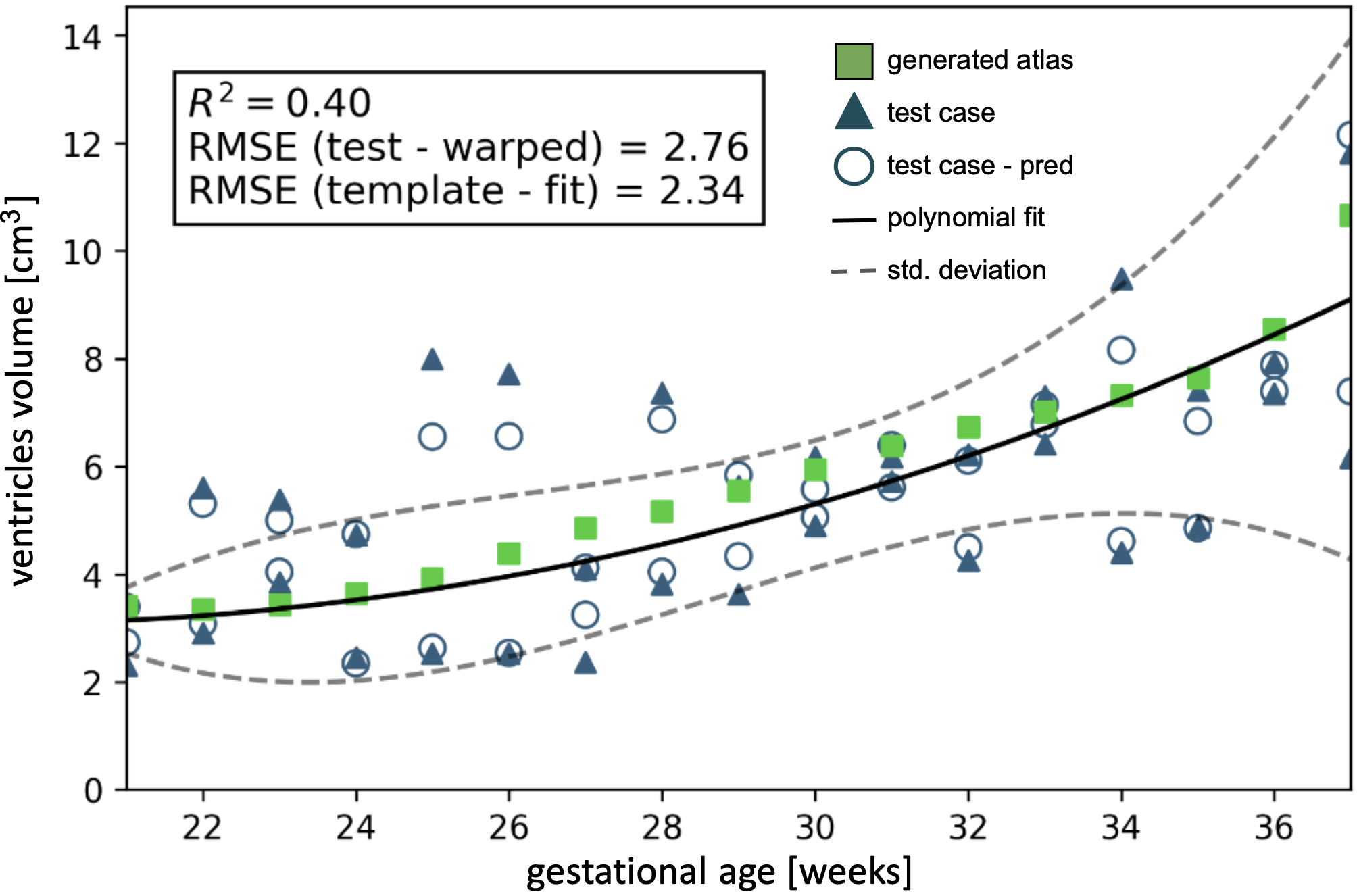}
        \put(0,72){\textbf{a)}}
    \end{overpic}
    \hspace{0.4em}
    \raisebox{0.1cm}{\begin{overpic}[width=0.41\linewidth]{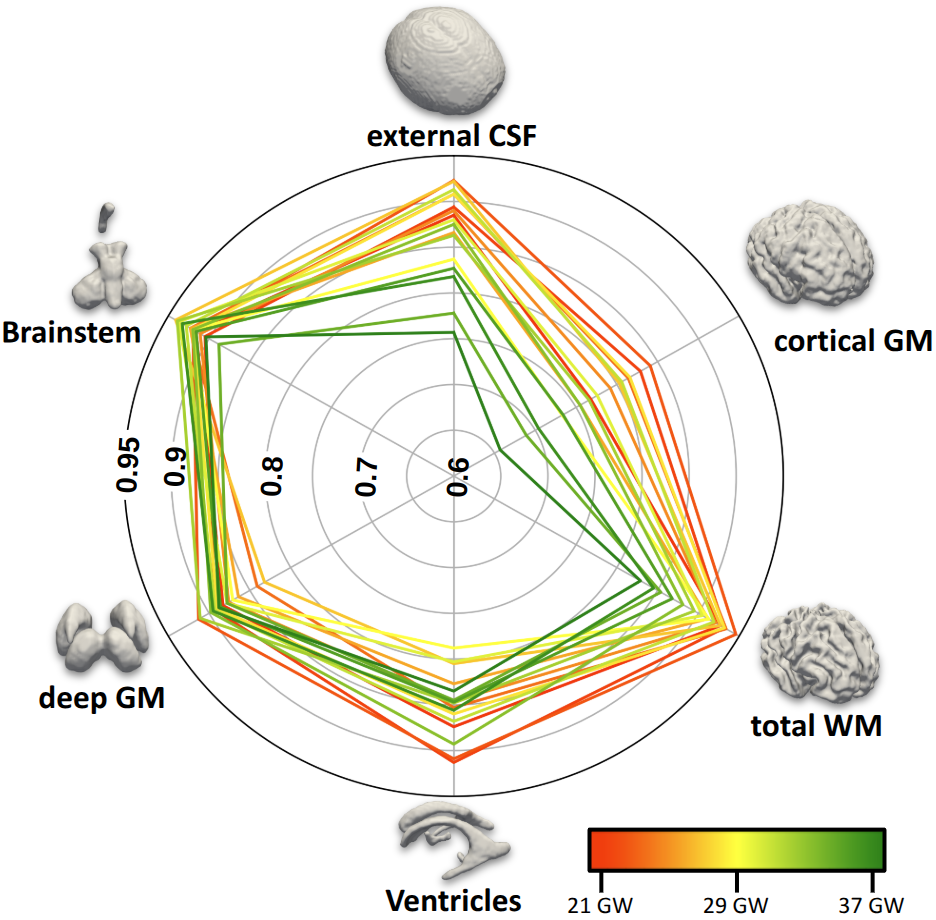}
        \put(-5,87){\textbf{b)}}
    \end{overpic}}
    \caption{a) Neurodevelopmental trajectory of ventricles volume for training dataset (solid line), generated atlas (squares), test labels (triangle), and predictions (circle). b) Radar plot illustrating segmentation performance across different GWs. Each axis represents an anatomical label and the distance indicates the DSC.}
    \label{fig:trajectory}
\end{figure}

\noindent Fig.~\ref{fig:trajectory} presents a radar plot illustrating the obtained DSC for six anatomical labels across GA for \textit{Ours-Disc}. The analysis highlights an imbalance in segmentation performance among brain structures and GWs. Specifically, the anatomical labels exhibit varying degrees of segmentation accuracy, influenced by factors such as the size and coherence of the structures. For instance, the segmented tissues include large anatomical labels (tWM), thin coherent structures (cGM), and thinly-split brain structures (Ven). Moreover, GA impacts segmentation performance (see Fig.~\ref{fig:trajectory}b). A notable decline in DSC can be observed for the cGM, which represents the outermost cortical layer. As GA increases, the DSC progressively decreases, starting at 83$\,\%$ in younger fetuses (<\,25\,GW), characterized by a relatively smooth cortical structure. In contrast, the DSC declines to 67$\,\%$ in older fetuses (>\,33\,GW), reflecting complex cortical developmental changes throughout gestation (see Fig.~\ref{fig:atlas_templates}). \\

\noindent \textbf{Registration Performance.} This aspect refers to the capacity to map existing knowledge (generated data) onto previously unseen subjects. Fig.~\ref{fig:atlas_templates}b illustrates the alignment between the atlas (Template - column 1) and a specific case from our test dataset at 35\,GW (Fix - column 2). The chosen case, being representative of subjects in the last trimester, displays individual gyri and sulci. These anatomical features must be accurately matched through the registration process. Our approach achieves a more precise alignment with individual anatomical deformations (see red box in Fig.~\ref{fig:atlas_templates}b). In comparison to \textit{Our-Reg}, the discriminator-guided registration (\textit{Our-Disc}) improves the framework’s outcomes, resulting in higher DSC values and lower HD, average deformation norm, and EFC (see Table~\ref{tab:metrics}).
\subsection{Neurodevelopmental Trajectories}
Segmentation maps enable volumetric quantification by multiplying the number of voxels per label with the corresponding voxel spacing. Utilizing a training dataset comprising 185 samples, we extracted volumes from six anatomical regions and modeled their trajectories across GA using polynomial regression. This approach establishes normative neurodevelopmental trajectories.
Fig.~\ref{fig:trajectory} illustrates the volumetry of the ventricular system of the training dataset, with the solid line representing the fitted trajectory. The volumetry of the ventricular atlas segmentation is denoted by the green square, demonstrating good alignment with the fit, indicating its consistency with expected developmental patterns (RMSE=2.34). Although, the test dataset, represented by blue triangles, displays deviations from the normative trend, the model's predictions (blue circles) generally follow the test labels (RMSE=2.76). This observation highlights that the framework captures developmental patterns in the generated template while matching the variability of individual development through the registration framework. Developmental patterns for the other anatomical regions are presented in Figure S1.

\section{Conclusion}
In this study, we introduced a continuous fetal brain atlas covering 21 to 37 weeks of gestation, developed using our in-house dataset containing 219 fetuses. The framework includes a registration approach enhanced by a conditional discriminator, ensuring anatomical accuracy in the generated templates, particularly during the third trimester. Furthermore, we demonstrated the framework's ability to adapt to previously unseen subjects exhibiting atypical neurodevelopment. In future work, we plan to expand the in-house dataset to include a broader age range and additional conditions. In addition, we will manually refine the segmentation maps and add more anatomical labels. The proposed architecture enables flexible extension to different tissue maps and modalities, mirroring the required flexibility in the clinical setup.  

\begin{credits}
\subsubsection{\ackname} 
The financial support by the Austrian Federal Ministry of Labour and Economy, the National Foundation for Research, Technology and Development and the Christian Doppler Research Association is gratefully acknowledged.
\subsubsection{\discintname}
The authors have no competing interests to declare that are relevant to the content of this article. 
\end{credits}


\newpage
\thispagestyle{empty}
\section*{Supplementary Material}
\renewcommand{\thefigure}{S\arabic{figure}}
\setcounter{figure}{0}  

\begin{figure}
    \centering
    \includegraphics[width=\linewidth]{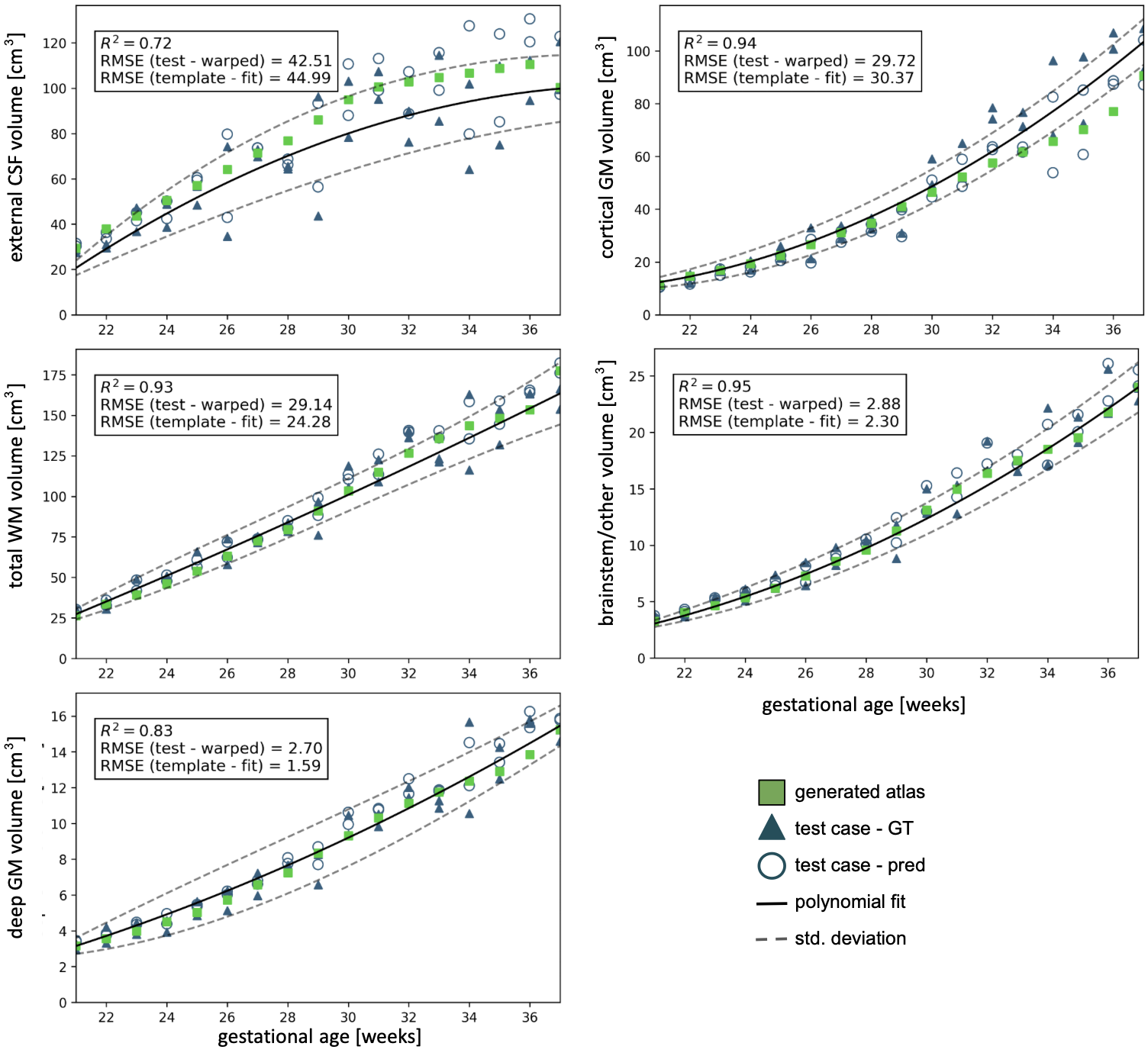}
    \caption{Volumetric trajectories of eCSF, cGM, total WM, brainstem, and deep GM, (from top left to bottom right) from 21 to 37\,GW. The trajectory of the training data, including its standard deviation, is modeled using a polynomial fit. The brain region-specific volumes are shown for the test labels ($\bigtriangleup$) and the predicted values ($\bigcirc$).}
    \label{fig:supp}
\end{figure}

\end{document}